\documentclass[%
 reprint,
nofootinbib,
 amsmath,amssymb,
 aps,
]{revtex4-2}
\usepackage{graphicx}
\usepackage{dcolumn}
\usepackage{bm}
\usepackage{hyperref}

\usepackage{orcidlink}

\usepackage{physics}

\usepackage[normalem]{ulem} 

\usepackage{aas_macros}

\usepackage{subfigure}

\usepackage{booktabs} 
\usepackage{siunitx}  
\usepackage{caption} 
\usepackage{enumitem}

\usepackage{multirow} 

\usepackage{tikz}
\usepackage{xcolor}
\definecolor{RED}{rgb}{1,0,0} 

\usepackage{ragged2e} 

\usepackage{ulem} 

\usepackage[utf8]{inputenc}
\newcommand{\ioka}[1]{\textcolor{black}{#1}}

\newcommand{\Nishiura}[1]{\textcolor{black}{#1}}

\newcommand{\nIshiura}[1]{\textcolor{black}{#1}}

\newcommand{\niShiura}[1]{\textcolor{black}{#1}}

\newcommand{\nisHiura}[1]{\textcolor{black}{#1}}

\newcommand{\rei}[1]{}

\newcommand{\Rei}[1]{}

\newcommand{\rEi}[1]{}

\begin{document}

\preprint{APS/123-QED}

\title{Unified kinetic theory of induced scattering: Compton, Brillouin, and Raman processes in magnetized electron and positron pair plasma}

\author{Rei Nishiura\orcidlink{0009-0003-8209-5030}}
 \email{nishiura@tap.scphys.kyoto-u.ac.jp}
 \affiliation{%
 Department of Physics, Kyoto University, Kyoto 606-8502, Japan}%
 \author{Shoma F. Kamijima\orcidlink{0000-0002-4821-170X}}%
 \email{shoma.kamijima@yukawa.kyoto-u.ac.jp}
\affiliation{%
 Center for Gravitational Physics and Quantum Information, 
 Yukawa Institute for Theoretical Physics, Kyoto University, Kyoto 606-8502, Japan}%
 \author{Kunihito Ioka\orcidlink{0000-0002-3517-1956}}%
 \email{kunihito.ioka@yukawa.kyoto-u.ac.jp}
\affiliation{%
 Center for Gravitational Physics and Quantum Information, 
 Yukawa Institute for Theoretical Physics, Kyoto University, Kyoto 606-8502, Japan}%

%
%
\date{\today}

\begin{abstract}
We \rei{present}\ioka{extend} a unified theoretical framework for induced (stimulated) scattering--parametric instabilities of electromagnetic waves, including induced Compton, stimulated Brillouin, and stimulated Raman scattering (SRS)--in strongly magnetized electron–positron ($e^{\pm}$) pair plasma. By solving the dispersion relations derived from kinetic theory, taking into account the ponderomotive force due to the beat of incident and scattered waves, we obtain analytical expressions for the linear growth rates of the ordinary, neutral, and charged modes of density fluctuations. Our results clarify which type of scattering dominates under different thermal coupling, resonance, and density conditions. In strong magnetic fields, scattering of perpendicularly polarized waves is generally suppressed, but by different powers of the cyclotron frequency. Moreover, SRS, which is forbidden in unmagnetized $e^{\pm}$ pair plasma, becomes possible in the charged mode. This framework enables a comprehensive evaluation of induced scattering in extreme astrophysical and laboratory plasma, such as fast radio burst (FRB) emission and propagation in magnetar magnetospheres.
\end{abstract}

\maketitle


\section{INTRODUCTION} 
Nonlinear interactions between electromagnetic (EM) waves and plasma serve as the foundation for many phenomena in both astrophysical and laboratory environments. Among nonlinear interactions, parametric instabilities are processes that coherently excite secondary waves. Since the energy and momentum of the EM pump wave are transferred to the secondary (daughter) waves, the process can be regarded as scattering in plasma. In addition, since some of the daughter waves dissipate, this also leads to damping of the pump wave. Such processes occur not only in astrophysical contexts—including the Sun~\citep{1963SPhD....7..988G,1966PhFl....9.1483B,1972JPlPh...8..197B,1978ApJ...224.1013D,1990JGR....9510525I,1993JGR....9813247J,1994JGR....9923431H,2001A&A...367..705D,2006PhPl...13l4501N,2015JPlPh..81a3202D,2017ApJ...842...63S,2022RvMPP...6...22N}, pulsars~\citep{1973PhFl...16.1480M,1976MNRAS.174...59B,1978MNRAS.185..297W,1982MNRAS.200..881W,1996AstL...22..399L}, and fast radio bursts (FRBs)~\citep{2008ApJ...682.1443L,2023MNRAS.522.2133I,2024PhRvE.110a5205I}—but also in laser plasma interactions~\citep{1973PhFl...16.1522K,1974PhRvL..33..209M,1975PhFl...18.1002F,1994PhPl....1.1626T,1996PhRvL..77.2483D} and free electron laser \citep{1979PhFl...22.1089K,1980PhFl...23.2376F} in laboratory plasma.
\begin{figure*}
\centering
\includegraphics[width=\textwidth]{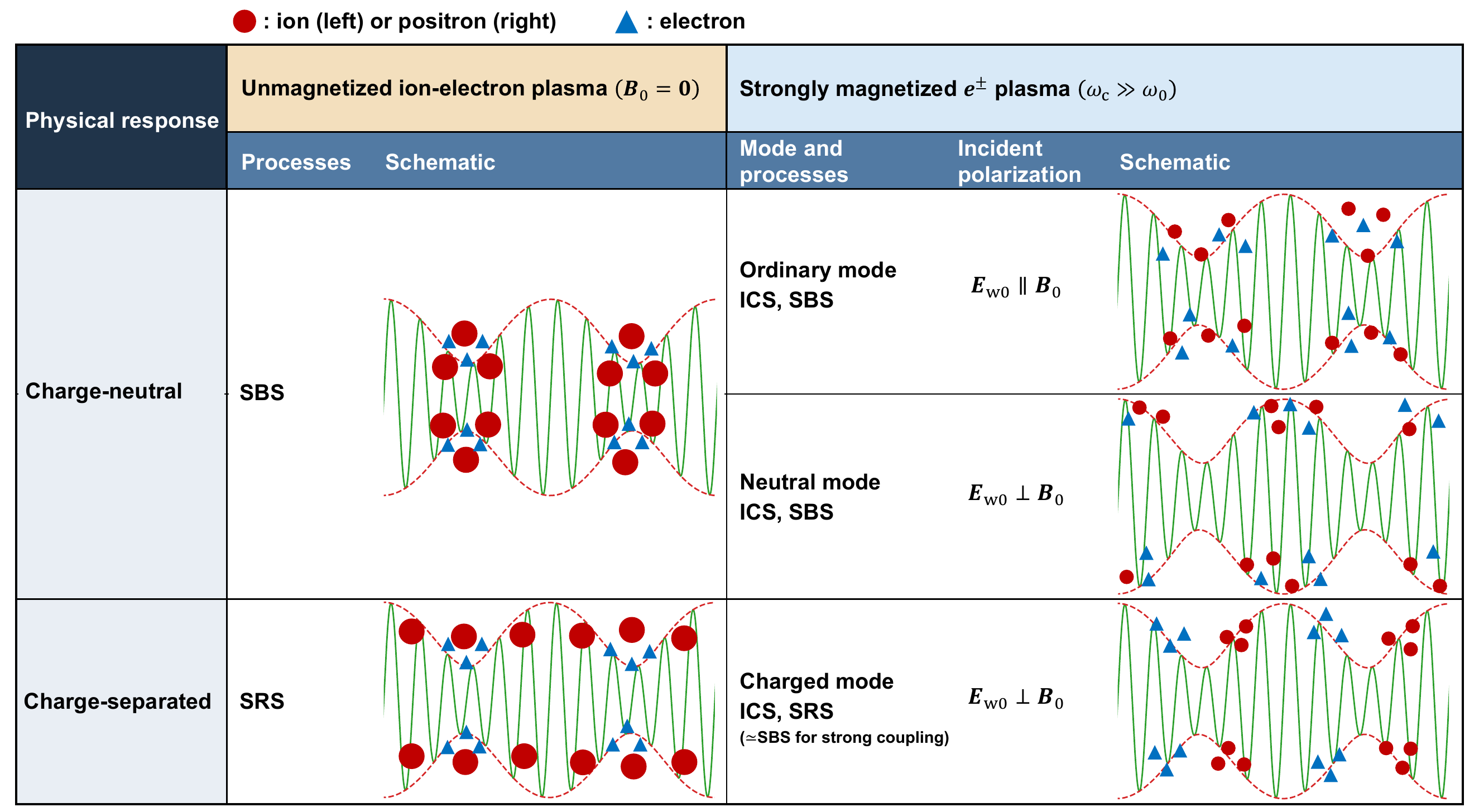}
\caption{\justifying
Schematic comparison between induced scattering in unmagnetized ion-electron plasma and in strongly magnetized $e^\pm$ plasma. ICS denotes induced Compton scattering, SBS denotes stimulated Brillouin scattering, and SRS denotes stimulated Raman scattering. The green curves represent the beat between the incident and scattered waves, and the red curves indicate the corresponding envelope. The horizontal positions where particles are concentrated indicate phase differences in the density response. The left panels assume no background magnetic field, whereas the right panels assume a strong background magnetic field. The row-by-row analogy refers only to the presence or absence of charge separation in the density response, not to identical wave modes or particle motions.
}
\label{fig:parametric_instability}
\end{figure*}

FRBs, first discovered in 2007, are the brightest radio transient in the Universe~\citep{2007Sci...318..777L,2013Sci...341...53T,2019A&ARv..27....4P}. Most FRBs are extragalactic, and their origin and emission mechanisms remain unresolved. However, in 2020, an FRB 20200428 was observed in coincidence with X-ray bursts from the Galactic magnetar SGR 1935+2154, providing compelling evidence for a magnetar origin~\citep{2020Natur.587...54C,2020Natur.587...59B,2020ATel13687....1Z,2020ApJ...898L..29M,2020ATel13686....1T,2020ATel13688....1R}. Since the observed properties of FRBs encode the effects of distant galaxies and the intergalactic medium, they have also applied as cosmological probes~\citep{2003ApJ...598L..79I,2004MNRAS.348..999I,2020Natur.581..391M}.

The theoretical challenges of FRBs can be broadly divided into two aspects, in both of which parametric instabilities can play critical roles: (1) explicating the physical conditions and processes for FRB generation, and (2) clarifying how an FRB, once generated, propagates from the source region to the observer~\citep{2021ApJ...922L...7B,2022PhRvL.128y5003B,2022MNRAS.515.2020Q,2022arXiv221013506C,2023ApJ...959...34B,2024ApJ...975..223B,2024ApJ...975..226H,2024PhRvD.109d3048N,2024A&A...690A.332S,2024PhRvE.110a5205I,2025PhRvD.111f3055N}. Regarding (1), emission models can be classified according to the generation region. The ``pulsar-like magnetosphere" model proposes that FRBs are generated within the strongly magnetized region up to several thousand kilometers from the magnetar surface~\citep{2014PhRvD..89j3009K,2016MNRAS.457..232C,2016MNRAS.462..941L,2017MNRAS.468.2726K,2017ApJ...836L..32Z,2018MNRAS.477.2470L,2018A&A...613A..61G,2018ApJ...868...31Y,2020ApJ...893L..26I,2020MNRAS.494.2385K,2021MNRAS.508L..32C,2021ApJ...922..166L,2024ApJ...972..124Q}. In contrast, the ``gamma-ray burst-like shock" model attributes FRB production to shocks formed when magnetar flares or outflows collide with the ambient medium~\citep{2014MNRAS.442L...9L,2016MNRAS.461.1498M,2017ApJ...842...34W,2020MNRAS.494.4627M,2019MNRAS.485.4091M,2020ApJ...896..142B,2021MNRAS.500.2704Y}. These scenarios differ by several orders of magnitude in the emission radius. Recent work has also suggested the outer magnetosphere regions, such as reconnection at the magnetospheric boundary or shocks of nonlinear waves~\citep{2020ApJ...897....1L,2022arXiv221013506C,2023ApJ...959...34B,2025PhRvL.134c5201V}.

From the perspective of emission mechanisms, FRB models are typically categorized into antenna and maser mechanisms~\citep{1969Ap&SS...4..464G,1975ARA&A..13..511G,1978ApJ...225..557M,2017RvMPP...1....5M,2018MNRAS.477.2470L,2023RvMP...95c5005Z}. Antenna models assume that coherent charge bunches—formed by some physical process—emit via spontaneous emission, as demonstrated by coherent curvature radiation~\citep{2014PhRvD..89j3009K,2016MNRAS.457..232C,2017MNRAS.468.2726K,2018MNRAS.477.2470L,2018A&A...613A..61G,2020MNRAS.494.2385K,2021MNRAS.508L..32C}. Maser models, in contrast, invoke stimulated emission as the origin of the high brightness, such as synchrotron maser emission near shock fronts~\citep{1992ApJ...390..454H,2014MNRAS.442L...9L,2016MNRAS.461.1498M,2017ApJ...842...34W,2020MNRAS.494.4627M,2019MNRAS.485.4091M,2020ApJ...896..142B,2021MNRAS.500.2704Y,2024PhRvL.132c5201I} or parametric instabilities~\citep{2021ApJ...922..166L,2024PhRvE.110a5205I,2025PhRvD.111f3055N}.

For (2), the propagation of FRBs is strongly influenced by wave amplitude, frequency, magnetic field strength, and plasma density. Theoretically, the propagation environment can be divided into three regimes. In the inner magnetosphere and near the neutron star surface, the wave amplitude is small ($a_{\mathrm{e}}\omega_0/\omega_{\mathrm{c}}\ll1$), and the plasma response is linear (see Eq.~\eqref{eq:non_rela_neutral_and_charged_Brillouin}). In the outer magnetosphere and near the light cylinder, nonlinear effects become significant as the amplitude exceeds unity ($a_{\mathrm{e}}\omega_0/\omega_{\mathrm{c}}>1$). Further out, the magnetic field of the magnetar declines and the strength parameter becomes small ($a_{\mathrm{e}}\ll1$), returning to a linear response (see Eq.~\eqref{eq:non_rela_ordinary_Brillouin}). Of particular importance is whether FRBs can escape the magnetosphere without severe damping from parametric instabilities with $e^\pm$ pair plasma. Recent observations suggest that FRB emission may occur within the magnetosphere~\citep{2022Natur.607..256C,2022NatAs...6..393N,2025Natur.637...48N}. Clarifying wave–plasma interactions in each propagation regime remains a central theoretical problem.

This work aims to provide a unified theoretical framework for induced scattering in strongly magnetized $e^\pm$ pair plasma (i.e., inside the magnetar magnetosphere). Here, induced scattering collectively refers to induced Compton scattering (ICS), stimulated Brillouin scattering (SBS), and stimulated Raman scattering (SRS). \ioka{As schematically summarized in Fig.~\ref{fig:parametric_instability}, }SBS and SRS are well-established parametric instabilities in electron–ion plasma~\citep{1974PhFl...17..778D,1979PhFl...22.1115C}. \ioka{In this classification, they are distinguished by the longitudinal plasma response resonant with the beat wave, namely an ion acoustic wave for SBS and a Langmuir wave for SRS.} In contrast, ICS is driven by Landau resonance between the beat of incident and scattered waves and the thermal motion of particles \footnote{ICS is sometimes referred to as the kinetic effect of SBS \citep{2017PhRvE..96e3204S}.}.



Previous studies on induced scattering in $e^\pm$ pair plasma have primarily been based on unmagnetized plasma theory \citep{2016PhRvL.116a5004E,2017PhRvE..96e3204S,2023MNRAS.522.2133I}. In such plasma, it has been considered that density fluctuations driven by the beat of the incident and scattered waves do not generate electrostatic waves. As a result, SRS and certain modes of modulation instabilities, which are excited in ion-electron plasma, are absent in $e^\pm$ pair plasma. It has also been pointed out that, as the amplitude of the incident wave increases, there is a continuous transition from ICS to SBS.

Parametric instabilities in magnetized $e^\pm$ pair plasma have been investigated through both theoretical and numerical approaches. Previous works typically assumed a circularly polarized Alfvén wave as the pump and employed one-fluid, two-fluid, or kinetic models \citep{1978A&A....66..139S,1998PhRvE..57..994M,1999PhRvE..59.4552M,2003PhRvE..67d6406M,2006EP&S...58.1213M,2012PhPl...19h2104L,2014PhPl...21c2102L,2014NPGeo..21..217M,2024PhRvE.110a5205I}. These studies identified nonlinear couplings among various modes, including decays into Langmuir or acoustic waves, modulation instability, and beat instability. However, most of these works relied on numerical eigenvalue analyses or Particle-in-Cell (PIC) simulations under specific parameter regimes, and there remains a need for a comprehensive analytic understanding of the linear growth rates and mode competition.

\citet{2025PhRvD.111f3055N} developed, for the first time, a theoretical framework for induced scattering in strongly magnetized $e^\pm$ pair plasma, and derived the linear growth rate of ICS analytically. Unlike previous studies, this framework explicitly formulates the coupling mechanism between the incident and scattered waves in terms of the ponderomotive force in a uniform magnetic field \citep{1968CzJPh..18.1280K,1977PhRvL..39..402C,1981PhFl...24.1238C,1981PhRvL..46..240H,10.1063/1.864196,1996GeoRL..23..327L}. In this formalism, the polarization of the pump wave can be chosen arbitrarily. This approach also enables, for the first time, the classification of density fluctuation modes into ordinary, neutral, and charged modes\ioka{, as schematically illustrated in Fig.~\ref{fig:parametric_instability}}. The ordinary mode is excited by the component of the incident EM wave electric field parallel to the background magnetic field. In contrast, the neutral and charged modes are excited by the component of the incident EM wave electric field perpendicular to the background field. In particular, \ioka{ as shown in Fig.~\ref{fig:parametric_instability}, }the charged mode is characterized by charge separation between electrons and positrons and excitation of electrostatic waves due to the ponderomotive force, which is fundamentally different from the unmagnetized case.

\ioka{Building on that framework, the present paper derives the SBS and SRS branches and provides the first systematic analytic classification of mode competition among ICS, SBS, and SRS. The companion PIC papers validate representative linear growth rates derived in the present paper and follow the nonlinear evolution beyond the scope of the present work \citep{2026arXiv260101169K,2026arXiv260118865N}.}

The structure of this paper is as follows. Sec.~\ref{sec:rironntekiwakugumi} provides an overview of the fundamental equations and dispersion relations for induced scattering in strongly magnetized $e^\pm$ pair plasma, as derived by \citet{2025PhRvD.111f3055N}. In Sec.~\ref{sec:kakumo-donokaiseki}, we analytically derive the linear growth rates and competition criteria for induced scattering in each instability mode—namely, the ordinary, neutral, and charged modes—and compare these results with direct numerical solutions of the dispersion relations. Sec.~\ref{sec:physical_interpretation} discusses the physical interpretation of induced scattering for each mode. Sec.~\ref{subsec:broadband_Brillouin} addresses the growth rates in the case of broadband incident waves. Finally, the Appendix provides detailed derivations of the linear growth rates for induced scattering in each instability mode.

The following conventions are used throughout this paper:
\begin{enumerate}[label=(\roman*)]
    \item The Centimeter-Gram-Second Gauss (CGS-Gauss) system of units is consistently employed.
    \item The italic symbol $e$ represents the magnitude of the electron charge, while the roman type $\mathrm{e}$ denotes the base of the natural logarithm, $\mathrm{exp}(1)$.
    \item \Nishiura{A hat over a vector symbol (e.g., $\hat{\bm{k}}$) indicates a unit vector in the direction of that vector.}
    \item \nIshiura{Multiple $\pm$ or $\mp$ signs are taken in the same order.}
\end{enumerate}

\section{Theoretical Framework}
\label{sec:rironntekiwakugumi}
Induced scattering in strongly magnetized $e^\pm$ pair plasma can be derived within a unified framework. In this paper, we treat not only ICS but also SBS and SRS in a unified manner, by starting from a common dispersion relation and calculating the imaginary part of the angular frequency with different approximations for each process.

\subsection{Basic Equations}
\label{subsec:basic_equations_Brillouin}
In this study, the analysis of induced scattering is performed under the following assumptions:
\begin{enumerate}[label=(\roman*)]
    \item The background magnetic field $\bm{B}_0$ is assumed to be spatially uniform and oriented along the $x$-axis, i.e., $\bm{B}_0 = (B_0, 0, 0)$.

    \item \rei{Both the incident and the scattered wave are assumed to be monochromatic. The incident wave is assumed to be linearly polarized, with the vector potential $\bm{A}_0 = A_0 \bm{\epsilon}_0$ characterized by wavevector $\bm{k}_0$ and angular frequency $\omega_0$.\footnote{For the treatment of a circularly polarized incident wave, see Appendix~\ref{ap:linear_to_circular}.} The scattered wave, $\bm{A}_1=A_1\bm{\epsilon}_1$, is characterized by $\bm{k}_1$ and $\omega_1$. Here, $\bm{\epsilon}_0$ and $\bm{\epsilon}_1$ denote the unit vectors specifying the polarization directions of the incident and scattered wave electric fields, respectively. The case of a broadband incident wave is discussed separately in Sec.~\ref{subsec:broadband_Brillouin}.}\nisHiura{Both the incident and the scattered wave generated through induced scattering are assumed to be monochromatic. 
The electromagnetic wave ($i=0,1$ for the incident and scattered waves, respectively) is described by the vector potential
\begin{equation}
\bm{A}_{\mathrm{w}i} \equiv \bm{A}_i \,
  \mathrm{e}^{\mathrm{i}(\bm{k}_i \cdot \bm{r} - \omega_i t)} + \text{c.c.},
  \label{eq:A_wave_general}
\end{equation}
where $\bm{k}_i$ and $\omega_i$ are the wave vector and angular frequency, respectively. 
The complex amplitude is written as 
$\bm{A}_i = A_i \bm{\epsilon}_i \mathrm{e}^{\mathrm{i}\psi_i}$, 
with real magnitude $A_i$, polarization unit vector $\bm{\epsilon}_i$, and initial phase $\psi_i$. 
Here, $i=0$ denotes the incident wave and $i=1$ the scattered wave. 
Throughout this paper the incident wave is assumed to be linearly polarized; 
the case of a circularly polarized incident wave is discussed separately in Appendix~\ref{ap:linear_to_circular}. 
The case of a broadband incident wave is discussed separately in Sec.~\ref{subsec:broadband_Brillouin}.
}

    \item Both the incident and scattered waves are considered transverse waves, while the density fluctuations are assumed to be longitudinal waves.\footnote{Under this assumption, phenomena such as the instability in which a fast magnetosonic wave decays into Alfvén waves, or two-plasmon decay, where the incident wave decays into two Langmuir waves, cannot be described.} In this case, the following relations hold:
    \begin{equation}
        \bm{k}_0 \cdot \bm{A}_0 = 0, \quad \bm{k}_1 \cdot \bm{A}_1 = 0,
        \label{eq:transverse_condition_Brillouin}
    \end{equation}
    \begin{equation}
        \widetilde{\bm{E}}(\bm{k}, \omega) = -\frac{4 \pi \mathrm{i}}{k^2} \bm{k} \, \widetilde{\rho}(\bm{k}, \omega).
        \label{eq:longitudinal_condition_Brillouin}
    \end{equation}
    Here, $\widetilde{\bm{E}}$ and $\widetilde{\rho}$ are the Fourier coefficients of the electric field and charge density \Nishiura{of the density fluctuations}, respectively. Eq.~\eqref{eq:longitudinal_condition_Brillouin} is derived from the Fourier component of the Maxwell equation, $\nabla\cdot\bm{E}=4\pi\rho$. The Fourier transform for a monochromatic quantity $X(\bm{r}, t)$ is defined as follows:
    \begin{equation}
        X(\bm{r}, t) = \mathrm{e}^{\mathrm{i}(\bm{k} \cdot \bm{r} - \omega t)} \widetilde{X}(\bm{k}, \omega) + \text{c.c.}
        \label{eq:Fourier_transformation_definition_Brillouin}
    \end{equation}

    \item The magnetic field of the incident EM wave is assumed to be sufficiently small compared to the background magnetic field, i.e., $|\delta\bm{B}| \ll |\bm{B}_0|$ (see Eqs. \eqref{eq:non_rela_neutral_and_charged_Brillouin} and \eqref{eq:definition_of_eta}).

    \item The charged particles are assumed to be nonrelativistic. As will be shown later, the nonrelativistic conditions are given by Eq.~\eqref{eq:non_rela_ordinary_Brillouin} for the ordinary mode instability and by Eq.~\eqref{eq:non_rela_neutral_and_charged_Brillouin} for the neutral and charged modes.

    \item The one-particle distribution functions of electrons and positrons, $f_{0\pm}$, are assumed to be spatially uniform and to depend only on velocity before scattering, that is, $f_{0\pm} = f_{0\pm}(\bm{v})$. The uniform electron and positron densities are equal,
    \begin{equation}
        n_{\mathrm{e}0} \equiv n_{0+} = n_{0-}.
        \label{eq:uniform_density_Brillouin}
    \end{equation}
\end{enumerate}

To obtain the linear growth rate of induced scattering, one must derive the dispersion relation for the scattered wave and compute the imaginary part of its angular frequency. The incident and scattered waves are assumed to obey the following wave equation:
\begin{equation}
\frac{\partial^{2} \bm{A}}{\partial t^{2}} - c^{2} \Delta \bm{A} = 4 \pi c \bm{j}.
\label{eq:wave_eq_vector_Brillouin}
\end{equation}
The Lorenz gauge condition, $\nabla \cdot \bm{A} + (1/c)(\partial \phi_{\text{e}} / \partial t) = 0$, has been applied to this wave equation. Furthermore, under the assumption that the EM waves are strictly transverse (see Eq. \eqref{eq:transverse_condition_Brillouin}), we can take $\bm{k} \cdot \bm{A} = 0$ and $\phi_{\text{e}} = 0$. As a result, the behavior of the EM wave can be fully described by the vector potential $\bm{A}$ alone.

The vector potential can be represented as the sum of the incident and scattered waves, \nisHiura{from Eq. \eqref{eq:A_wave_general},}
\begin{equation}
\nisHiura{\begin{aligned}
\bm{A}(\bm{r}, t) 
  &= \bm{A}_{\mathrm{w}0} + \bm{A}_{\mathrm{w}1} \\
  &= \bm{A}_{0} \, \mathrm{e}^{\mathrm{i}\left(\bm{k}_{0} \cdot \bm{r} - \omega_{0} t\right)} 
   + \bm{A}_{1} \, \mathrm{e}^{\mathrm{i}\left(\bm{k}_{1} \cdot \bm{r} - \omega_{1} t\right)} 
   + \text{c.c.}.
\end{aligned}}
\label{eq:incident_and_scattered_wave_Brillouin}
\end{equation}
The source term in the wave equation~\eqref{eq:wave_eq_vector_Brillouin} is given by the plasma current density $\bm{j}$,
\begin{equation}
\bm{j} = \sum_{q= \pm e} q n_{\pm}(\bm{r}, t) \bm{v}_{\pm}(\bm{r}, t).
\label{eq:current_oomoto_Brillouin}
\end{equation}
Here, $n_{\pm}$ are the number densities of positrons and electrons, respectively, and $\bm{v}_{\pm}$ denote the velocities of positrons and electrons induced by the EM wave. Note that $\bm{v}_{\pm}$ should be distinguished from the velocity coordinate $\bm{v}$. \rei{The current density~\eqref{eq:current_oomoto_Brillouin} plays a central role in driving plasma instabilities.} Furthermore, plasma particles respond to the EM wave according to the following equation of motion:
\begin{equation}
\frac{\dd \bm{v}_{\pm}}{\dd t} = \pm \frac{e}{m_{\text{e}}}\left(\bm{E} + \frac{\bm{v}_{\pm} \times \bm{B}_{0}}{c}\right).
\label{eq:EOM_oomoto_induced_Compton_magnetized_Brillouin}
\end{equation}

The number densities of positrons ($+$) and electrons ($-$) can be expanded as
\begin{equation}
    n_{\pm}(\bm{r}, t) = n_{\mathrm{e}0} + \delta n_{\pm}(\bm{r}, t),
\end{equation}
where $\delta n_{\pm}$ represents the density fluctuation. In order to evaluate the source current in Eq.~\eqref{eq:current_oomoto_Brillouin}, it is necessary to obtain the density fluctuations $\delta n_{\pm}$. These density fluctuations are excited by the beat (ponderomotive) field arising from the incident and scattered waves \ioka{(see Fig. \ref{fig:parametric_instability})}, and possess the following wavevector and frequency components:
\begin{equation}
\omega = \omega_{1} - \omega_{0}, \quad \bm{k} = \bm{k}_{1} - \bm{k}_{0}.
\label{eq:energy_momentum_conservation_Compton_Brillouin}
\end{equation}
\nisHiura{In what follows, we focus on Stokes scattering ($\omega < 0$).} The evolution of the density fluctuations is governed by the collisionless Boltzmann equation for the distribution function $f_{\pm}$:
\begin{equation}
\frac{\partial f_{\pm}}{\partial t} + \bm{v} \cdot \bm{\nabla} f_{\pm} + \bm{F} \cdot \frac{\partial f_{\pm}}{\partial \bm{p}} = 0.
\end{equation}
Here, the force $\bm{F}$ in the Boltzmann equation consists of the ponderomotive force and the Lorentz force\footnote{
The Lorentz force considered here arises from the electrostatic wave generated by the density fluctuation and possesses the same frequency and wavevector components as the beat. On the other hand, the Lorentz force components with the frequency and wavevector of the incident or scattered waves themselves are much higher frequency than the beat component ($\omega_{1} \sim \omega_{0} \gg |\omega|$), and they averaged out over time, leaving the ponderomotive force oscillating at the beat frequency.
}:
\begin{equation}
\bm{F} = -\bm{\nabla} \phi_{\pm} \pm e \left( \bm{E} + \frac{\bm{v} \times \bm{B}_{0}}{c} \right).
\end{equation}
\rei{Among these, the ponderomotive force $-\bm{\nabla} \phi_{\pm}$ and the force due to the electric field fluctuation $\pm e\bm{E}$ are treated as perturbations.} \Nishiura{This force $\bm{F}$ is treated as perturbations.} The fluctuation of the distribution function is then expressed as
\begin{equation}
f_{\pm}(\bm{r}, \bm{v}, t) = f_{0\pm}(\bm{v}) + \delta f_{\pm}(\bm{r}, \bm{v}, t).
\end{equation}
Furthermore, the electric field generated by plasma density fluctuations follows the Maxwell equation,
\begin{equation}
\bm{\nabla} \cdot \bm{E} = \sum_{q = \pm e} 4 \pi q n_{\text{e} 0} \int \delta f_{\pm} \dd^{3} \bm{v}.
\end{equation}
When the background magnetic field is uniform along the $x$-direction, the ponderomotive potential is given as follows \citep{1968CzJPh..18.1280K,1977PhRvL..39..402C,1981PhFl...24.1238C,1981PhRvL..46..240H,10.1063/1.864196,1996GeoRL..23..327L}:
\begin{equation}
\nisHiura{\begin{aligned}
\phi_{\pm} = \frac{e^{2}}{2 m_{\text{e}}} \Biggl\langle
&\frac{|\bm{E}_{\mathrm{w}\|}|^{2}}{\omega_{0}^{2}}
 - \frac{|\bm{E}_{\mathrm{w}\perp}|^{2}}{\omega_{\text{c}}^{2} - \omega_{0}^{2}} \\
&+~\mathrm{i} \frac{\omega_{\text{c}\pm}\,
  \hat{\bm{B}}_0\!\cdot\!(\bm{E}_{\mathrm{w}\perp}^{*}\!\times\!\bm{E}_{\mathrm{w}\perp})}
  {\omega_{0} \left(\omega_{0}^{2} - \omega_{\text{c}}^{2}\right)}
\Biggr\rangle,
\end{aligned}}
\label{eq:ponderomotive_potential_Brillouin}
\end{equation}
where $\omega_{1} \sim \omega_{0} \gg |\omega|$, and the time average is taken over a timescale longer than $\omega_0^{-1}$ and shorter than $|\omega|^{-1}$. The cyclotron frequencies are defined as
\begin{equation}
\omega_{\text{c}\pm} \equiv \pm \frac{e B_0}{m_{\text{e}} c}, \quad \omega_{\text{c}} \equiv \omega_{\text{c}+} = -\omega_{\text{c}-}.
\end{equation}
With these, the Boltzmann equation including the ponderomotive potential becomes a closed system, allowing the calculation of density fluctuations. Finally, by combining Eqs.~\eqref{eq:wave_eq_vector_Brillouin} and \eqref{eq:current_oomoto_Brillouin} and computing the Fourier coefficients for $(\bm{k}_1, \omega_1)$, the dispersion relation for the scattered wave can be derived \citep{2025PhRvD.111f3055N}.

\subsection{Dispersion Relation for a Maxwellian Distribution}
When the unperturbed component of the $e^\pm$ pair plasma \Nishiura{$f_{0\pm}(\bm{v})$} follows a Maxwellian distribution, the dispersion relation for induced scattering can be analytically derived under the assumptions described in Sec.~\ref{subsec:basic_equations_Brillouin}. If the electric field component of the incident wave is polarized along the direction of the background magnetic field, only the ordinary mode is excited. On the other hand, if the electric field is polarized perpendicular to the background magnetic field, both neutral and charged modes can be excited simultaneously \citep{2025PhRvD.111f3055N}.

In this study, we analyze the behavior of each instability in regions where the background magnetic field is sufficiently strong. Specifically, we impose the conditions that the cyclotron frequency is much larger than the frequencies of the EM waves,
\begin{equation}
\omega_{0},~\omega_1 \ll \omega_{\mathrm{c}},
\label{eq:ziba_tsuyoi}
\end{equation}
and that the particle gyro-radius, $v_{\text{th}}/\omega_{\text{c}}$, is much smaller than the spatial scale of the density fluctuation,\rei{ i.e., the strong magnetization condition,}
\begin{equation}
\frac{k_{\perp}v_{\text{th}}}{\omega_{\mathrm{c}}}\ll 1,
\label{eq:zika_tsuyoi}
\end{equation}
throughout the analysis (see Eqs.~\eqref{eq:thermal_velocity_Brillouin} and \eqref{eq:definition_of_kperp_Brillouin} for the definitions of the physical quantities). \nisHiura{We do not specify the relative magnitude between the wave frequencies ($\omega_{0}, \omega_{1}$) and the plasma frequency $\omega_{\mathrm{p}}$.}

The dispersion relations for each mode are given as follows~\citep{2025PhRvD.111f3055N}.

\textbf{Ordinary mode}\Nishiura{~(See Eqs. (36) and (40) in \citep{2025PhRvD.111f3055N}.)}:
\begin{equation}
\begin{aligned}
c^{2} k_{1}^{2} - \omega_{1}^{2} + \omega_{\text{p}}^{2} 
= \frac{1}{2}\omega_{\text{p}}^2 a_{\mathrm{e}}^2 \mu^2
\left(\frac{c}{v_{\text{th}}}\right)^2\left[1+\zeta Z(\zeta)\right].
\end{aligned}
\label{eq:dispersion_relation_no_magnetic_Brillouin}
\end{equation}

\textbf{Neutral mode}\Nishiura{~(See Eqs. (95) and (40) in \citep{2025PhRvD.111f3055N}.)}:
\begin{equation}
\begin{aligned}
c^{2} k_{1}^{2} - \omega_{1}^{2}
- \omega_{\mathrm{p}}^{2}\left(\frac{\omega_{1}}{\omega_{\mathrm{c}}}\right)^{2}
= \frac{1}{2} \omega_{\mathrm{p}}^{2} a_{\text{e}}^{2} \mu^{2}
\left(\frac{\omega_{0}}{\omega_{\mathrm{c}}}\right)^{4}
\\
\times\left(\frac{c}{v_{\text{th}}}\right)^{2}
[1 + \zeta Z(\zeta)].
\end{aligned}
\label{eq:dispersion_relation_neutral_Brillouin}
\end{equation}

\textbf{Charged mode}\Nishiura{~(See Eqs. (58) and (40) in \citep{2025PhRvD.111f3055N}.)}:
\begin{equation}
\begin{aligned}
c^{2} k_{1}^{2} - \omega_{1}^{2}
- \omega_{\text{p}}^{2}\left(\frac{\omega_{1}}{\omega_{\mathrm{c}}}\right)^{2} 
= \frac{1}{2\varepsilon_{\text{L}}} \omega_{\text{p}}^2 a_{\mathrm{e}}^2 
\left(\frac{\omega_{0}}{\omega_{\mathrm{c}}}\right)^{2} 
\\
\times (1 - \mu^2) \left|\bm{n} \cdot \hat{\bm{B}}_0\right|^2
\left(\frac{c}{v_{\text{th}}}\right)^2 \left[1 + \zeta Z(\zeta)\right].
\end{aligned}
\label{eq:dispersion_relation_charged_Brillouin}
\end{equation}

The definitions of the fundamental physical quantities used in this study are summarized below. The uniform component of the $e^\pm$ pair plasma is assumed to follow the Maxwellian distribution:
\begin{equation}
f_{0+}=f_{0-}= \frac{1}{(\pi v_{\text{th}}^2)^{\frac{3}{2}}} \exp\left(-\frac{v_{\|}^2 + v_{\perp}^2}{v_{\text{th}}^2}\right),
\label{eq:Maxwellian_distribution_induced_Compton_Brillouin}
\end{equation}
where the three-dimensional thermal velocity is defined as
\begin{equation}
v_{\text{th}} \equiv \sqrt{\frac{2 k_{\text{B}} T_{\text{e}}}{m_{\text{e}}}}.
\label{eq:thermal_velocity_Brillouin}
\end{equation}
The plasma frequency $\omega_{\mathrm{p}}$ is determined from Eq.~\eqref{eq:uniform_density_Brillouin} as
\begin{equation}
  \omega_{\mathrm{p}} \equiv \sqrt{\frac{8\pi e^2 n_{\mathrm{e}0}}{m_{\mathrm{e}}}}.
  \label{eq:plasma_frequency_induced_Compton_magnetized_Brillouin}
\end{equation}
The strength parameter $a_{\mathrm{e}}$ is defined as:
\begin{equation}
  \nisHiura{a_{\mathrm{e}} \equiv 
\frac{e\left|\bm{A}_{\mathrm{w}0}\right|_{\mathrm{max}}}{m_{\mathrm{e}} c^2}
\xrightarrow[\;]{\text{lin.\ pol.}}
\frac{2 e A_0}{m_{\mathrm{e}} c^2}.}
  \label{eq:strength_parameter_no_magnetic_Brillouin}
\end{equation}
\rei{The factor of 2 in Eq.~\eqref{eq:strength_parameter_no_magnetic_Brillouin} arises from the definition of the incident wave amplitude as $2A_0$ for linear polarization, as described in Eq.~\eqref{eq:incident_and_scattered_wave_Brillouin}}\nisHiura{Here, ``lin.\ pol.'' denotes the assumption of a linearly polarized incident wave, for which the peak amplitude satisfies 
$\left|\bm{A}_{\mathrm{w}0}\right|_{\mathrm{max}} = 2A_0$ from Eq. \eqref{eq:A_wave_general}}~\footnote{For circular polarization, the amplitude is $\sqrt{2}A_0$, which introduces a correction factor to the linear growth rate. See Appendix~\ref{ap:linear_to_circular} for further discussion.}.
The parallel and perpendicular components of the wavenumber are given by
\begin{equation}
  k_{\parallel} \equiv k_x,
\end{equation}
\begin{equation}
  k_{\perp} \equiv \sqrt{k_y^2 + k_z^2}.
  \label{eq:definition_of_kperp_Brillouin}
\end{equation}
The plasma dispersion function $Z(\zeta)$ is defined by
\begin{equation}
  Z(\zeta) \equiv \frac{1}{\sqrt{\pi}} \int_{-\infty}^{\infty} \frac{1}{z - \zeta} e^{-z^2} \, \mathrm{d}z,
  \label{eq:definition_of_plasma_dispersion_function}
\end{equation}
and the argument $\zeta$ in Eqs.~\eqref{eq:dispersion_relation_no_magnetic_Brillouin}--\eqref{eq:dispersion_relation_charged_Brillouin} is given by\rei{ the frequency $\omega$, parallel wavenumber $k_{\parallel}$, and thermal velocity $v_{\mathrm{th}}$ as}
\begin{equation}
  \zeta \equiv \frac{\omega}{k_{\parallel} v_{\mathrm{th}}}.
  \label{eq:definition_of_zeta_Brillouin}
\end{equation}
The longitudinal dielectric function $\varepsilon_{\mathrm{L}}$ is defined as\Nishiura{~(See Eqs. (21) and (40) in \citep{2025PhRvD.111f3055N}.)}
\begin{equation}
  \varepsilon_{\mathrm{L}} \simeq 1 + \frac{2 \omega_{\mathrm{p}}^2}{k^2 v_{\mathrm{th}}^2} \left[1 + \zeta Z(\zeta)\right].
  \label{eq:longitudinal_electric_susceptibility_Brillouin_Maxwell}
\end{equation}

The angular parameters used in this study are defined as follows.
The coefficient $\mu$, representing the cosine of the angle between the electric field components of the incident wave $\bm{A}_0$ and the scattered wave $\bm{A}_1$, is defined by
\begin{equation}
\nisHiura{\mu \equiv \frac{\left|\bm{A}_{1} \cdot \bm{A}_{0}^{*}\right|}{A_{1} A_{0}},\quad 0\leq\mu\leq 1,}
\label{eq:definition_mu_induced_magnetized_Brillouin}
\end{equation}
The unit vector $\bm{n}$, which is perpendicular to the polarization planes of the incident wave $\bm{A}_0$ and the scattered wave $\bm{A}_1$, is defined as
\begin{equation}
\bm{n} \equiv \frac{\bm{A}_1 \times \bm{A}_0^{*}}{\left|\bm{A}_1 \times \bm{A}_0^{*}\right|}.
\label{eq:definition_of_n_induced_Compton_Brillouin}
\end{equation}
The angle $\theta_{kB}$ between the background magnetic field and the wave vector $\bm{k}$ of the density fluctuation is defined by
\begin{equation}
 \cos \theta_{kB} \equiv \hat{\bm{B}}_{0} \cdot \frac{\bm{k}}{k},\quad 0\leq\theta_{kB}\leq\pi,
 \label{eq:Brillouin_theta_kb_definition}
\end{equation}
The cosine of the angle between the wave vectors of the incident and scattered waves, $\nu$, is given by
\begin{equation}
\nu \equiv \frac{\bm{k_{0}} \cdot \bm{k_{1}}}{k_{0} k_{1}},\quad -1\leq\nu\leq1.
\label{eq:definition_of_nu_induced_Brillouin}
\end{equation}

When the oscillatory velocity of particles driven by the incident wave is non-relativistic, that is, $|\bm{v}_{\pm}|/c \ll 1$, each instability mode is subject to an upper limit on the incident wave amplitude \citep{2025PhRvD.111f3055N}. Specifically, in the case of the ordinary mode, the strength parameter $a_{\text{e}}$ serves as the dimensionless amplitude of the incident wave and must satisfy
\begin{equation}
    a_{\text{e}} \ll 1.
    \label{eq:non_rela_ordinary_Brillouin}
\end{equation}
On the other hand, for the neutral and charged modes, $a_{\text{e}} \omega_0/\omega_{\text{c}}$ acts as the \nIshiura{dimensionless }amplitude of the incident wave, and the following condition must be satisfied:
\begin{equation}
    a_{\text{e}} \frac{\omega_0}{\omega_{\text{c}}} \ll 1.
    \label{eq:non_rela_neutral_and_charged_Brillouin}
\end{equation}

In the following section, we analytically derive the linear growth rates of the instabilities associated with the ordinary mode (Eq.~\eqref{eq:dispersion_relation_no_magnetic_Brillouin}), neutral mode (Eq.~\eqref{eq:dispersion_relation_neutral_Brillouin}), and charged mode (Eq.~\eqref{eq:dispersion_relation_charged_Brillouin}), based on their respective dispersion relations. We also investigate in detail how these instabilities compete with each other. The linear growth rate of the energy for the scattered wave and the density fluctuation is defined as follows:
\begin{equation}
    t^{-1}\equiv2~\text{Im}~\omega\equiv2\gamma=2~\text{Im}~\omega_1.
\end{equation}

\section{Analysis of Each Mode}
\label{sec:kakumo-donokaiseki}
\begin{table*}[htbp]
  \centering
  \captionsetup{
    skip=1em,
    justification=raggedright
  }
  \caption{Correspondence between coupling conditions and dominant instabilities. ICS denotes induced Compton scattering, SBS denotes stimulated Brillouin scattering, and SRS denotes stimulated Raman scattering. For the dominant instabilities of the charged mode, further subdivisions in the density-temperature space are given in Fig.~\ref{fig:charged_regime}.}
  \label{tab:roadmap_coupling}
  \renewcommand{\arraystretch}{1.5}
  \sisetup{table-number-alignment=left}
  \begin{tabular}{lll c| ll}
    \toprule
    Mode & Coupling Condition & & Resonance Condition & Dominant Instability & Growth Rate \\
    \midrule
    Ordinary 
      & $|\Im \zeta| \ll 1$ (weak)  
      & & -- 
      & ICS 
      & Eq.~\eqref{eq:maximum_growth_rate_no_magnetic_Compton_Brillouin} ($|\zeta|\ll1$) \\
    & $|\Im \zeta| \gg 1$ (strong)
      & & -- 
      & SBS 
      & Eq.~\eqref{eq:induced_Brillouin_nomagnetic_maximum_growth_rate} ($|\zeta|\gg1$) \\
    \midrule
    Neutral  
      & $|\Im \zeta| \ll 1$ (weak)  
      & & -- 
      & ICS 
      & Eq.~\eqref{eq:growth_rate_induced_Compton_subdominant_Brillouin} ($|\zeta|\ll1$) \\
    & $|\Im \zeta| \gg 1$ (strong)
      & & -- 
      & SBS 
      & Eq.~\eqref{eq:induced_Brillouin_neutral_maximum_growth_rate} ($|\zeta|\gg1$) \\
    \midrule
    \multirow{3}{*}{Charged} 
      & \multirow{2}{*}{$|\Im \zeta| \ll 1$ (weak)}  
      & \multirow{2}{*}{\Large$\left\{\vphantom{\dfrac{1}{1}}\right.$} 
      & $|\mathrm{Re}\,\omega| \simeq \omega_{\mathrm p} \gg k_{\parallel} v_{\mathrm{th}}$ 
      & SRS 
      & Eq.~\eqref{eq:growth_rate_of_induced_Raman_weak_coupling} ($|\zeta|\gg1$)\footnote{For the small-angle SRS, see Eq.~\eqref{eq:stimulated_Raman_growth_rate_small_angle}.} \\
      & 
      & 
      & otherwise 
      & ICS 
      & Eq.~\eqref{eq:maximum_growth_rate_magnetic2_Brillouin}, \eqref{eq:induced_Compton_Debye_screening_Brillouin} ($|\zeta|\ll1$) \\
      & $|\Im \zeta| \gg 1$ (strong) 
      & & -- 
      & SRS = SBS 
      & Eq.~\eqref{eq:induced_Brillouin_charged_maximum_growth_rate} ($|\zeta|\gg1$) \\
    \bottomrule
  \end{tabular}
\end{table*}

In this study, we analyze induced scattering in strongly magnetized $e^\pm$ pair plasma using a unified kinetic framework. Each instability is driven by a distinct physical resonance mechanism or scattering process. We systematically classify their linear growth rates and identify the parameter regimes in which each instability dominates.

An overview of the classification of instabilities is summarized in Tab.~\ref{tab:roadmap_coupling}. On top of that, for the charged mode, an upper limit on the plasma density arises due to the excitation condition for SRS, as described by Eq.~\eqref{eq:induced_Raman_condition_upper_limit}. Therefore, only for the charged mode, a more detailed classification in the density–temperature parameter space is required. This \Nishiura{additional} classification is presented in the latter part of this paper (see Fig.~\ref{fig:charged_regime}).

As summarized in Tab.~\ref{tab:roadmap_coupling}, ICS, SBS, and SRS are classified as the dominant instabilities according to the respective coupling and resonance conditions. The coupling strength is defined by the relation between the thermal fluctuation frequency, $k_{\parallel} v_{\mathrm{th}}$, and the growth rate, $\gamma \equiv |\Im \omega|$, as follows:
\begin{equation}
\gamma \ll k_{\parallel} v_{\mathrm{th}} \quad \Leftrightarrow \quad |\Im \zeta| \ll 1 \quad \text{(weak coupling)},
\label{eq:weak_coupling_condition}
\end{equation}
\begin{equation}
\gamma \gg k_{\parallel} v_{\mathrm{th}} \quad \Leftrightarrow \quad |\Im \zeta| \gg 1 \quad \text{(strong coupling)}.
\label{eq:strong_coupling_condition}
\end{equation}
Here, $\zeta$ is defined in Eq.~\eqref{eq:definition_of_zeta_Brillouin}. The physical interpretation of these coupling regimes is discussed in Sec.~\ref{subsub:transition_weak_to_strong_coupling}.

For analytic estimation of the maximum growth rates of each instability, we employ the asymptotic expansion of the plasma dispersion function \eqref{eq:definition_of_plasma_dispersion_function}, depending on the magnitude of $|\zeta|$ \citep{1961pdf..book.....F}:
\begin{equation}
Z(\zeta) = 
\begin{cases}
\mathrm{i} \sqrt{\pi}\mathrm{e}^{-\zeta^{2}} - 2 \zeta + \dfrac{4}{3} \zeta^{3} - \cdots \\
\hspace{3.5em} \text{for } |\zeta| \ll 1, \\[2ex]
\mathrm{i} \sqrt{\pi}\mathrm{e}^{-\zeta^{2}} \sigma - \zeta^{-1}\left(1 + \dfrac{1}{2 \zeta^{2}} + \cdots\right) \\
\hspace{3.5em} \text{for } |\zeta| \gg 1,
\end{cases}
\label{eq:plasma_dispersion_function_Brillouin}
\end{equation}
where $\sigma = 0, 1, 2$ for $\text{Im}~\zeta > 0, = 0, < 0$, respectively.

The classification in Tab.~\ref{tab:roadmap_coupling} shows that, for all of the ordinary, neutral, and charged modes, the expansion for $|\zeta| \ll 1$ under the weak coupling condition ($|\Im \zeta| \ll 1$) yields ICS due to Landau resonance between the beat EM wave and thermal particles. In contrast, applying the $|\zeta| \gg 1$ expansion under the strong coupling condition ($|\Im \zeta| \gg 1$) suppresses the Landau resonance term, and SBS becomes dominant. For the charged mode, SRS dominates only when the weak coupling condition ($|\Im \zeta| \ll 1$) is satisfied and both the resonance condition $|\mathrm{Re}\, \omega| \simeq \omega_{\mathrm p}$ and the non-damping condition for the Langmuir wave ($\omega_{\mathrm p} \gg k_{\parallel} v_{\mathrm{th}}$) hold. Thus, the $|\zeta| \gg 1$ expansion is \Nishiura{appropriate} for weak coupling SRS.

\subsection{Ordinary mode}
The dispersion relation for the ordinary mode, as given by Eq.~\eqref{eq:dispersion_relation_no_magnetic_Brillouin}, is expressed as follows:
\begin{equation}
\begin{aligned}
c^{2} k_{1}^{2} - \omega_{1}^{2} + \omega_{\text{p}}^{2} = 
\frac{1}{2} \omega_{\text{p}}^2 a_{\mathrm{e}}^2 \mu^2 
\left(\frac{c}{v_{\text{th}}}\right)^2 \left[1 + \zeta Z(\zeta)\right].
\end{aligned}
\end{equation}
\ioka{The physical picture is illustrated in Fig. \ref{fig:parametric_instability}. }Two types of instabilities can arise from this dispersion relation: ICS and SBS. The properties of ICS have already been analyzed in detail by \citep{2025PhRvD.111f3055N}. In this section, we explicitly present the linear growth rate and growth wavenumber and the angular parameter at maximum growth for both ICS and SBS. Details of the derivations are provided in Appendix~\ref{sec:ordinary_mode_detail_derivation}.

\subsubsection{Induced Compton Scattering (Ordinary mode)}

ICS in the ordinary mode is described as an instability arising from Landau resonance under the weak coupling condition, Eq.~\eqref{eq:weak_coupling_condition}. This instability corresponds to the regime where the exponential term $\mathrm{i} \sqrt{\pi} \mathrm{e}^{-\zeta^2}$ dominates in the asymptotic expansion of the plasma dispersion function $Z(\zeta)$ given by Eq.~\eqref{eq:plasma_dispersion_function_Brillouin}. The linear growth rate is given by \Nishiura{(See Eq. (44) in \citep{2025PhRvD.111f3055N})}
\begin{equation}
t_{\mathrm{C} \|}^{-1}(\mu)=\sqrt{\frac{\pi}{32 \mathrm{e}}} \frac{\omega_{\mathrm{p}}^{2} a_{\mathrm{e}}^{2}\mu^2}{\omega_{0}} \frac{m_{\mathrm{e}} c^{2}}{k_{\mathrm{B}} T_{\mathrm{e}}}.
\label{eq:maximum_growth_rate_no_magnetic_Compton_Brillouin}
\end{equation}
The wavenumber of the density fluctuation at maximum growth is expressed as\footnote{\Nishiura{This expression can be derived from Eq. (47) in \citep{2025PhRvD.111f3055N} by neglecting terms of order $(v_{\mathrm{th}}/c)^2$ and $a_{\text{e}}^2$.}}
\begin{equation}
\begin{aligned}
k_{\text{max}} \simeq\sqrt{2(1 - \nu)}k_0\left(1-\sqrt{\frac{1 - \nu}{2} \cos^2 \theta_{kB} \frac{k_{\mathrm{B}} T_{\mathrm{e}}}{m_{\mathrm{e}} c^2}}\right).
\end{aligned}
\label{eq:maximum_growth_wave_number_Compton_ordinary}
\end{equation}
ICS becomes dominant when the \Nishiura{incident wave amplitude} $a_{\mathrm{e}}$ satisfies the following condition:
\begin{equation}
a_{\mathrm{e}} \ll 4\left(\frac{2\text{e}}{\pi}\right)^{\frac{1}{4}} \frac{\omega_0}{\omega_{\mathrm{p}}}\left(\frac{k_{\mathrm{B}} T_{\mathrm{e}}}{m_{\mathrm{e}} c^2}\right)^{\frac{3}{4}}\frac{\left\{(1-\nu) \cos ^2 \theta_{kB}\right\}^{\frac{1}{4}}}{|\mu|}
\label{eq:strength_parameter_limit_para2_Brillouin}
\end{equation}
This requirement is obtained by substituting the linear growth rate, Eq.~\eqref{eq:maximum_growth_rate_no_magnetic_Compton_Brillouin}, into the weak coupling condition, Eq.~\eqref{eq:weak_coupling_condition}.

Maximum growth is realized when a parallel-polarized EM wave is incident perpendicular to the background magnetic field and undergoes sidescattering. A more detailed physical interpretation is provided in Sec.~\ref{subsubsec:Ordinary_sidescattering}.

\subsubsection{Stimulated Brillouin Scattering (Ordinary mode)}
SBS in the ordinary mode is the dominant instability under the strong coupling condition, Eq.~\eqref{eq:strong_coupling_condition}. In this regime, the plasma dispersion function $Z(\zeta)$ can be approximated by the asymptotic form for $|\zeta| \gg 1$, as shown in Eq.~\eqref{eq:plasma_dispersion_function_Brillouin}, yielding \ioka{Eq. \eqref{eq:plasma_dispersion_function_yuurikannsuu_Brillouin} in Appendix~\ref{subsec:strong_coupling_Brillouin_ordinary_derivation}.}
Using this limit, the dispersion relation is reduced to \ioka{Eq. \eqref{eq:dispersion_relation_induced_Brillouin_nomagnetic_ap}.} \rei{the following form:
\begin{equation}
\omega^{2}\left\{\omega - \frac{c^{2}\left(k^{2} + 2 \bm{k}_{0}\cdot\bm{k}\right)}{2 \omega_{0}}\right\} = \frac{a_{\text{e}}^{2} \omega_{\mathrm{p}}^{2} \mu^{2}c^2 k_{\|}^{2}}{8 \omega_{0}}.
\label{eq:dispersion_relation_induced_Brillouin_nomagnetic}
\end{equation}
A detailed derivation is given in Appendix~\ref{subsec:strong_coupling_Brillouin_ordinary_derivation}.} The linear growth rate obtained from this dispersion relation is
\begin{equation}
\left(t_{\mathrm{B} \|}^{\max}\right)^{-1} = \frac{\sqrt{3}}{2^{\frac{4}{3}}}\left(\frac{a_{\text{e}}^{2}\omega_{\mathrm{p}}^{2}\omega_{0}}{2}\right)^{\frac{1}{3}}.
\label{eq:induced_Brillouin_nomagnetic_maximum_growth_rate}
\end{equation}
The wavenumber of the density fluctuation at maximum growth is given by
\begin{equation}
k_{\text{max}} \sim \sqrt{2}k_0.
\label{eq:maximum_wave_number_Brillouin_ordinary}
\end{equation}
\Nishiura{Here, the representative case of $90^\circ$ sidescattering is considered. See Appendix~\ref{subsec:Derivation_of_SBS_ordinary_growth_rate} for details.}
The angle parameter giving the maximum growth rate, as Eq.~\eqref{eq:maximum_growth_angle_condition}, may not precisely reflect the actual value. \nIshiura{Therefore, the maximum linear growth rate in Eq. \eqref{eq:induced_Brillouin_nomagnetic_maximum_growth_rate} may be modified by a factor of a few. See Appendix \ref{subsec:angle_dependence_ordinary_Brillouin} for details.}

SBS becomes dominant when the \Nishiura{incident wave amplitude} $a_{\mathrm{e}}$ satisfies the following range:
\begin{equation}
7.0~\frac{\omega_{0}}{\omega_{\mathrm{p}}}\left(\frac{k_{\mathrm{B}}T_{\mathrm{e}}}{m_{\mathrm{e}} c^{2}}\right)^{\frac{3}{4}} \ll a_{\mathrm{e}} \ll 1.
\label{eq:strong_coupling_condition_for_strength_parameter_Ordinary}
\end{equation}
This range is obtained by applying the strong coupling condition, Eq.~\eqref{eq:strong_coupling_condition}, to the expression for the growth rate, Eq.~\eqref{eq:induced_Brillouin_nomagnetic_maximum_growth_rate}, together with Eq.~\eqref{eq:non_rela_ordinary_Brillouin}.

Maximum growth occurs when a parallel-polarized EM wave is incident perpendicular to the background magnetic field and undergoes sidescattering. Further discussion of the physical interpretation is provided in Sec.~\ref{subsubsec:Ordinary_sidescattering}.

\subsubsection{Summary of the Ordinary Mode}

The instability based on the dispersion relation of the ordinary mode is classified by the magnitude of the strength parameter of the incident wave. Under the weak coupling condition, Eq.~\eqref{eq:strength_parameter_limit_para2_Brillouin}, ICS becomes the dominant process. In contrast, under the strong coupling condition, Eq.~\eqref{eq:strong_coupling_condition_for_strength_parameter_Ordinary}, SBS dominates. The physical interpretation of this transition is discussed in Sec.~\ref{subsub:transition_weak_to_strong_coupling}. The representative analytical forms of the linear growth rate for the ordinary mode, evaluated at the angle parameter that gives the maximum for SBS, are summarized as follows, according to Eqs.~\eqref{eq:maximum_growth_rate_no_magnetic_Compton_Brillouin} and \eqref{eq:induced_Brillouin_nomagnetic_maximum_growth_rate}:
\begin{equation}
\begin{aligned}
    \left(t_{\|}^{\max}\right)^{-1} \sim 
    \begin{cases}
        \sqrt{\frac{\pi}{128\, \mathrm{e}}}\,
        \frac{\omega_{\mathrm{p}}^{2} a_{\mathrm{e}}^{2}}{\omega_{0}}\,
        \frac{m_{\mathrm{e}} c^{2}}{k_{\mathrm{B}} T_{\mathrm{e}}},
        \\
        \hspace{1cm}
        a_{\mathrm{e}} \ll 4.6~\frac{\omega_{0}}{\omega_{\mathrm{p}}}
        \left(\frac{k_{\mathrm{B}} T_{\mathrm{e}}}{m_{\mathrm{e}} c^{2}}\right)^{\frac{3}{4}},
        \\[2ex]
        \frac{\sqrt{3}}{2^{\frac{4}{3}}}
        \left(
            \frac{a_{\mathrm{e}}^{2} \omega_{\mathrm{p}}^{2} \omega_{0}}{2}
        \right)^{\frac{1}{3}},
        \\
        \hspace{1cm}
            7.0~\frac{\omega_{0}}{\omega_{\mathrm{p}}}
            \left(\frac{k_{\mathrm{B}} T_{\mathrm{e}}}{m_{\mathrm{e}} c^{2}}\right)^{\frac{3}{4}}
            \ll a_{\mathrm{e}} \ll 1.
    \end{cases}
\end{aligned}
\label{eq:induced_scattering_growth_rate_summary_nomagnetic}
\end{equation}

The wavenumber at maximum growth is given by Eqs.~\eqref{eq:maximum_growth_wave_number_Compton_ordinary} and \eqref{eq:maximum_wave_number_Brillouin_ordinary} as
\begin{equation}
\begin{aligned}
    k_{\text{max}} \sim 
    \begin{cases}
        \sqrt{2}\,k_0
        \left(
            1 - \sqrt{
                \frac{
                    k_{\mathrm{B}} T_{\mathrm{e}}
                }{
                    8 m_{\mathrm{e}} c^2
                }
            }
        \right),
        \\
        \hspace{1cm}
            a_{\mathrm{e}} \ll 4.6~\frac{\omega_{0}}{\omega_{\mathrm{p}}}
            \left(
                \frac{k_{\mathrm{B}} T_{\mathrm{e}}}{m_{\mathrm{e}} c^{2}}
            \right)^{\frac{3}{4}},
        \\[2ex]
        \sqrt{2}\,k_0,
        \\
        \hspace{1cm}
            7.0~\frac{\omega_{0}}{\omega_{\mathrm{p}}}
            \left(
                \frac{k_{\mathrm{B}} T_{\mathrm{e}}}{m_{\mathrm{e}} c^{2}}
            \right)^{\frac{3}{4}}
            \ll a_{\mathrm{e}} \ll 1.
    \end{cases}
\end{aligned}
\label{eq:induced_scattering_growth_wave_vector_summary_nomagnetic}
\end{equation}

\Nishiura{The transition point between the weak and strong coupling regimes is defined as the incident wave amplitude $a_{\text{e}}$ at which the growth rates in both coupling regimes become equal, as described by Eq.~\eqref{eq:induced_scattering_growth_rate_summary_nomagnetic},}
\begin{equation}
    a_{\mathrm{e},\text{trans}}\simeq3.7~\frac{\omega_{0}}{\omega_{\mathrm{p}}}\left(\frac{k_{\mathrm{B}}T_{\mathrm{e}}}{m_{\mathrm{e}} c^{2}}\right)^{\frac{3}{4}}.
    \label{eq:transition_point_for_ordinary_mode}
\end{equation}

Compared to the unmagnetized $e^\pm$ pair plasma, the order of the maximum linear growth rate for both ICS and SBS in the ordinary mode remains the same. A more detailed physical interpretation is given in Sec.~\ref{subsubsec:scattering_suppression}.

\subsubsection{Numerical Evaluation}

The linear growth rate of induced scattering in the ordinary mode can be calculated numerically by solving the dispersion relation, Eq.~\eqref{eq:dispersion_relation_no_magnetic_Brillouin}. In this study, we investigate how the maximum linear growth rate varies as a function of the strength parameter, $a_{\mathrm{e}}$, as defined by Eq.~\eqref{eq:strength_parameter_no_magnetic_Brillouin}.
\begin{figure*}
\centering
\includegraphics[width=\textwidth]{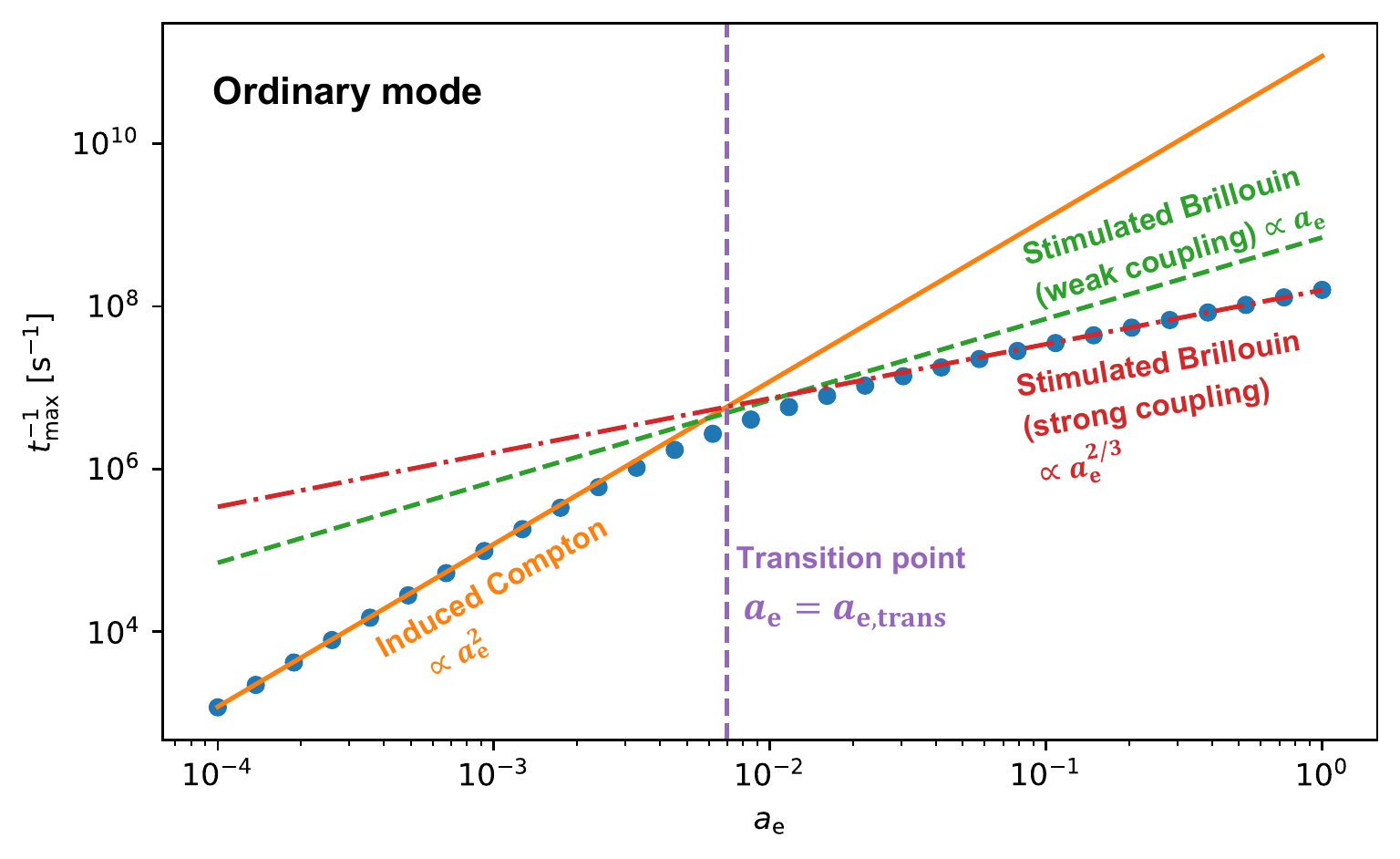}
\caption[Concept of Parametric Decay Instability]{\justifying The maximum linear growth rate of induced scattering excited by the ordinary mode as a function of the strength parameter of the incident EM wave. The orange solid line represents the growth rate of ICS, given by Eq.~\eqref{eq:maximum_growth_rate_no_magnetic_Compton_Brillouin}. The red dashed-dotted line corresponds to the growth rate of SBS in the strong coupling regime, given by Eq.~\eqref{eq:induced_Brillouin_nomagnetic_maximum_growth_rate}. The green dotted line indicates the growth rate of SBS in the weak coupling regime, given by Eq.~\eqref{eq:stimulated_Brillouin_weak_ordinary}. The purple vertical dotted line marks the value of the strength parameter at the transition from weak to strong coupling, as described by Eq.~\eqref{eq:transition_point_for_ordinary_mode}. The blue dots show numerical solutions of the dispersion relation, Eq.~\eqref{eq:dispersion_relation_no_magnetic_Brillouin}. The following parameters are used: electron thermal velocity $v_{\mathrm{th}}/c=10^{-3}$, plasma frequency $\omega_{\mathrm{p}}/\omega_0=10^{-2}$, and incident wave frequency $\omega_0=2\pi \times 10^9~\mathrm{Hz}$.}
\label{fig:stimulated_Brillouin_ordinary_growth_rate}
\end{figure*}

Fig.~\ref{fig:stimulated_Brillouin_ordinary_growth_rate} shows the maximum linear growth rate of the scattered wave as a function of the strength parameter~\eqref{eq:strength_parameter_no_magnetic_Brillouin}. The numerical results indicate that, as the strength parameter increases, there is a continuous transition from ICS to strong-coupling SBS across the transition point, $a_{\mathrm{e},\text{trans}}$, given by Eq.~\eqref{eq:transition_point_for_ordinary_mode}.

In the weak coupling regime, ICS becomes the dominant instability rather than SBS. For the ordinary mode, since the electric field component of the incident wave is aligned with the direction of the background magnetic field, the behavior is similar to that in the absence of a background field~\citep{2016PhRvL.116a5004E,2017PhRvE..96e3204S,2023MNRAS.522.2133I}. According to the fluid theory for unmagnetized plasma, the maximum linear growth rate of SBS in the weak coupling regime is expressed as~\citep{1978ApJ...224.1013D,1978ApJ...219..700G,1993JGR....9819049J,kruer2019physics,2023MNRAS.522.2133I}
\begin{equation}
\left(t_{\mathrm{B}\|}^{\text{weak}}\right)^{-1} \sim a_{\mathrm{e}} \omega_{\mathrm{p}} \left(\frac{m_{\mathrm{e}} c^{2}}{k_{\mathrm{B}} T_{\mathrm{e}}}\right)^{\frac{1}{4}}.
\label{eq:stimulated_Brillouin_weak_ordinary}
\end{equation}
As illustrated in Fig.~\ref{fig:stimulated_Brillouin_ordinary_growth_rate}, however, a kinetic approach for $e^\pm$ pair plasma reveals that ICS is the dominant instability in the weak coupling regime, even though the analytic value of the SBS growth rate is higher. This result is consistent with previous studies for unmagnetized $e^\pm$ pair plasma~\citep{2017PhRvE..96e3204S}. Further physical interpretation is provided in Sec.~\ref{subsub:transition_weak_to_strong_coupling}.

\subsection{Neutral mode}
This section discusses the instability associated with the neutral mode. The corresponding dispersion relation is given as follows (see Eq.~\eqref{eq:dispersion_relation_neutral_Brillouin}):
\begin{equation}
\begin{aligned}
c^{2} k_{1}^{2} - \omega_{1}^{2} 
&\ - \omega_{\mathrm{p}}^{2} \left( \frac{\omega_{1}}{\omega_{\mathrm{c}}} \right)^{2} \\
&= \frac{1}{2} \omega_{\mathrm{p}}^{2} a_{\text{e}}^{2} \mu^{2}
   \left( \frac{\omega_{0}}{\omega_{\mathrm{c}}} \right)^{4}
   \left( \frac{c}{v_{\text{th}}} \right)^{2}
   \left[ 1 + \zeta Z(\zeta) \right].
\end{aligned}
\label{eq:dispersion_relation_neutral_Brillouin2}
\end{equation}
\ioka{The physical picture is illustrated in Fig. \ref{fig:parametric_instability}. }As summarized in Tab.~\ref{tab:roadmap_coupling}, this dispersion relation supports two types of instabilities: ICS and SBS. The ICS in the neutral mode has already been analyzed in detail~\citep{2025PhRvD.111f3055N}, and its main results are presented below\ioka{, while the detailed derivation of the SBS is provided in Appendix~\ref{sec:neutral_mode_detail_derivation}}.

\subsubsection{Induced Compton Scattering (Neutral mode)}

In the neutral mode, ICS appears as the dominant instability under the weak coupling condition, Eq.~\eqref{eq:weak_coupling_condition}, and is governed by Landau resonance\Nishiura{~(see Tab.~\ref{tab:roadmap_coupling})}. This corresponds to the regime in which the exponential term, $\mathrm{i} \sqrt{\pi} \mathrm{e}^{-\zeta^2}$, in the asymptotic expansion of the plasma dispersion function $Z(\zeta)$ (Eq.~\eqref{eq:plasma_dispersion_function_Brillouin}) is dominant. The linear growth rate is given by \Nishiura{(see Eq. (96) in \citep{2025PhRvD.111f3055N})}
\begin{equation}
\begin{aligned}
\left(t_{\mathrm{C,neutral}}^{\text{max}}\right)^{-1} 
&= \sqrt{\frac{\pi}{32 \mathrm{e}}} 
   \frac{m_{\mathrm{e}} c^{2}}{k_{\mathrm{B}} T_{\mathrm{e}}} 
   \frac{a_{\mathrm{e}}^{2} \omega_{\mathrm{p}}^{2}}{\omega_{0}} \\
&\quad \times \left(\frac{\omega_{0}}{\omega_{\mathrm{c}}}\right)^{4}
   \left(1 + \frac{\omega_{\text{p}}^2}{\omega_{\text{c}}^2}\right)^{-1}.
\end{aligned}
\label{eq:growth_rate_induced_Compton_subdominant_Brillouin}
\end{equation}
The wavenumber of the density fluctuation at maximum growth is expressed as\footnote{\Nishiura{This expression can be derived from Eq. (98) in \citep{2025PhRvD.111f3055N} by neglecting terms of order $(v_{\mathrm{th}}/c)^2$ and $a_{\text{e}}^2$.}}
\begin{equation}
\begin{aligned}
k_{\mathrm{max}} 
&\simeq \sqrt{2(1 - \nu)}\,k_0 \\
&\times \left[1 
- \sqrt{\frac{1 - \nu}{2}\,\cos^2\theta_{kB}
    \,\frac{k_{\mathrm{B}}T_{\mathrm{e}}}{m_{\mathrm{e}}c^2}
    \left(1 + \frac{\omega_{\mathrm{p}}^2}{\omega_{\mathrm{c}}^2}\right)
}\right].
\end{aligned}
\label{eq:maximum_geowth_wave_vector_neutral_Compton}
\end{equation}
ICS becomes dominant if the incident wave amplitude~\Nishiura{$a_{\text{e}} \omega_0/\omega_{\text{c}}$} satisfies
\begin{equation}
\begin{aligned}
a_{\mathrm{e}}\frac{\omega_0}{\omega_{\mathrm{c}}}
&\ll 4
\left(\frac{2\mathrm{e}}{\pi}\right)^{\!\tfrac{1}{4}}
\left(\frac{k_{\mathrm{B}}T_{\mathrm{e}}}{m_{\mathrm{e}}c^2}\right)^{\!\tfrac{3}{4}}
\frac{\omega_{\mathrm{c}}}{\omega_{\mathrm{p}}}\\
&\quad\times
\left(1 + \frac{\omega_{\mathrm{p}}^2}{\omega_{\mathrm{c}}^2}\right)^{\!\tfrac{3}{4}}
\left[(1-\nu)\cos^2\theta_{kB}\right]^{\!\tfrac{1}{4}},
\end{aligned}
\label{eq:strength_parameter_limit_perp2-3_Brillouin}
\end{equation}
as shown by substituting the growth rate into the weak coupling condition, Eq.~\eqref{eq:weak_coupling_condition}~(see also Eq. (100) in \citep{2025PhRvD.111f3055N}).  
The angle parameter for maximum growth is given by~(see Eq. (97) in \citep{2025PhRvD.111f3055N})
\begin{equation}
\nisHiura{\mu=1.}
\label{eq:maximum_growth_angle_condition_neutral_Compton}
\end{equation}
This corresponds to the case where the electric field of the scattered wave is parallel to that of the incident wave\footnote{\Nishiura{The maximum growth rate is independent of the scattering angles such as $\nu$ and $\theta_{kB}$. In contrast, these angular parameters affect the range of the incident wave amplitude $a_{\text{e}} \omega_0/\omega_{\text{c}}$ where ICS becomes dominant, as described by Eq.~\eqref{eq:strength_parameter_limit_perp2-3_Brillouin}.}}.

\subsubsection{Stimulated Brillouin Scattering (Neutral mode)}

In the neutral mode, SBS becomes the dominant instability under the strong coupling condition given by Eq.~\eqref{eq:strong_coupling_condition}. As in the ordinary mode, applying the asymptotic expansion of the plasma dispersion function $Z(\zeta)$ for $|\zeta|\gg1$ (see Eq.~\eqref{eq:plasma_dispersion_function_Brillouin}) and manipulating Eq.~\eqref{eq:plasma_dispersion_function_yuurikannsuu_Brillouin}, the dispersion relation\Nishiura{~in Eq. \eqref{eq:dispersion_relation_neutral_Brillouin2}} can be expressed as \ioka{Eq. \eqref{eq:dispersion_relation_induced_Brillouin_Neutral_ap} in Appendix \ref{sec:neutral_mode_detail_derivation}.}
\rei{\begin{equation}
\begin{aligned}
\omega^{2}\left\{
\omega \frac{c^2}{v_{\text{A}}^2}
- \frac{c^{2}\left(k^{2} + 2 \bm{k}_{0} \cdot \bm{k}\right)}{2 \omega_{0}}
\right\}
= \frac{\omega_{\mathrm{p}}^{2} a_{\mathrm{e}}^{2} \mu^{2} c^{2} k_{\|}^{2}}{8 \omega_{0}}
\left(\frac{\omega_{0}}{\omega_{\mathrm{c}}}\right)^{4}.
\end{aligned}
\label{eq:dispersion_relation_induced_Brillouin_Neutral}
\end{equation}}
\rei{The Alfvén velocity is defined by
\begin{equation}
v_{\text{A}}
\equiv \frac{c}{\sqrt{1 + \frac{\omega_{\text{p}}^2}{\omega_{\text{c}}^2}}},
\label{eq:Alfven_velocity_Brillouin}
\end{equation}
and details of the derivation are provided in Appendix~\ref{sec:neutral_mode_detail_derivation}.} The linear growth rate is then given by
\begin{equation}
\left(t_{\mathrm{B,neutral}}^{\text{max}}\right)^{-1}
= \sqrt{3}
\left(
    \frac{a_{\mathrm{e}}^{2} \omega_{\mathrm{p}}^{2} \omega_{0}}{2}
\right)^{\!\tfrac{1}{3}}
\left(\frac{\omega_{0}}{\omega_{\mathrm{c}}}\right)^{\!\tfrac{4}{3}}.
\label{eq:induced_Brillouin_neutral_maximum_growth_rate}
\end{equation}
This result is consistent with the growth rate derived using the MHD approach~\citep{2024PhRvE.110a5205I}. The wavenumber corresponding to maximum growth is given by
\begin{equation}
k = 2k_0.
\label{eq:induced_Brillouin_neutral_wave_vector2}
\end{equation}
For SBS to be dominant, the incident wave amplitude~\Nishiura{$a_{\text{e}} \omega_0/\omega_{\text{c}}$} must satisfy
\begin{equation}
\begin{aligned}
8.3~\frac{\omega_{\text{c}}}{\omega_{\mathrm{p}}}
\left(\frac{k_{\mathrm{B}}T_{\mathrm{e}}}{m_{\mathrm{e}} c^{2}}\right)^{\!\tfrac{3}{4}}
\left(1+\frac{\omega_{\text{p}}^2}{\omega_{\text{c}}^2}\right)^{\!\tfrac{3}{4}}
\ll
a_{\mathrm{e}}\frac{\omega_{0}}{\omega_{\mathrm{c}}}
\ll 1.
\end{aligned}
\label{eq:strong_coupling_condition_neutral_Brillouin}
\end{equation}
This constraint is obtained by substituting the strong coupling condition~\eqref{eq:strong_coupling_condition} into the growth rate~\eqref{eq:induced_Brillouin_neutral_maximum_growth_rate}, together with Eq.~\eqref{eq:non_rela_neutral_and_charged_Brillouin}. The maximum growth is achieved for the angle parameters
\begin{equation}
\nisHiura{\left(\mu,~\nu,~ \cos \theta_{kB}\right) = (1,~-1,~\pm1).}
\label{eq:maximum_growth_angle_condition_neutral_Brillouin}
\end{equation}
This corresponds to the case where an EM wave propagates along the direction of the background magnetic field and undergoes $180^{\circ}$ backward scattering, with the electric field of the scattered wave parallel to that of the incident wave. For arbitrary incidence angles, the growth rate is reduced by a factor of a few.

\subsubsection{Summary of Neutral Mode}

In summary, the instabilities arising from the neutral mode are classified according to the coupling regime. When the weak coupling condition in Eq.~\eqref{eq:strength_parameter_limit_perp2-3_Brillouin} is satisfied, ICS becomes the dominant instability. In contrast, when the strong coupling condition in Eq.~\eqref{eq:strong_coupling_condition_neutral_Brillouin} applies, SBS is dominant. The linear growth rate at the angle parameter~\eqref{eq:maximum_growth_angle_condition_neutral_Brillouin} that gives the maximum for both ICS and SBS are given by Eqs.~\eqref{eq:growth_rate_induced_Compton_subdominant_Brillouin} and \eqref{eq:induced_Brillouin_neutral_maximum_growth_rate} as follows,
\begin{equation}
\left(t_{\text{neutral}}^{\text{max}}\right)^{-1} \sim 
\begin{cases}
\sqrt{\frac{\pi}{32 \text{e}}} \frac{a_{\mathrm{e}}^{2} \omega_{\mathrm{p}}^{2}}{\omega_{0}} \frac{m_{\mathrm{e}} c^{2}}{k_{\mathrm{B}} T_{\mathrm{e}}}
\left(\frac{\omega_{0}}{\omega_{\mathrm{c}}}\right)^{4}
\left(1+\frac{\omega_{\text{p}}^2}{\omega_{\text{c}}^2}\right)^{-1}, 
\\
\quad a_{\mathrm{e}} \frac{\omega_{0}}{\omega_{\mathrm{c}}} 
\ll 5.5~\frac{\omega_{\mathrm{c}}}{\omega_{\mathrm{p}}}
\left(\frac{k_{\mathrm{B}} T_{\mathrm{e}}}{m_{\mathrm{e}} c^{2}}\right)^{\frac{3}{4}}
\left(1+\frac{\omega_{\text{p}}^2}{\omega_{\text{c}}^2}\right)^{\frac{3}{4}}, 
\\[2ex]
\sqrt{3}\left(\frac{a_{\mathrm{e}}^{2} \omega_{\mathrm{p}}^{2} \omega_{0}}{2}\right)^{\frac{1}{3}}
\left(\frac{\omega_{0}}{\omega_{\mathrm{c}}}\right)^{\frac{4}{3}}, 
\\
\quad8.3~\frac{\omega_{\mathrm{c}}}{\omega_{\mathrm{p}}}
\left(\frac{k_{\mathrm{B}} T_{\mathrm{e}}}{m_{\mathrm{e}} c^{2}}\right)^{\frac{3}{4}}
\left(1+\frac{\omega_{\text{p}}^2}{\omega_{\text{c}}^2}\right)^{\frac{3}{4}}
\ll a_{\mathrm{e}} \frac{\omega_{0}}{\omega_{\mathrm{c}}} \ll 1.
\end{cases}
\label{eq:growth_rate_neutral_matome_Brillouin}
\end{equation}
The wavenumber corresponding to the maximum growth is given by Eqs.~\eqref{eq:maximum_geowth_wave_vector_neutral_Compton} and \eqref{eq:induced_Brillouin_neutral_wave_vector2} as follows,
\begin{equation}
k_{\text{max}} \sim 
\begin{cases}
2k_0\left(1-\sqrt{\frac{k_{\mathrm{B}} T_{\mathrm{e}}}{m_{\mathrm{e}} c^2}\left(1 + \frac{\omega_{\mathrm{p}}^2}{\omega_{\mathrm{c}}^2}\right)}\right), 
\\
\qquad a_{\mathrm{e}} \frac{\omega_{0}}{\omega_{\mathrm{c}}} 
\ll 5.5~\frac{\omega_{\mathrm{c}}}{\omega_{\mathrm{p}}}
\left(\frac{k_{\mathrm{B}} T_{\mathrm{e}}}{m_{\mathrm{e}} c^{2}}\right)^{\frac{3}{4}}
\left(1+\frac{\omega_{\text{p}}^2}{\omega_{\text{c}}^2}\right)^{\frac{3}{4}},
\\[2ex]
2k_0, 
\\
\qquad 8.3~\frac{\omega_{\mathrm{c}}}{\omega_{\mathrm{p}}}
\left(\frac{k_{\mathrm{B}} T_{\mathrm{e}}}{m_{\mathrm{e}} c^{2}}\right)^{\frac{3}{4}}
\left(1+\frac{\omega_{\text{p}}^2}{\omega_{\text{c}}^2}\right)^{\frac{3}{4}}
\ll a_{\mathrm{e}} \frac{\omega_{0}}{\omega_{\mathrm{c}}} \ll 1.
\end{cases}
\end{equation}
\Nishiura{The transition point between the weak and strong coupling regimes is defined as the incident wave amplitude $a_{\text{e}}\omega_0/\omega_{\text{c}}$ at which the growth rates in both coupling regimes become equal, as described by Eq.~\eqref{eq:growth_rate_neutral_matome_Brillouin},}
\begin{equation}
    a_{\mathrm{e},\text{trans}}\frac{\omega_{0}}{\omega_{\mathrm{c}}}\simeq4.4~\frac{\omega_{\text{c}}}{\omega_{\mathrm{p}}}
    \left(\frac{k_{\mathrm{B}}T_{\mathrm{e}}}{m_{\mathrm{e}} c^{2}}\right)^{\frac{3}{4}}
    \left(1+\frac{\omega_{\text{p}}^2}{\omega_{\text{c}}^2}\right)^{\frac{3}{4}}.
    \label{eq:transition_point_for_neutral_mode}
\end{equation}

In strongly magnetized $e^\pm$ pair plasma, the linear growth rate of the neutral mode is significantly suppressed compared to the case without a background magnetic field. Specifically, the growth rate for ICS acquires a suppression factor of $(\omega_{0}/\omega_{\mathrm{c}})^4$, while for SBS the suppression is $(\omega_{0}/\omega_{\mathrm{c}})^{4/3}$. For ICS, an additional subluminal effect $(1+\omega_{\mathrm{p}}^2/\omega_{\mathrm{c}}^2)^{-1}$ arises due to the phase velocity of the EM wave falling below the speed of light. However, when $\omega_{\mathrm{p}} \ll \omega_{\mathrm{c}}$, this effect \Nishiura{is negligible}\footnote{
The exact correction factor associated with the subluminal effect is $1+\omega_{\mathrm{c}}^2\omega_{\mathrm{p}}^2/(\omega_{\mathrm{c}}^2-\omega_1^2)^2$ (see Eq.~(69) in \citep{2025PhRvD.111f3055N}). In the limit of a vanishing background magnetic field, $\omega_{\mathrm{c}} \rightarrow 0$, this correction reduces to unity.
}. The physical interpretation of these growth rate scalings is discussed in detail in Sec.~\ref{subsubsec:scattering_suppression}.

\subsubsection{Numerical Evaluation}
The linear growth rate of induced scattering for the neutral mode is obtained by numerically solving the dispersion relation in Eq.~\eqref{eq:dispersion_relation_neutral_Brillouin2}. In this study, we investigate how the maximum linear growth rate varies as a function of the normalized amplitude of the incident EM wave, defined with respect to the background magnetic field\Nishiura{~by using Eq. \eqref{eq:strength_parameter_no_magnetic_Brillouin} and $\omega_0=k_0v_{\text{A}}$},
\begin{equation}
    \eta \equiv \frac{\delta B}{B_0}
    = a_{\text{e}} \frac{\omega_{0}}{\omega_{\mathrm{c}}}
    \left(1+\frac{\omega_{\text{p}}^2}{\omega_{\text{c}}^2}\right)^{\frac{1}{2}},
    \label{eq:definition_of_eta}
\end{equation}
where $\delta B = 2k_0A_0$ denotes the peak amplitude of the magnetic field component of the incident EM wave for linear polarization~(see Appendix \ref{ap:linear_to_circular} for a circular polarized incident wave). The factor $2$ in the right-hand side arises because the EM wave is defined as Eq.~\eqref{eq:A_wave_general}. \ioka{The Alfvén velocity is defined by
\begin{equation}
v_{\text{A}}
\equiv \frac{c}{\sqrt{1 + \frac{\omega_{\text{p}}^2}{\omega_{\text{c}}^2}}}.
\label{eq:Alfven_velocity_Brillouin}
\end{equation}}
\begin{figure*}
\centering
\includegraphics[width=\textwidth]{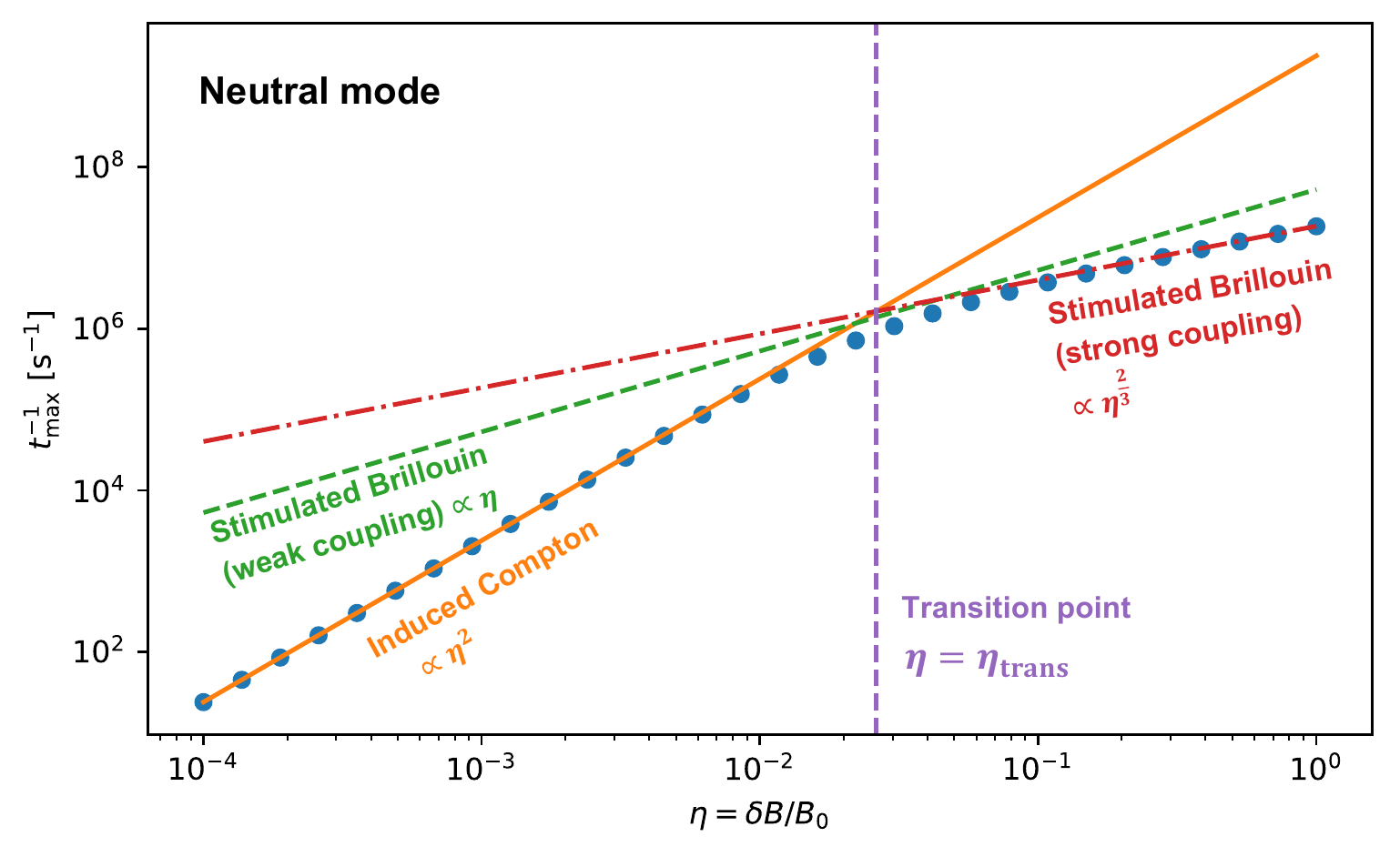}
\caption[Concept of Parametric Decay Instability]{\justifying The dependence of the maximum linear growth rate of induced scattering for the neutral mode on the incident wave amplitude \eqref{eq:definition_of_eta}. The orange solid line represents the growth rate of ICS, as given by Eq.~\eqref{eq:growth_rate_induced_Compton_subdominant_Brillouin}. The red dashed-dotted line corresponds to the SBS growth rate in the strong coupling regime, as given by Eq.~\eqref{eq:induced_Brillouin_neutral_maximum_growth_rate}. The green dotted line denotes the SBS growth rate in the weak coupling regime, given by Eq.~\eqref{eq:stimulated_Brillouin_weak_neutral}. The purple vertical dashed line indicates the incident wave amplitude at the transition point from weak to strong coupling, as given by Eq.~\eqref{eq:Transition_point_eta_neutral}. Blue dots represent the numerically obtained solutions of the dispersion relation in Eq.~\eqref{eq:dispersion_relation_neutral_Brillouin2}. The parameters used are $v_{\mathrm{th}}/c=10^{-4}$, $\omega_{\mathrm{p}}/\omega_0=10^{-2}$, $\omega_{\mathrm{c}}/\omega_0=10^{2}$, and $\omega_0=2\pi\times 10^9~\mathrm{Hz}$.}
\label{fig:stimulated_Brillouin_neutral_growth_rate}
\end{figure*}

Fig.~\ref{fig:stimulated_Brillouin_neutral_growth_rate} shows the maximum linear growth rate of the scattered wave as a function of the incident wave amplitude, defined in Eq.~\eqref{eq:definition_of_eta}. The transition point between weak and strong coupling, denoted as $\eta_{\text{trans}}$, is defined as follows using Eqs.~\eqref{eq:transition_point_for_neutral_mode} and \eqref{eq:definition_of_eta}:
\begin{equation}
\begin{aligned}
    \eta_{\text{trans}}
    &\equiv a_{\mathrm{e},\text{trans}}\frac{\omega_{0}}{\omega_{\mathrm{c}}}\left(1+\frac{\omega_{\text{p}}^2}{\omega_{\text{c}}^2}\right)^{\frac{1}{2}} \\
    &\simeq 4.4~\frac{\omega_{\text{c}}}{\omega_{\mathrm{p}}}
    \left(\frac{k_{\mathrm{B}}T_{\mathrm{e}}}{m_{\mathrm{e}} c^{2}}\right)^{\frac{3}{4}}
    \left(1+\frac{\omega_{\text{p}}^2}{\omega_{\text{c}}^2}\right)^{\frac{5}{4}}.
\end{aligned}    
\label{eq:Transition_point_eta_neutral}
\end{equation}
Numerical calculations show that as the amplitude of the incident wave increases, the dominant instability transitions continuously from ICS in the weak coupling regime to SBS in the strong coupling regime at the transition point $\eta_{\text{trans}}$.

In the weak coupling regime, as in the ordinary mode, SBS does not become the dominant instability. The maximum linear growth rate of SBS in this regime, predicted by the parametric instability based on the MHD approach for strongly magnetized plasma, is expressed as follows~\citep{1978ApJ...224.1013D,1978ApJ...219..700G,1993JGR....9819049J,2024PhRvE.110a5205I}:
\begin{equation}
\left(t_{\mathrm{B},\text{MHD}}^{\text{weak}}\right)^{-1}
\sim a_{\mathrm{e}} \omega_{\mathrm{p}}
\left(\frac{\omega_{0}}{\omega_{\mathrm{c}}}\right)^{2}
\left(\frac{m_{\mathrm{e}} c^{2}}{k_{\mathrm{B}} T_{\mathrm{e}}}\right)^{\frac{1}{4}}.
\label{eq:stimulated_Brillouin_weak_neutral}
\end{equation}
However, as illustrated in Fig.~\ref{fig:stimulated_Brillouin_neutral_growth_rate}, the kinetic approach for $e^\pm$ pair plasma indicates that ICS, rather than SBS, dominates in the weak coupling regime. The physical interpretation is the same as that for the ordinary mode and is discussed in Sec.~\ref{subsub:transition_weak_to_strong_coupling}.

\subsection{Charged mode}
This section analyzes the instability of the charged mode in a manner similar to the ordinary and neutral modes. \ioka{The physical picture is illustrated in Fig. \ref{fig:parametric_instability}. }The dispersion relation for the scattered wave, based on Eq.~\eqref{eq:dispersion_relation_charged_Brillouin} and the longitudinal dielectric function in Eq.~\eqref{eq:longitudinal_electric_susceptibility_Brillouin_Maxwell}, is expressed as follows:
\begin{equation}
\begin{aligned}
c^{2} k_{1}^{2} &- \omega_{1}^{2}
- \omega_{\mathrm{p}}^{2}\left(\frac{\omega_{1}}{\omega_{\mathrm{c}}}\right)^{2}
\simeq 
 \frac{1}{2} a_{\mathrm{e}}^{2} \omega_{\mathrm{p}}^{2} 
\left(\frac{c}{v_{\text{th}}}\right)^{2} \left(\frac{\omega_{0}}{\omega_{\mathrm{c}}}\right)^{2} 
 \\
& \times \left(1 - \mu^{2}\right) \left|\bm{n} \cdot \hat{\bm{B}}_{0}\right|^{2}\frac{1 + \zeta Z(\zeta)}{1 + \dfrac{2 \omega_{\mathrm{p}}^{2}}{k^{2} v_{\text{th}}^{2}}[1 + \zeta Z(\zeta)]}.
\end{aligned}
\label{eq:dispersion_relation_charged_Brillouin2}
\end{equation}
As summarized in Tab.~\ref{tab:roadmap_coupling}, the charged mode exhibits not only ICS and SBS but also SRS, in contrast to the ordinary and neutral modes. For each instability, approximate solutions for the growth rate can be derived in the following two limits.\\
\textbf{Noncollective limit (low density regime):}
\begin{equation}
\left(\frac{\omega_{\mathrm{p}}}{k v_{\mathrm{th}}}\right)^{2}
= \frac{1}{8\pi^2}\left(\frac{\lambda}{\lambda_{\text{De}}}\right)^2 \ll 1.
\label{eq:limitation_of_plasma_frequency_Brilluin1}
\end{equation}
\textbf{Collective limit (intermediate and high density regimes):}
\begin{equation}
\left(\frac{\omega_{\mathrm{p}}}{k v_{\mathrm{th}}}\right)^{2}
= \frac{1}{8\pi^2} \left(\frac{\lambda}{\lambda_{\text{De}}}\right)^2 \gg 1.
\label{eq:plasma_density_dekai_debye_screening_Brillouin}
\end{equation}
The Debye length is defined as
\begin{equation}
\lambda_{\text{De}} \equiv
\left(\frac{k_{\text{B}}T_{\text{e}}}{8 \pi e^2 n_{\text{e}0}}\right)^{1/2}
= \frac{v_{\text{th}}}{\sqrt{2} \omega_{\text{p}}}.
\label{eq:definition_of_Debye_length_induced_Compton}
\end{equation}

The noncollective limit corresponds to the regime where the wavelength of density fluctuations is much shorter than the Debye length (or plasma density is low), so the particle dynamics are governed by thermal motion. In contrast, the collective limit describes the regime where the wavelength of density fluctuations is much longer than the Debye length (or plasma density is high), and the motion is governed by collective plasma behavior.

As will be discussed later, the collective limit defined by Eq.~\eqref{eq:plasma_density_dekai_debye_screening_Brillouin} can be further subdivided into intermediate and high density regimes, depending on the density scale (see Fig.~\ref{fig:charged_regime}). Each region exhibits different dominant instabilities and scaling relations for the growth rate. This section clarifies the characteristic features of these regimes. Detailed derivations of the linear growth rates and angle conditions for each regime are presented in Appendix~\ref{sec:charged_mode_detail_derivation}.

\subsubsection{Induced Compton Scattering (Charged mode)}
For charged mode, ICS under the weak coupling condition (see Tab.~\ref{tab:roadmap_coupling}) exhibits distinct behavior for the growth rate in the noncollective limit~\eqref{eq:limitation_of_plasma_frequency_Brilluin1} and the collective limit~\eqref{eq:plasma_density_dekai_debye_screening_Brillouin}.

\paragraph{Low density regime (noncollective limit)}
In the dispersion relation~\eqref{eq:dispersion_relation_charged_Brillouin2} for the charged mode, the longitudinal dielectric function~\eqref{eq:longitudinal_electric_susceptibility_Brillouin_Maxwell} can be approximated as $\varepsilon_{\mathrm{L}} \simeq 1$ in the noncollective limit~\eqref{eq:limitation_of_plasma_frequency_Brilluin1}. Thus, the dispersion relation simplifies as follows,
\begin{equation}
\begin{aligned}
c^{2} k_{1}^{2} - \omega_{1}^{2}
-& \omega_{\text{p}}^{2}\left(\frac{\omega_{1}}{\omega_{\mathrm{c}}}\right)^2
\simeq 
 \frac{1}{2} a_{\mathrm{e}}^{2} \omega_{\mathrm{p}}^{2} 
\left(\frac{c}{v_{\text{th}}}\right)^{2} 
\left(\frac{\omega_{0}}{\omega_{\mathrm{c}}}\right)^{2} \\
& \times (1 - \mu^{2})|\bm{n} \cdot \hat{\bm{B}}_{0}|^{2}[1 + \zeta Z(\zeta)].
\end{aligned}
\end{equation}

As indicated in Tab.~\ref{tab:roadmap_coupling}, ICS dominates the instability under the weak coupling condition~\eqref{eq:weak_coupling_condition}.\footnote{In contrast, under the strong coupling condition~\eqref{eq:strong_coupling_condition}, the dominant instability is the SBS, which is equivalent to the strong-coupling SRS, as discussed later.} When the exponential term $\mathrm{i} \sqrt{\pi} \mathrm{e}^{-\zeta^2}$ in the plasma dispersion function~\eqref{eq:plasma_dispersion_function_Brillouin} is dominant, the linear growth rate is given by \nIshiura{Eqs. (78) and (79) in~\citep{2025PhRvD.111f3055N},}
\begin{equation}
\begin{aligned}
\left(t_{\mathrm{C,charged}}^{\max }\right)^{-1} =
\sqrt{\frac{\pi}{32 \mathrm{e}}}
\left(\frac{\omega_0}{\omega_{\text{c}}}\right)^2
\frac{\omega_{\mathrm{p}}^{2} a_{\mathrm{e}}^{2}}{\omega_{0}}
\frac{m_{\mathrm{e}} c^{2}}{k_{\mathrm{B}} T_{\mathrm{e}}}
\left(1+\frac{\omega_{\text{p}}^2}{\omega_{\text{c}}^2}\right)^{-1}.
\end{aligned}
\label{eq:maximum_growth_rate_magnetic2_Brillouin}
\end{equation}
The wavenumber at which maximum growth occurs is given by \nIshiura{(see Eqs. (74) and (82) in \citep{2025PhRvD.111f3055N})}
\begin{equation}
\begin{aligned}
k_{\max} &\simeq {} 
\sqrt{2(1-\nu)}\,k_{0} \\
& \times \!\Biggl\{
1-\sqrt{\frac{1-\nu}{2}\,
      \cos^{2}\theta_{kB}\,
      \frac{k_{\mathrm{B}}T_{\mathrm{e}}}{m_{\mathrm{e}}c^{2}}
      \Bigl(1+\frac{\omega_{\mathrm{p}}^{2}}{\omega_{\mathrm{c}}^{2}}\Bigr)}
\Biggr\}.
\end{aligned}
\label{eq:maximum_geowth_wave_vector_charged_Compton}
\end{equation}
For ICS to dominate, the incident wave amplitude~\Nishiura{$a_{\text{e}} \omega_0/\omega_{\text{c}}$} must satisfy
\begin{equation}
\begin{aligned}
a_{\mathrm{e}}\frac{\omega_{0}}{\omega_{\mathrm{c}}}
\ll {} &
4\!\left(\frac{2\text{e}}{\pi}\right)^{\frac{1}{4}}
\!\left(\frac{k_{\mathrm{B}}T_{\mathrm{e}}}{m_{\mathrm{e}}c^{2}}\right)^{\frac{3}{4}}
\frac{\omega_{0}}{\omega_{\mathrm{p}}}
\left(1+\frac{\omega_{\mathrm{p}}^{2}}{\omega_{\mathrm{c}}^{2}}\right)^{\frac{3}{4}} \\
& \times 
\Bigl[(1-\nu)\cos^{2}\theta_{kB}\Bigr]^{\frac{1}{4}}.
\end{aligned}
\label{eq:strength_parameter_limit_perp2-2_Brillouin}
\end{equation}
This condition is derived by substituting the growth rate~\eqref{eq:maximum_growth_rate_magnetic2_Brillouin} into the weak coupling condition~\eqref{eq:weak_coupling_condition} \nIshiura{(see Eq. (100) in \citep{2025PhRvD.111f3055N})}.

The maximum growth rate is achieved for the angular parameters \nIshiura{(see Eqs. (78) and (79) in \citep{2025PhRvD.111f3055N})},
\begin{equation}
    \left(\mu,~\left|\bm{n} \cdot \hat{\bm{B}}_{0}\right|\right)=\left(0,1\right),
    \label{eq:maximum_growth_angle_condition_charged_nonDebye}
\end{equation}
which correspond to a mutually orthogonal configuration of the background magnetic field, the electric field of the incident wave, and the electric field of the scattered wave.

When the angular variables satisfy Eq.~\eqref{eq:maximum_growth_angle_condition_charged_nonDebye}, the validity condition for the low density regime is obtained by substituting the maximum growth wavenumber~\eqref{eq:maximum_geowth_wave_vector_charged_Compton} into the noncollective limit~\eqref{eq:limitation_of_plasma_frequency_Brilluin1}. In the non-relativistic limit $v_{\text{th}}/c \ll 1$, the condition can be expressed as
\begin{equation}
\begin{aligned}
\frac{\omega_{\mathrm{p}}}{\omega_{0}}
\ll \sqrt{4(1-\nu)\frac{ k_{\mathrm{B}} T_{\mathrm{e}}}{m_{\mathrm{e}} c^{2}}\left(1+\frac{\omega_{\text{p}}^2}{\omega_{\text{c}}^2}\right)}.
\end{aligned}
\label{eq:charged_mode_instability_thin_condition_Brillouin_Brillouin}
\end{equation}

\paragraph{Intermediate and high density regimes (collective limit)}
In the charged mode, the dispersion relation given by Eq.~\eqref{eq:dispersion_relation_charged_Brillouin2} requires a treatment that includes the longitudinal dielectric function in the collective limit~\eqref{eq:plasma_density_dekai_debye_screening_Brillouin}. The detailed analysis of ICS in this regime has already been presented in \citep{2025PhRvD.111f3055N}. As indicated in Tab.~\ref{tab:roadmap_coupling}, ICS remains dominant under the weak coupling condition~\eqref{eq:weak_coupling_condition}, but in the collective limit~\eqref{eq:plasma_density_dekai_debye_screening_Brillouin}, Debye screening modifies the instability\footnote{For the strong coupling regime, as discussed later, the instability is dominated by the strong-coupling SRS (or strong-coupling SBS).}.

When the exponential term $\mathrm{i} \sqrt{\pi} \mathrm{e}^{-\zeta^2}$ in the asymptotic expansion of the plasma dispersion function~\eqref{eq:plasma_dispersion_function_Brillouin} is dominant, the linear growth rate is given by\footnote{\label{fn:growth_correction}\niShiura{As noted in footnote 8 of \citep{2025PhRvD.111f3055N}, Eq.~\eqref{eq:induced_Compton_Debye_screening_Brillouin} is obtained by applying the $|\zeta|\ll1$ expansion of the plasma dispersion function, Eq.~\eqref{eq:definition_of_plasma_dispersion_function}, as given in Eq.~\eqref{eq:plasma_dispersion_function_Brillouin}. When the approximation is not used and the coefficient is instead computed numerically, the maximum growth rate becomes about $34\%$ larger than Eq.~\eqref{eq:induced_Compton_Debye_screening_Brillouin}. In Fig.~\ref{fig:Debye_Compton_charged_growth_rate}, the analytic curve (orange solid) incorporates this numerical correction, which brings it into close agreement with the numerical solution of the dispersion relation.}}
 \nIshiura{(see Eq. (72) in \citep{2025PhRvD.111f3055N})}
\begin{equation}
\begin{aligned}
\left(t_{\mathrm{C,charged}}^{\max}\right)^{-1}
&= \sqrt{\frac{32 \text{e}}{\pi}}
   \frac{k_{\mathrm{B}}T_{\mathrm{e}}}{m_{\mathrm{e}}c^{2}}
   \left(\frac{\omega_{0}}{\omega_{\mathrm{c}}}\right)^{2}
\\
&\quad \times
   \left(\frac{\omega_{0}}{\omega_{\mathrm{p}}}\right)^{4}
   \frac{a_{\mathrm{e}}^{2}\omega_{\mathrm{p}}^{2}}{\omega_{0}}
   \left(1 + \frac{\omega_{\mathrm{p}}^{2}}{\omega_{\mathrm{c}}^{2}}\right).
\end{aligned}
\label{eq:induced_Compton_Debye_screening_Brillouin}
\end{equation}
The wavenumber corresponding to maximum growth is given by \nIshiura{(see Eq. (74) in \citep{2025PhRvD.111f3055N})}
\begin{equation}
    k_{\text{max}} \simeq
    2k_0 \left(1 -
    \sqrt{\frac{k_{\mathrm{B}} T_{\mathrm{e}}}{m_{\mathrm{e}} c^2}
    \left(1 + \frac{\omega_{\mathrm{p}}^2}{\omega_{\mathrm{c}}^2}\right)}
    \right).
    \label{eq:wave_vector_charged_Brilliouin}
\end{equation}
For ICS to be dominant, the incident wave amplitude~\Nishiura{$a_{\text{e}} \omega_0/\omega_{\text{c}}$} must satisfy
\begin{equation}
\begin{aligned}
 a_{\mathrm{e}} \frac{\omega_{0}}{\omega_{\mathrm{c}}} \ll
 \left(\frac{\pi}{\text{e}}\right)^{\frac{1}{4}}
 \frac{\omega_{\mathrm{p}}}{\omega_{0}}
 \left(\frac{m_{\mathrm{e}} c^{2}}{k_{\mathrm{B}} T_{\mathrm{e}}}\right)^{\frac{1}{4}}
 \left(1+\frac{\omega_{\text{p}}^2}{\omega_{\text{c}}^2}\right)^{-\frac{1}{4}}.
\end{aligned}
\label{eq:weak_coupling_condition_for_Debye_Compton}
\end{equation}
This condition is derived by substituting the growth rate given by Eq.~\eqref{eq:induced_Compton_Debye_screening_Brillouin} into the weak coupling condition~\eqref{eq:weak_coupling_condition} \nIshiura{(see Eq. (75) in \citep{2025PhRvD.111f3055N})}. 

The maximum growth angle parameters are \nIshiura{(see Eq. (73) in \citep{2025PhRvD.111f3055N})}
\begin{equation}
    \left(\mu,~\nu,~\cos\theta_{kB},~\left|\bm{n} \cdot \hat{\bm{B}}_{0}\right|\right) = \left(0, -1, \pm1, 1\right).
    \label{eq:maximum_growth_angle_condition_charged_Debye}
\end{equation}
This corresponds to a geometry where the background magnetic field, the incident wave, and the scattered wave electric field are mutually orthogonal, and the incident wave propagates along the background magnetic field to produce $180^\circ$ backscattering. For arbitrary incidence angles, the growth rate is reduced by a factor of a few. 

The validity condition for the intermediate and high density regimes can be expressed, by substituting the maximum growth wavenumber~\eqref{eq:wave_vector_charged_Brilliouin} into the collective limit~\eqref{eq:plasma_density_dekai_debye_screening_Brillouin}, as
\begin{equation}
\sqrt{\frac{8 k_{\mathrm{B}} T_{\mathrm{e}}}{m_{\mathrm{e}} c^{2}}\left(1+\frac{\omega_{\text{p}}^2}{\omega_{\text{c}}^2}\right)} \ll \frac{\omega_{\mathrm{p}}}{\omega_{0}}.
\label{eq:charged_mode_instability_thin_condition_Brillouin_Compton}
\end{equation}

In the collective limit, ICS and SRS may coexist. In particular, SRS tends to be dominant in the intermediate density regime. The detailed competition between the two processes is discussed in Appendix~\ref{subsec:raman_compton_competition_charged}. The physical reason for simultaneous excitation of ICS and SRS is further addressed in Sec.~\ref{subsub:transition_weak_to_strong_coupling}.

\subsubsection{Stimulated Brillouin Scattering (Charged mode)}
\label{subsubsec:charged_Brillouin}
SBS becomes the dominant instability in the charged mode under the strong coupling condition~\eqref{eq:strong_coupling_condition}, as shown in Tab.~\ref{tab:roadmap_coupling}. This holds for both the noncollective limit~\eqref{eq:limitation_of_plasma_frequency_Brilluin1} and the collective limit~\eqref{eq:definition_of_Debye_length_induced_Compton}, and is degenerate with the strong coupling limit of SRS. See Sec.~\ref{subsec:degeneracy_Brillouin_Raman_charged_mode} for further discussion of this physical equivalence. 

Following the approach for the ordinary and neutral modes, the plasma dispersion function $Z(\zeta)$ is approximated by its asymptotic expansion for $|\zeta| \gg 1$ in Eq.~\eqref{eq:plasma_dispersion_function_Brillouin}. Substituting Eq.~\eqref{eq:plasma_dispersion_function_yuurikannsuu_Brillouin} into the dispersion relation~\eqref{eq:dispersion_relation_charged_Brillouin2} yields \ioka{Eq. \eqref{eq:dispersion_relation_neutral_tochuu_Raman} in Appendix~\ref{subsec:strong_coupling_Brillouin_charged}.}\rei{the following:
\begin{equation}
\begin{aligned}
&\left(\omega_{\mathrm{p}}^{2}\cos^{2} \theta_{kB}-\omega^{2} \right)
\left\{\omega\frac{c^2}{v_{\text{A}}^2} - \frac{c^{2}\left(k^{2} + 2 \bm{k}_{0} \cdot \bm{k}\right)}{2 \omega_{0}}\right\} \\
&= -\frac{1}{8} \frac{a_{\mathrm{e}}^{2} \omega_{\mathrm{p}}^{2} c^{2} k^{2}}{\omega_{0}}
\left(\frac{\omega_{0}}{\omega_{\mathrm{c}}}\right)^{2}\cos^{2} \theta_{kB}\left(1 - \mu^{2}\right)\left|\bm{n} \cdot \hat{\bm{B}}_{0}\right|^{2}.
\end{aligned}
\label{eq:dispersion_relation_tochuu_charged_Brillouin}
\end{equation}
Details of this derivation are provided in Appendix~\ref{subsec:strong_coupling_Brillouin_charged}.}

The growth rate is given by
\begin{equation}
\left(t_{\mathrm{B},\text{charged}}^{\max}\right)^{-1}  = \sqrt{3}\left(\frac{a_{\text{e}}^{2}\omega_{\mathrm{p}}^{2}\omega_{0}}{2}\right)^{\frac{1}{3}}\left(\frac{\omega_{0}}{\omega_{\mathrm{c}}}\right)^{\frac{2}{3}}.
\label{eq:induced_Brillouin_charged_maximum_growth_rate}
\end{equation}
The wavenumber corresponding to maximum growth is given by
\begin{equation}
k_{\text{max}} = 2k_0,
\label{eq:induced_Brillouin_charged_wave_vector}
\end{equation}
and for SBS to be dominant, the strength parameter must satisfy
\begin{equation}
8.3~\frac{\omega_{0}}{\omega_{\mathrm{p}}}
\left(\frac{ k_{\mathrm{B}} T_{\mathrm{e}}}{m_{\mathrm{e}} c^{2}}\right)^{\frac{3}{4}}
\left(1+\frac{\omega_{\text{p}}^2}{\omega_{\text{c}}^2}\right)^{\frac{3}{4}}
\ll a_{\mathrm{e}} \frac{\omega_{0}}{\omega_{\mathrm{c}}} \ll 1.
\label{eq:strong_coupling_condition_charged_Brillouin}
\end{equation}
This condition is derived by substituting the strong coupling condition~\eqref{eq:strong_coupling_condition} into the growth rate~\eqref{eq:induced_Brillouin_charged_maximum_growth_rate} together with Eq.~\eqref{eq:non_rela_neutral_and_charged_Brillouin}. The maximum growth angle is achieved for
\begin{equation}
    \left(\mu,~\nu,~\cos\theta_{kB},~\left|\bm{n} \cdot \hat{\bm{B}}_{0}\right|\right)=\left(0,-1,\pm1,1\right).
    \label{eq:maximum_growth_angle_condition_charged_Brillouin}
\end{equation}

\subsubsection{Stimulated Raman Scattering (Charged mode)}
\label{subsubsec:stimulated_Raman}

As shown in Tab.~\ref{tab:roadmap_coupling}, SRS can be excited in the charged mode under both weak and strong coupling conditions. Under strong coupling, SRS is degenerate with SBS, and both the derivation and the resulting growth rate are identical. The detailed physical equivalence is discussed in Sec.~\ref{subsec:degeneracy_Brillouin_Raman_charged_mode}. Therefore, in the following, we focus on SRS in the weak coupling regime \eqref{eq:weak_coupling_condition}.

SRS exhibits qualitatively different behavior depending on the plasma density regime. In the low density regime, the scattering angle is strongly constrained, and only small-angle scattering is allowed. In contrast, SRS is not excited in the high density regime. As shown in Appendix~\ref{subsec:raman_compton_competition_charged}, SRS always dominates over ICS in the intermediate density regime.

\paragraph{Intermediate and high density regimes}
As indicated in Tab.~\ref{tab:roadmap_coupling}, SRS in the weak coupling regime is dominant when the weak coupling condition~\eqref{eq:weak_coupling_condition} holds and the resonance condition and the requirement of negligible Landau damping for the Langmuir wave are satisfied,
\begin{equation}
    |\mathrm{Re}\,\omega| \simeq \omega_{\mathrm{p}}\gg k_{\parallel} v_{\mathrm{th}}.
    \label{eq:resonance_condition_raman}
\end{equation}
As with SBS in the neutral mode, we apply the asymptotic expansion of the plasma dispersion function $Z(\zeta)$ for $|\zeta|\gg1$ in Eq.~\eqref{eq:plasma_dispersion_function_Brillouin} and rewrite the dispersion relation~\eqref{eq:dispersion_relation_charged_Brillouin2} to obtain Eq.~\eqref{eq:dispersion_relation_neutral_tochuu_Raman}. Under the resonance condition in Eq.~\eqref{eq:resonance_condition_raman}, the growth rate is given by (see Appendix~\ref{subsec:weak_coupling_Raman_charged} for derivation)\footnote{
\nIshiura{The growth rate obtained in this study is in good agreement with the numerical solutions of the dispersion relation (see Fig.~\ref{fig:stimulated_Raman_charged_growth_rate}). In contrast, the result presented in \citep{2003PhRvE..67d6406M} differs by a factor of $(\omega_{0}/\omega_{\mathrm{p}})^{\frac{1}{2}}$ (see Eq.~(40) in \citep{2003PhRvE..67d6406M}). This discrepancy is likely due to a typographical error in \citep{2003PhRvE..67d6406M}.}
}
\begin{equation}
\left(t_{\text{R}}^{\text{max}}\right)^{-1}  = a_{\mathrm{e}} \frac{\omega_{0}}{\omega_{\mathrm{c}}}\left(\omega_{0} \omega_{\mathrm{p}}\right)^{\frac{1}{2}}.
\label{eq:growth_rate_of_induced_Raman_weak_coupling}
\end{equation}
The wavenumber corresponding to maximum growth is expressed as
\begin{equation}
k_{\text{max}} \simeq k_{0}\left(1 + \sqrt{1 - 2 \frac{\omega_{\mathrm{p}}}{\omega_{0}}}\right).
\label{eq:maximum_growth_wave_number_Raman5}
\end{equation}
For SRS to be dominant in the weak coupling regime, the incident wave amplitude~\Nishiura{$a_{\text{e}} \omega_0/\omega_{\text{c}}$} must satisfy
\begin{equation}
a_{\mathrm{e}} \frac{\omega_{0}}{\omega_{\mathrm{c}}} \ll 5.7~
\left(
    \frac{k_{\mathrm{B}} T_{\mathrm{e}}}{m_{\mathrm{e}} c^2}
\right)^{\frac{1}{2}}
\left(
    1 + \frac{\omega_{\mathrm{p}}^2}{\omega_{\mathrm{c}}^2}
\right)^{\frac{1}{2}}
\left(
    \frac{\omega_0}{\omega_{\mathrm{p}}}
\right)^{\frac{1}{2}}.
\label{eq:weak_coupling_condition_for_weak_Raman}
\end{equation}
This condition is obtained by substituting the growth rate~\eqref{eq:growth_rate_of_induced_Raman_weak_coupling} into the weak coupling condition~\eqref{eq:weak_coupling_condition}\nIshiura{, using $\omega_{\text{p}}\ll\omega_0$}. The angle parameters that yield the maximum growth are
\begin{equation}
    \left(\mu,~\nu,~\cos\theta_{kB},~\left|\bm{n} \cdot \hat{\bm{B}}_{0}\right|\right)=\left(0,-1,\pm1,1\right).
    \label{eq:maximum_growth_angle_condition_charged_Raman}
\end{equation}
If the incident angle is arbitrary, the growth rate decreases by a factor of a few. When the condition in Eq.~\eqref{eq:maximum_growth_angle_condition_charged_Raman} is satisfied, the condition for the intermediate density regime is obtained by substituting the wavenumber~\eqref{eq:maximum_growth_wave_number_Raman5} into Eq.~\eqref{eq:resonance_condition_raman}. Assuming $\omega_{\mathrm{p}}\ll\omega_0$, this leads to the inequality~\eqref{eq:charged_mode_instability_thin_condition_Brillouin_Compton}. Additionally, requiring the wavenumber in Eq.~\eqref{eq:maximum_growth_wave_number_Raman5} to be real yields
\begin{equation}
1 - 2 \frac{\omega_{\mathrm{p}}}{\omega_{0}} > 0.
\label{eq:induced_Raman_condition_upper_limit}
\end{equation}
Therefore, the conditions~\eqref{eq:charged_mode_instability_thin_condition_Brillouin_Compton} and~\eqref{eq:induced_Raman_condition_upper_limit} must be satisfied simultaneously. This defines the intermediate density regime as
\begin{equation}
\sqrt{\frac{8 k_{\mathrm{B}} T_{\mathrm{e}}}{m_{\mathrm{e}} c^{2}}\left(1+\frac{\omega_{\text{p}}^2}{\omega_{\text{c}}^2}\right)} \ll \frac{\omega_{\mathrm{p}}}{\omega_{0}} < \frac{1}{2}.
\label{eq:induced_Raman_condition_Brillouin}
\end{equation}

On the other hand, the high density regime is defined by
\begin{equation}
\niShiura{\max\left\{\frac{1}{2},~\sqrt{\frac{8 k_{\mathrm{B}} T_{\mathrm{e}}}{m_{\mathrm{e}} c^{2}}\left(1+\frac{\omega_{\text{p}}^2}{\omega_{\text{c}}^2}\right)}\right\} < \frac{\omega_{\mathrm{p}}}{\omega_{0}},}
\label{eq:charged_mode_instability_dense_condition_Brillouin}
\end{equation}
Within this regime, weak coupling SRS cannot be excited.%
\footnote{If the plasma temperature is sufficiently high that the intermediate density regime does not exist, i.e.,
\begin{equation}
\frac{1}{2}<\sqrt{\frac{8 k_{\mathrm{B}} T_{\mathrm{e}}}{m_{\mathrm{e}} c^{2}}\left(1+\frac{\omega_{\text{p}}^2}{\omega_{\text{c}}^2}\right)}.
\nonumber
\end{equation}
then, even in the region where $1/2<\omega_{\mathrm{p}}/\omega_{0}$, SRS may still be driven in the low density regime due to small angle scattering, as described in the next section (see Fig.~\ref{fig:charged_regime}).}

\paragraph{Low density regime (small angle scattering)}
As described in the previous section, weak coupling SRS achieves maximum growth in the case of $180^\circ$ backward scattering, represented by Eq.~\eqref{eq:maximum_growth_angle_condition_charged_Raman}. However, such backward scattering is only realized in the intermediate density regime, as given by Eq.~\eqref{eq:induced_Raman_condition_Brillouin}, and is not allowed in the low density regime described by Eq.~\eqref{eq:charged_mode_instability_thin_condition_Brillouin}.

On the other hand, even within the low density regime, it is well known for ion-electron plasma that SRS can be excited if restricted to small angle scattering \citep{1994ApJ...422..304T,2008ApJ...682.1443L,2021Univ....7...56L}. In a strong background magnetic field, the dependence of the growth rate on the angle parameters differs from the non-magnetized case. Nevertheless, a similar analytical approach can be applied. A detailed derivation is provided in Appendix~\ref{subsec:small_angle_Raman_charged} and~\ref{subsec:small_angle_Raman_angle_max_charged}. The main results are summarized below.

\nIshiura{We consider the resonance between EM waves and Langmuir waves propagating at a general angle $\theta_{kB}$ with respect to the background magnetic field}\ioka{, as detailed in Appendix~\ref{subsec:small_angle_Raman_charged}.}\rei{, as expressed by
\begin{equation}
\omega = -\omega_{\mathrm{p}} \left|\cos \theta_{kB}\right|.
\end{equation}}
Suppression of Landau damping for the Langmuir wave \ioka{imposes an upper limit on the wavenumber of the density fluctuation in Eq. \eqref{eq:upper_limit_wave_number_small_angle}. Combining this condition with the wavenumber determined by the energy-momentum conservation condition in Eq. \eqref{eq:energy_momentum_conservation_Compton_Brillouin} gives an upper limit on the scattering angle, Eq.~\eqref{eq:max_1-nu_Raman}. Thus, in the low density regime, weak-coupling SRS is allowed only for small-angle scattering.}

With this restriction, the maximum growth rate of SRS is given from Eq. \eqref{eq:SRS_small_angele_Unified} by
\begin{equation}
\begin{aligned}
\left(t_{\text{R}}^{\text{max}}\right)^{-1} \sim{}& 0.30\,a_{\mathrm{e}}
\frac{\omega_0}{\omega_{\mathrm{c}}}
\left(\omega_0 \omega_{\mathrm{p}}\right)^{\tfrac12}
\left(\frac{m_{\mathrm{e}} c^{2}}{32 k_{\mathrm{B}} T_{\mathrm{e}}}\right)^{\tfrac12}
\frac{\omega_{\mathrm{p}}}{\omega_0}
\left(1+\frac{\omega_{\mathrm{p}}^{2}}{\omega_{\mathrm{c}}^{2}}\right)^{-\tfrac12}.
\end{aligned}
\label{eq:stimulated_Raman_growth_rate_small_angle}
\end{equation}
For this instability to be driven, the incident wave amplitude $a_{\text{e}}\omega_0/\omega_{\text{c}}$ must satisfy the following condition:
\begin{equation}
a_{\mathrm{e}} \frac{\omega_0}{\omega_{\mathrm{c}}} 
\ll 5.1 \left( \frac{\omega_0}{\omega_{\mathrm{p}}} \right)^{\frac{1}{2}} 
\left( \frac{32 k_{\mathrm{B}} T_{\mathrm{e}}}{m_{\mathrm{e}} c^2} \right)^{\frac{1}{2}} 
\left( 1 + \frac{\omega_{\mathrm{p}}^2}{\omega_{\mathrm{c}}^2} \right)^{\frac{1}{2}}.
\label{eq:weak_coupling_small_Raman}
\end{equation}
This condition is obtained by substituting Eq.~\eqref{eq:stimulated_Raman_growth_rate_small_angle} into the weak coupling condition, Eq.~\eqref{eq:weak_coupling_condition_induced_Raman_scattering1}\niShiura{, using Eqs. \eqref{eq:costheta_kB_angleparameter} and \eqref{eq:phi_polarization_small_angle}}. In the low density regime, both the small-angle SRS and the ICS given by Eq.~\eqref{eq:maximum_growth_rate_magnetic2_Brillouin} can be simultaneously driven. The detailed physical reason for the simultaneous excitation of ICS and SRS is discussed in Sec.~\ref{subsub:transition_weak_to_strong_coupling}.

\phantomsection
\paragraph{Competition between ICS and SRS in the intermediate density regime}
\label{par:competition_Compton_Raman}
In intermediate density regime~\eqref{eq:induced_Raman_condition_Brillouin}, both ICS and SRS can be driven simultaneously. However, the growth rate of SRS always exceeds that of ICS. A detailed proof of this result is provided in Appendix~\ref{subsec:raman_compton_competition_charged}.

\subsubsection{Summary of Charged Mode}
\begin{figure*}
\centering
\includegraphics[width=\textwidth]{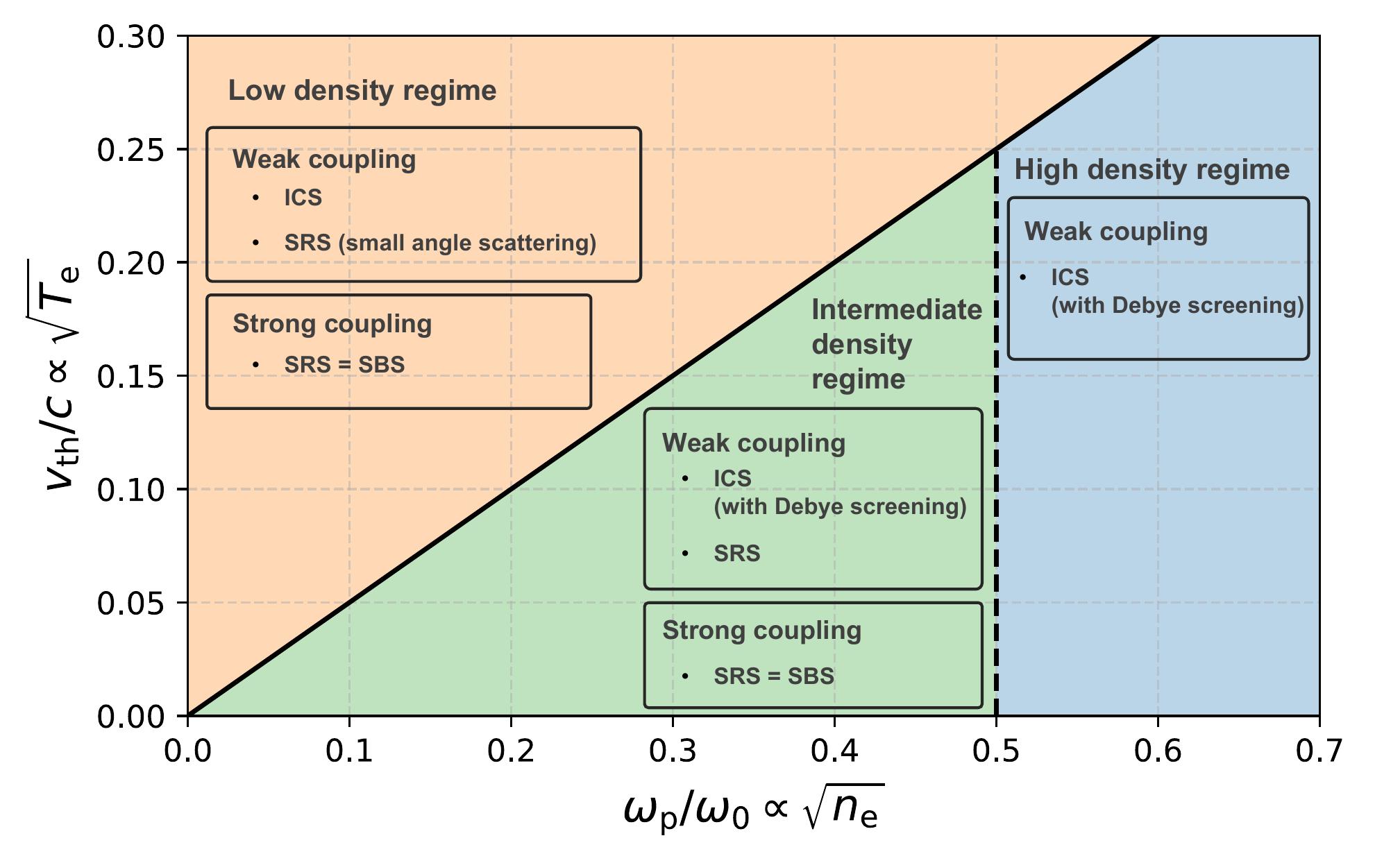}
\caption[Concept of Parametric Decay Instability]{\justifying The further subdivision of the charged mode instability classification, as summarized in Tab.~\ref{tab:roadmap_coupling}, according to plasma density and temperature. The horizontal axis shows the dimensionless plasma frequency, while the vertical axis represents the dimensionless thermal velocity. The solid black diagonal line denotes the boundary between the low density and intermediate/high density regimes, as defined by Eqs.~\eqref{eq:charged_mode_instability_thin_condition_Brillouin} and \eqref{eq:induced_Raman_condition_Brillouin}. The black dashed line indicates the boundary between the intermediate  and high density regimes, following Eqs.~\eqref{eq:induced_Raman_condition_Brillouin} and \eqref{eq:charged_mode_instability_dense_condition_Brillouin}. The dominant induced scattering process in each density regime (ICS: induced Compton scattering, SRS: stimulated Raman scattering, SBS: stimulated Brillouin scattering) is indicated within the diagram. For simplicity, the subluminal effect $(1+\omega_{\mathrm{p}}^2/\omega_{\mathrm{c}}^2)$ appearing in Eq.~\eqref{eq:Alfven_velocity_Brillouin} is neglected by assuming it is of order unity~\citep{2025PhRvD.111f3055N}. Note also that, in the high density regime, strong coupling does not occur because, as discussed in Eq.~\eqref{eq:weak_coupling_Debye_screening_henkei}, the incident wave amplitude~\Nishiura{$a_{\text{e}} \omega_0/\omega_{\text{c}}$} always satisfies the weak coupling condition as long as it remains within the linear regime.}
\label{fig:charged_regime}
\end{figure*}
Instabilities in the charged mode are initially classified according to coupling and resonance conditions, as summarized in Tab.~\ref{tab:roadmap_coupling}. For the charged mode, a more detailed classification is provided in the density–temperature plane, as illustrated in Fig.~\ref{fig:charged_regime}. This map delineates the nature of scattering processes in the low, intermediate, and high density regimes. In the following, we summarize the specific behaviors in each density regime based on the analytic results for the growth rates derived above.

\paragraph{Low density regime}
According to Tab.~\ref{tab:roadmap_coupling}, both ICS and SRS are excited in the weak coupling condition, while SBS (which is degenerate with strong coupling SRS) is driven in the strong coupling condition. The low density regime is defined as follows:
\begin{equation}
\begin{aligned}
\frac{\omega_{\mathrm{p}}}{\omega_{0}} \ll
\sqrt{\frac{8 k_{\mathrm{B}} T_{\mathrm{e}}}{m_{\mathrm{e}} c^{2}}
\left(1+\frac{\omega_{\text{p}}^2}{\omega_{\text{c}}^2}\right)}.
\end{aligned}
\label{eq:charged_mode_instability_thin_condition_Brillouin}
\end{equation}
In this regime, as indicated by Eq.~\eqref{eq:max_1-nu_Raman}, SRS is only excited through small-angle scattering, and backscattering is prohibited.

The linear growth rates at the angle parameter where both ICS and SBS attain their maximum growth, corresponding to the condition in Eq.~\eqref{eq:maximum_growth_angle_condition_charged_Brillouin}, are summarized as follows, using Eqs.~\eqref{eq:maximum_growth_rate_magnetic2_Brillouin} and \eqref{eq:induced_Brillouin_charged_maximum_growth_rate}:
\begin{equation}
\left(t_{\text{charged}}^{\text{max}}\right)^{-1} \sim
\begin{cases}
\sqrt{\frac{\pi}{32 \text{e}}}\,
\frac{a_{\mathrm{e}}^{2}\omega_{\mathrm{p}}^{2}}{\omega_{0}}
\frac{m_{\mathrm{e}} c^{2}}{k_{\mathrm{B}} T_{\mathrm{e}}}
\left(\frac{\omega_{0}}{\omega_{\mathrm{c}}}\right)^{2}
\left(1+\frac{\omega_{\mathrm{p}}^{2}}{\omega_{\mathrm{c}}^{2}}\right)^{-1},\\
\quad
a_{\mathrm{e}}\frac{\omega_{0}}{\omega_{\mathrm{c}}}
\ll
5.5~\frac{\omega_{0}}{\omega_{\mathrm{p}}}
\left(\frac{k_{\mathrm{B}} T_{\mathrm{e}}}{m_{\mathrm{e}} c^{2}}\right)^{\frac{3}{4}}
\left(1+\frac{\omega_{\mathrm{p}}^{2}}{\omega_{\mathrm{c}}^{2}}\right)^{\frac{3}{4}},\\
\sqrt{3}\left(\frac{a_{\mathrm{e}}^{2}\omega_{\mathrm{p}}^{2}\omega_{0}}{2}\right)^{\frac{1}{3}}
\left(\frac{\omega_{0}}{\omega_{\mathrm{c}}}\right)^{\frac{2}{3}},\\
\quad
8.3~\frac{\omega_{0}}{\omega_{\mathrm{p}}}
\left(\frac{k_{\mathrm{B}} T_{\mathrm{e}}}{m_{\mathrm{e}} c^{2}}\right)^{\frac{3}{4}}
\left(1+\frac{\omega_{\mathrm{p}}^{2}}{\omega_{\mathrm{c}}^{2}}\right)^{\frac{3}{4}}
\ll
a_{\mathrm{e}}\frac{\omega_{0}}{\omega_{\mathrm{c}}}\ll1 .
\end{cases}
\label{eq:growth_rate_charged_matome_Brillouin_low}
\end{equation}
The corresponding wavenumber for maximum growth is given by Eqs.~\eqref{eq:maximum_geowth_wave_vector_charged_Compton} and \eqref{eq:induced_Brillouin_charged_wave_vector} as follows:
\begin{equation}
k_{\text{max}}\sim
\begin{cases}
2k_0\left(
1-\sqrt{\frac{k_{\mathrm B}T_{\mathrm e}}{m_{\mathrm e}c^{2}}
\left(1+\frac{\omega_{\mathrm p}^{2}}{\omega_{\mathrm c}^{2}}\right)}
\right), \\
\quad a_{\mathrm e}\frac{\omega_{0}}{\omega_{\mathrm c}}
\ll
5.5~\frac{\omega_{0}}{\omega_{\mathrm p}}
\left(\frac{k_{\mathrm B}T_{\mathrm e}}{m_{\mathrm e}c^{2}}\right)^{\frac{3}{4}}
\left(1+\frac{\omega_{\mathrm p}^{2}}{\omega_{\mathrm c}^{2}}\right)^{\frac{3}{4}},\\
2k_0, \\
\quad8.3~\frac{\omega_{0}}{\omega_{\mathrm p}}
\left(\frac{8k_{\mathrm B}T_{\mathrm e}}{m_{\mathrm e}c^{2}}\right)^{\frac{3}{4}}
\left(1+\frac{\omega_{\mathrm p}^{2}}{\omega_{\mathrm c}^{2}}\right)^{\frac{3}{4}}
\ll
a_{\mathrm e}\dfrac{\omega_{0}}{\omega_{\mathrm c}}\ll1.
\end{cases}
\end{equation}
\Nishiura{The transition point between the weak and strong coupling regimes is defined as the incident wave amplitude $a_{\text{e}}\omega_0/\omega_{\text{c}}$ at which the growth rates in both coupling regimes become equal, as described by Eq.~\eqref{eq:growth_rate_charged_matome_Brillouin_low},}
\begin{equation}
    a_{\mathrm{e},\text{trans}}\frac{\omega_{0}}{\omega_{\mathrm{c}}}
    \simeq 4.4~\frac{\omega_0}{\omega_{\mathrm{p}}}
    \left(\frac{k_{\mathrm{B}}T_{\mathrm{e}}}{m_{\mathrm{e}} c^{2}}\right)^{\frac{3}{4}}
    \left(1+\frac{\omega_{\text{p}}^2}{\omega_{\text{c}}^2}\right)^{\frac{3}{4}}.
    \label{eq:transition_point_for_charged_mode}
\end{equation}

Furthermore, SRS is also excited in this regime, but only small-angle scattering is permitted. The maximum growth rate is expressed from Eq.~\eqref{eq:stimulated_Raman_growth_rate_small_angle} as follows:
\begin{equation}
    \left(t_{\text{R}}^{\text{max}}\right)^{-1} \sim 0.30\, a_{\mathrm{e}} \frac{\omega_0}{\omega_{\mathrm{c}}}
    \left( \omega_0 \omega_{\mathrm{p}} \right)^{\frac{1}{2}}
    \left( \frac{m_{\mathrm{e}} c^2}{32 k_{\mathrm{B}} T_{\mathrm{e}}} \right)^{\frac{1}{2}}
    \frac{\omega_{\mathrm{p}}}{\omega_0}
    \left( 1 + \frac{\omega_{\mathrm{p}}^2}{\omega_{\mathrm{c}}^2} \right)^{-\frac{1}{2}}.
    \label{eq:stimulated_Raman_small_angle_growth_rate}
\end{equation}
The weak coupling condition for this regime is given from Eq.~\eqref{eq:weak_coupling_small_Raman} by:
\begin{equation}
a_{\mathrm{e}} \frac{\omega_0}{\omega_{\mathrm{c}}}
\ll
5.1 \left( \frac{\omega_0}{\omega_{\mathrm{p}}} \right)^{\frac{1}{2}}
    \left( \frac{32 k_{\mathrm{B}} T_{\mathrm{e}}}{m_{\mathrm{e}} c^2} \right)^{\frac{1}{2}}
    \left( 1 + \frac{\omega_{\mathrm{p}}^2}{\omega_{\mathrm{c}}^2} \right)^{\frac{1}{2}}.
\end{equation}

In a strongly magnetized $e^\pm$ pair plasma, the linear growth rates of the charged mode in the low density regime are strongly suppressed compared to the unmagnetized case. For ICS, the suppression factor is $(\omega_{0}/\omega_{\mathrm{c}})^2$, while for SBS, the suppression factor is $(\omega_{0}/\omega_{\mathrm{c}})^{2/3}$. In addition, for ICS, the so-called subluminal effect, $(1+\omega_{\mathrm{p}}^2/\omega_{\mathrm{c}}^2)^{-1}$, arises due to the phase velocity of the EM wave becoming less than the speed of light, although this effect is of order unity when $\omega_{\mathrm{p}} \ll \omega_{\mathrm{c}}$. Notably, SRS in the weak coupling regime does not occur in unmagnetized $e^\pm$ pair plasma \citep{2017PhRvE..96e3204S}. When a background magnetic field is present, however, our results demonstrate that SRS can also be excited in $e^\pm$ pair plasma. A more detailed physical discussion on the order-of-magnitude suppression of these growth rates is provided in Sec.~\ref{subsubsec:scattering_suppression}.

\paragraph{Intermediate density regime}

According to Tab.~\ref{tab:roadmap_coupling}, both ICS and SRS are excited under the weak coupling condition. For the intermediate density regime, defined as
\begin{equation}
\sqrt{\frac{8 k_{\mathrm{B}} T_{\mathrm{e}}}{m_{\mathrm{e}} c^{2}}\left(1+\frac{\omega_{\text{p}}^2}{\omega_{\text{c}}^2}\right)} \ll \frac{\omega_{\mathrm{p}}}{\omega_{0}} < \frac{1}{2},
\end{equation}
the discussion in Appendix~\ref{subsec:raman_compton_competition_charged} demonstrates that the maximum linear growth rate of SRS, given by Eq.~\eqref{eq:growth_rate_of_induced_Raman_weak_coupling}, always exceeds that of ICS, given by Eq.~\eqref{eq:induced_Compton_Debye_screening_Brillouin}. Therefore, SRS is the dominant instability in this regime. Under the strong coupling condition, SRS (degenerate with SBS in the strong coupling limit) is also excited.

The linear growth rates at the angle parameter where both weak and strong coupling SRS attain their maximum growth, corresponding to the condition in Eq.~\eqref{eq:maximum_growth_angle_condition_charged_Raman}, are summarized as follows, using Eqs.~\eqref{eq:growth_rate_of_induced_Raman_weak_coupling} and \eqref{eq:induced_Brillouin_charged_maximum_growth_rate}:
\begin{equation}
\begin{aligned}
\left(t_{\text{charged}}\right)^{-1} = 
\begin{cases} 
a_{\mathrm{e}} \left(\omega_{0} \omega_{\mathrm{p}}\right)^{\frac{1}{2}} \frac{\omega_{0}}{\omega_{\mathrm{c}}},
\\
\quad a_{\mathrm{e}} \frac{\omega_{0}}{\omega_{\mathrm{c}}} \ll 5.7~
\left(
    \frac{k_{\mathrm{B}} T_{\mathrm{e}}}{m_{\mathrm{e}} c^2}
\right)^{\frac{1}{2}}
\left(
    1 + \frac{\omega_{\mathrm{p}}^2}{\omega_{\mathrm{c}}^2}
\right)^{\frac{1}{2}}
\left(
    \frac{\omega_0}{\omega_{\mathrm{p}}}
\right)^{\frac{1}{2}},
\\
\sqrt{3}\left(\frac{\omega_{\mathrm{p}}^{2} a_{\mathrm{e}}^{2} \omega_{0}}{2}\right)^{\frac{1}{3}}\left(\frac{\omega_{0}}{\omega_{\mathrm{c}}}\right)^{\frac{2}{3}},
\\
\quad8.3~\frac{\omega_{0}}{\omega_{\mathrm{p}}}
\left(\frac{ k_{\mathrm{B}} T_{\mathrm{e}}}{m_{\mathrm{e}} c^{2}}\right)^{\frac{3}{4}}
\left(1+\frac{\omega_{\text{p}}^2}{\omega_{\text{c}}^2}\right)^{\frac{3}{4}}
\ll a_{\mathrm{e}} \frac{\omega_{0}}{\omega_{\mathrm{c}}} \ll 1. 
\end{cases}
\end{aligned}
\label{eq:growth_rate_charged_matome_Brillouin_intermediate}
\end{equation}
The corresponding maximum growth wavenumber is given by Eqs.~\eqref{eq:maximum_growth_wave_number_Raman5} and \eqref{eq:induced_Brillouin_charged_wave_vector}:
\begin{equation}
\begin{aligned}
k_{\text{max}} \sim 
\begin{cases}
k_{0}\left(1 + \sqrt{1 - 2 \frac{\omega_{\mathrm{p}}}{\omega_{0}}}\right),
\\
\quad a_{\mathrm{e}} \frac{\omega_{0}}{\omega_{\mathrm{c}}} \ll 5.7~
\left(
    \frac{k_{\mathrm{B}} T_{\mathrm{e}}}{m_{\mathrm{e}} c^2}
\right)^{\frac{1}{2}}
\left(
    1 + \frac{\omega_{\mathrm{p}}^2}{\omega_{\mathrm{c}}^2}
\right)^{\frac{1}{2}}
\left(
    \frac{\omega_0}{\omega_{\mathrm{p}}}
\right)^{\frac{1}{2}},
\\
2k_0,
\\
\quad8.3~\frac{\omega_{0}}{\omega_{\mathrm{p}}}
\left(\frac{ k_{\mathrm{B}} T_{\mathrm{e}}}{m_{\mathrm{e}} c^{2}}\right)^{\frac{3}{4}}
\left(1+\frac{\omega_{\text{p}}^2}{\omega_{\text{c}}^2}\right)^{\frac{3}{4}}
\ll a_{\mathrm{e}} \frac{\omega_{0}}{\omega_{\mathrm{c}}} \ll 1.
\end{cases}
\end{aligned}
\end{equation}
\Nishiura{The transition point between the weak and strong coupling regimes is defined as the incident wave amplitude $a_{\text{e}}\omega_0/\omega_{\text{c}}$ at which the growth rates in both coupling regimes become equal, as described by Eq.~\eqref{eq:growth_rate_charged_matome_Brillouin_intermediate},}
\begin{equation}
    a_{\mathrm{e},\text{trans}}\frac{\omega_{0}}{\omega_{\mathrm{c}}}\simeq2.6~\left(\frac{\omega_{\mathrm{p}}}{\omega_{0}}\right)^{\frac{1}{2}}.
    \label{eq:transition_point_for_Raman_mode}
\end{equation}

In strongly magnetized $e^\pm$ pair plasma, the linear growth rates of the charged mode in the intermediate density regime are also strongly suppressed compared to the unmagnetized case. Specifically, the suppression factor for SRS in the weak coupling regime is $(\omega_{0}/\omega_{\mathrm{c}})$ compared to SRS in unmagnetized ion-electron plasma, while the suppression for SRS in the strong coupling regime is $(\omega_{0}/\omega_{\mathrm{c}})^{2/3}$. The physical interpretation and scaling of these suppression effects are discussed in detail in Sec.~\ref{subsubsec:scattering_suppression}.

\paragraph{High density regime}
According to Tab.~\ref{tab:roadmap_coupling}, both ICS and SRS are in principle driven under the weak coupling condition. However, for the high density regime,
\begin{equation}
\niShiura{\max\left\{\frac{1}{2},~\sqrt{\frac{8 k_{\mathrm{B}} T_{\mathrm{e}}}{m_{\mathrm{e}} c^{2}}\left(1+\frac{\omega_{\text{p}}^2}{\omega_{\text{c}}^2}\right)}\right\} < \frac{\omega_{\mathrm{p}}}{\omega_{0}},}
\end{equation}
the resonance condition for the excitation of plasma (Langmuir) waves, given by Eq.~\eqref{eq:induced_Raman_condition_upper_limit}, is not satisfied. As a result, only ICS is excited in this regime. The maximum linear growth rate for ICS under the condition of maximum growth, as expressed by Eq.~\eqref{eq:maximum_growth_angle_condition_charged_Debye}, is summarized as follows (see Eq.~\eqref{eq:induced_Compton_Debye_screening_Brillouin}):
\begin{equation}
\begin{aligned}
\left(t_{\mathrm{charged}}^{\max }\right)^{-1} =\ 
&\sqrt{\dfrac{32\mathrm{e}}{\pi}}\,
\dfrac{k_{\mathrm{B}} T_{\mathrm{e}}}{m_{\mathrm{e}} c^{2}}\,
\left(\dfrac{\omega_{0}}{\omega_{\mathrm{c}}}\right)^{2}
\left(\dfrac{\omega_{0}}{\omega_{\mathrm{p}}}\right)^{4} \\[8pt]
&\times
\dfrac{a_{\mathrm{e}}^{2}\omega_{\mathrm{p}}^{2}}{\omega_{0}}
\left(1+\dfrac{\omega_{\text{p}}^2}{\omega_{\text{c}}^2}\right).
\end{aligned}
\label{eq:growth_rate_charged_matome_Brillouin_high}
\end{equation}
The corresponding wave number for maximum growth, as given by Eq.~\eqref{eq:wave_vector_charged_Brilliouin}, is
\begin{equation}
    k_{\text{max}}=2k_0\left(1-\sqrt{\frac{k_{\mathrm{B}} T_{\mathrm{e}}}{m_{\mathrm{e}} c^2}
    \left(1 + \frac{\omega_{\mathrm{p}}^2}{\omega_{\mathrm{c}}^2}\right)}\right).
\end{equation}

In a strongly magnetized $e^\pm$ pair plasma, the linear growth rate for ICS in the high density regime is significantly suppressed compared to the unmagnetized case. Unlike SBS or SRS, the ICS growth rate is suppressed both by the gyroradius effect, represented by the factor $(\omega_{0}/\omega_{\mathrm{c}})^{2}$, and by Debye screening:
\begin{equation}
    \frac{\mathrm{e}}{2\pi} \left( \frac{\omega_{0}}{\omega_{\mathrm{p}}} \right)^{4} 
    \left(\frac{8k_{\mathrm{B}} T_{\mathrm{e}}}{m_{\mathrm{e}} c^{2}}\right)^{2}\left(1 + \frac{\omega_{\mathrm{p}}^2}{\omega_{\mathrm{c}}^2}\right)^2
    = \frac{2\mathrm{e}}{\pi} \left( \frac{4\pi\lambda_{\text{De}}}{\lambda_0} \right)^{4}.
    \label{eq:suppression_effect_Debye_screening_Brillouin}
\end{equation}
The effect of Debye screening on ICS was first demonstrated in \citet{2025PhRvD.111f3055N}. A detailed discussion on the order-of-magnitude suppression of these growth rates is provided in Sec.~\ref{subsubsec:scattering_suppression}.

\subsubsection{Numerical Evaluation}

The linear growth rate of induced scattering in the charged mode can be evaluated by numerically solving the dispersion relation expressed as Eq.~\eqref{eq:dispersion_relation_charged_Brillouin2}. In this study, following the approach used for the neutral mode, we systematically vary the dimensionless amplitude of the incident EM wave, as defined by Eq.~\eqref{eq:definition_of_eta}, and examine how the maximum linear growth rate responds. For simplicity, we do not address the small-angle SRS in the low density regime. The behavior of the charged mode varies significantly depending on the plasma density, so we consider the low, intermediate, and high density regimes, as defined by Eqs.~\eqref{eq:charged_mode_instability_thin_condition_Brillouin}, \eqref{eq:induced_Raman_condition_Brillouin}, and \eqref{eq:charged_mode_instability_dense_condition_Brillouin}, respectively. The parameters adopted for each regime are listed in Tab.~\ref{tab:Charged_parameter}.
\begin{table}[htbp]
  \centering
  \captionsetup{
    skip=1em,          
    justification=raggedright 
  }
  \caption{Model parameters for charged mode}
  \label{tab:Charged_parameter}
  \renewcommand{\arraystretch}{1.5}
  \sisetup{table-number-alignment=left}
  \begin{tabular}{lSSS}
    \toprule
    Parameter & {Low density} & {Intermediate density} & {High density} \\
    \midrule
    $v_{\text{th}}/c$ & {$10^{-4}$} & {$10^{-4}$} & {$10^{-4}$} \\
    $\omega_{\text{p}}/\omega_0$ & {$5\times10^{-5}$} & {$2\times10^{-3}$} & {$10^{0}$} \\
    $\omega_{\text{c}}/\omega_0$ & {$10^{2}$} & {$10^{2}$} & {$10^{2}$} \\
    $\omega_0$~[\si{\hertz}] & {$2\pi \times 10^9$} & {$2\pi \times 10^9$} & {$2\pi \times 10^9$} \\
    \bottomrule
  \end{tabular}
\end{table}

\paragraph{Low density regime}
\begin{figure*}
\centering
\includegraphics[width=\textwidth]{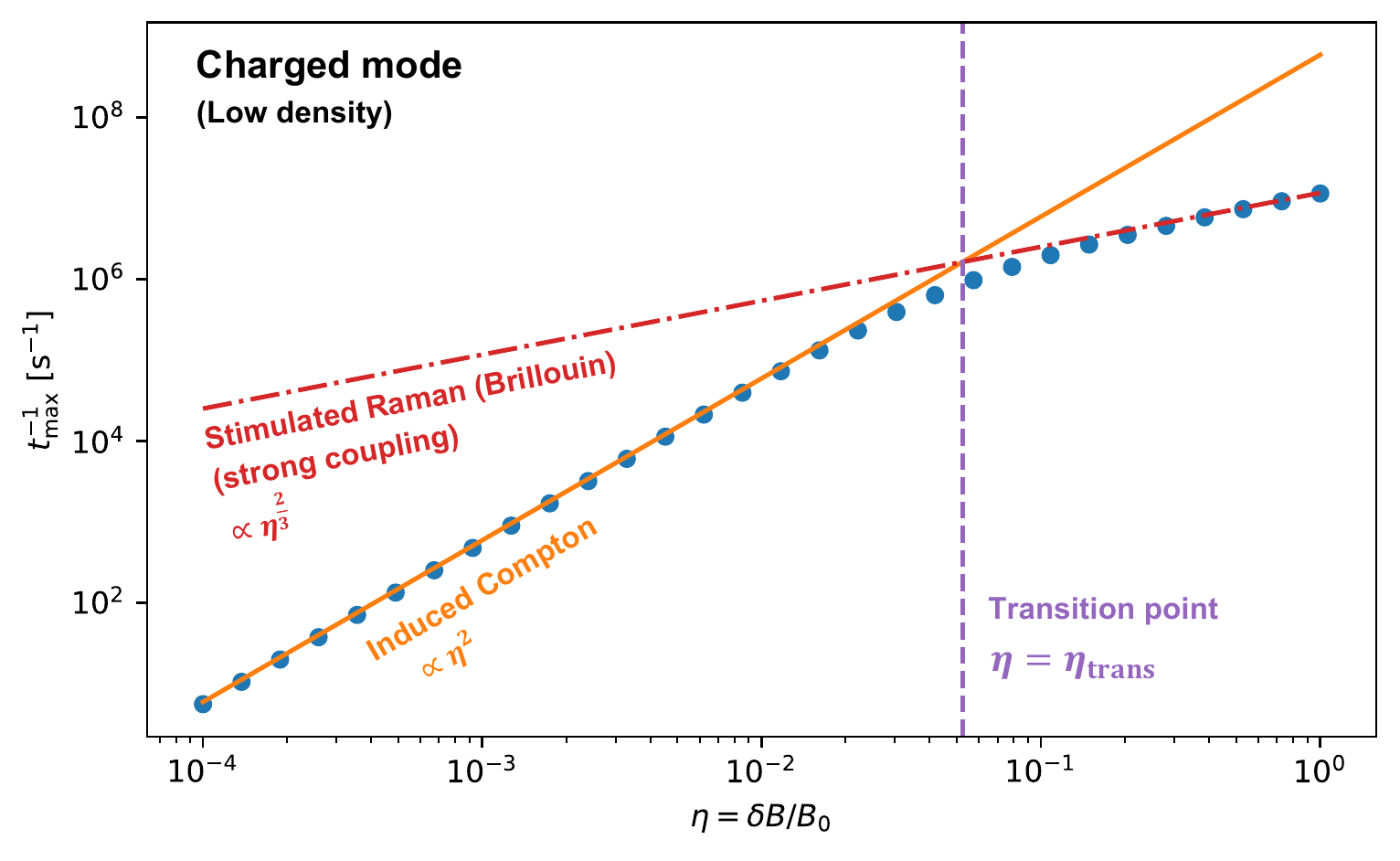}
\caption[Concept of Parametric Decay Instability]{\justifying Dependence of the maximum linear growth rate of induced scattering excited by the charged mode on the incident wave amplitude \eqref{eq:definition_of_eta} in the low density regime \eqref{eq:charged_mode_instability_thin_condition_Brillouin}. The orange solid curve represents the linear growth rate of ICS, given by Eq.~\eqref{eq:maximum_growth_rate_magnetic2_Brillouin}. The red dashed-dotted curve shows the linear growth rate of SBS in the strong coupling regime, as given by Eq.~\eqref{eq:induced_Brillouin_charged_maximum_growth_rate}. The purple vertical dotted line indicates the incident wave amplitude at the transition from the weak to strong coupling regime, as given by Eq.~\eqref{eq:Transition_point_eta_charged}. Blue dots represent the results obtained from the numerical solution of the dispersion relation, Eq.~\eqref{eq:dispersion_relation_charged_Brillouin2}.}
\label{fig:stimulated_Brillouin_charged_growth_rate}
\end{figure*}
Fig.~\ref{fig:stimulated_Brillouin_charged_growth_rate} illustrates the dependence of the maximum linear growth rate of the scattered wave on the amplitude of the incident wave \eqref{eq:definition_of_eta}, under the low density conditions specified in Tab.~\ref{tab:Charged_parameter}. The transition between the weak and strong coupling regimes is defined as follows, according to Eqs.~\eqref{eq:transition_point_for_charged_mode} and \eqref{eq:definition_of_eta}:
\begin{equation}
\begin{aligned}
    \eta_{\text{trans}}\simeq4.4~\frac{\omega_0}{\omega_{\mathrm{p}}}
    \left(\frac{k_{\mathrm{B}}T_{\mathrm{e}}}{m_{\mathrm{e}} c^{2}}\right)^{\frac{3}{4}}
    \left(1+\frac{\omega_{\text{p}}^2}{\omega_{\text{c}}^2}\right)^{\frac{5}{4}}.
\end{aligned}    
\label{eq:Transition_point_eta_charged}
\end{equation}
Numerical calculations show that, as the amplitude of the incident wave increases, there is a continuous transition from ICS in the weak coupling regime to SBS in the strong coupling regime (degenerate with SRS in the strong coupling limit), across the transition point given by Eq.~\eqref{eq:Transition_point_eta_charged}. SBS does not appear in the weak coupling regime. These qualitative behaviors are the same as those found for the ordinary and neutral modes.

\paragraph{Intermediate density regime}
\begin{figure*}
\centering
\includegraphics[width=\textwidth]{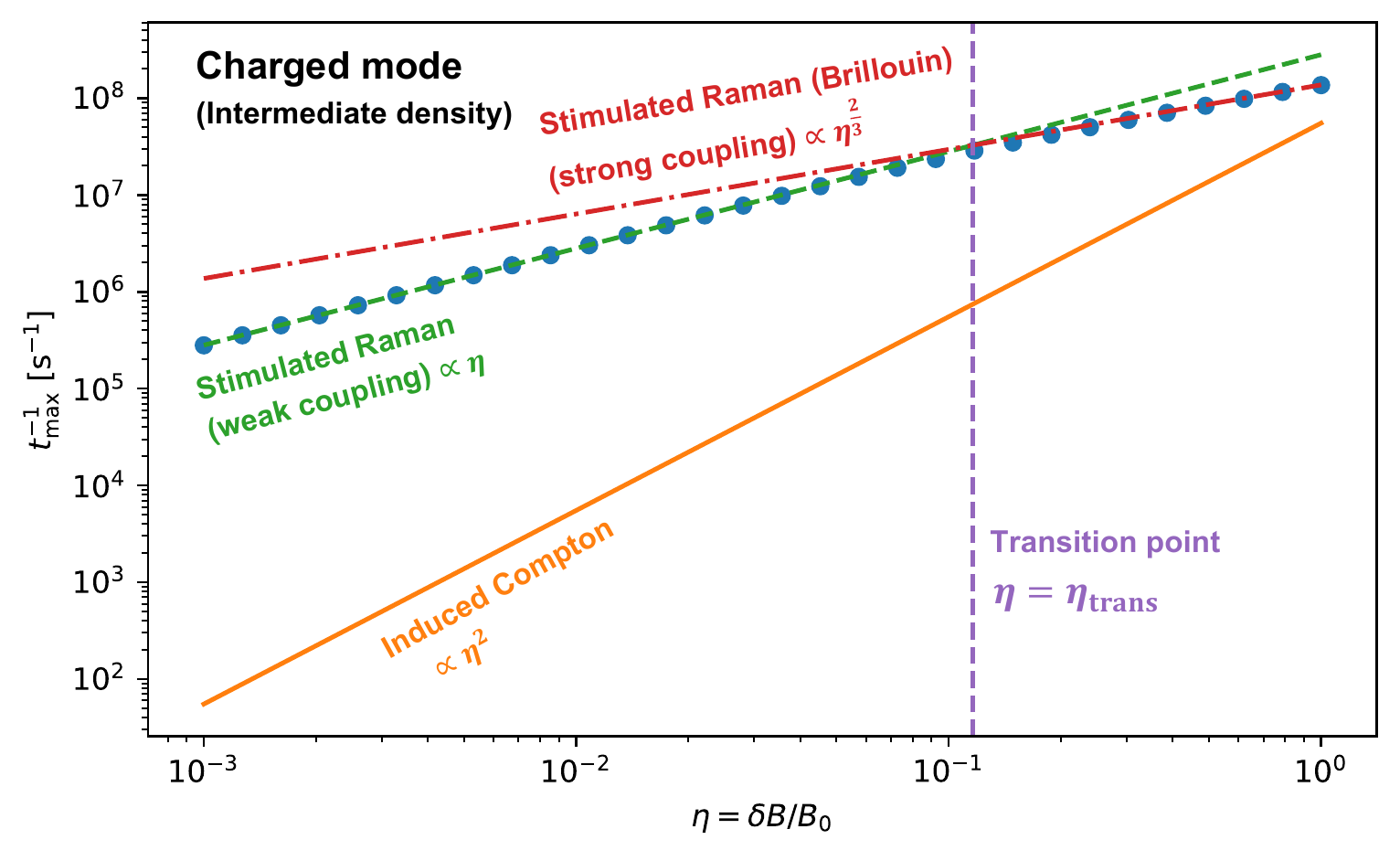}
\caption[Concept of Parametric Decay Instability]{\justifying The dependence of the maximum linear growth rate of induced scattering in the charged mode on the incident wave amplitude \eqref{eq:definition_of_eta} for the intermediate density regime \eqref{eq:induced_Raman_condition_Brillouin}. The orange solid line represents the growth rate of ICS as given by Eq.~\eqref{eq:maximum_growth_rate_magnetic2_Brillouin}, while the red dash-dotted line corresponds to the growth rate of SBS in the strong coupling regime as given by Eq.~\eqref{eq:induced_Brillouin_charged_maximum_growth_rate}. The purple vertical dotted line indicates the transition point of the incident wave amplitude between the weak and strong coupling regimes, as described by Eq.~\eqref{eq:Transition_point_eta_charged_Raman}. The blue dots show the results obtained from numerical solutions of the dispersion relation in Eq.~\eqref{eq:dispersion_relation_charged_Brillouin2}.}
\label{fig:stimulated_Raman_charged_growth_rate}
\end{figure*}
Fig.~\ref{fig:stimulated_Raman_charged_growth_rate} shows the dependence of the maximum linear growth rate of induced scattering in the charged mode under the intermediate density regime specified in Tab.~\ref{tab:Charged_parameter}. The transition point between the weak and strong coupling regimes is defined as follows, using Eqs.~\eqref{eq:transition_point_for_Raman_mode} and \eqref{eq:definition_of_eta}:
\begin{equation}
\begin{aligned}
    \eta_{\text{trans}} \simeq 2.6~\left(\frac{\omega_{\mathrm{p}}}{\omega_0}\right)^{\frac{1}{2}}
    \left(1+\frac{\omega_{\mathrm{p}}^2}{\omega_{\mathrm{c}}^2}\right)^{\frac{1}{2}}.
\end{aligned}    
\label{eq:Transition_point_eta_charged_Raman}
\end{equation}
The numerical results demonstrate that increasing the incident wave amplitude leads to a continuous transition from weak-coupling SRS to strong-coupling SRS at the transition point given by Eq.~\eqref{eq:Transition_point_eta_charged_Raman}. In the weak coupling regime, as shown in Appendix~\ref{subsec:raman_compton_competition_charged}, the growth rate of SRS always exceeds that of ICS.

\paragraph{High density regime}
\begin{figure*}
\centering
\includegraphics[width=\textwidth]{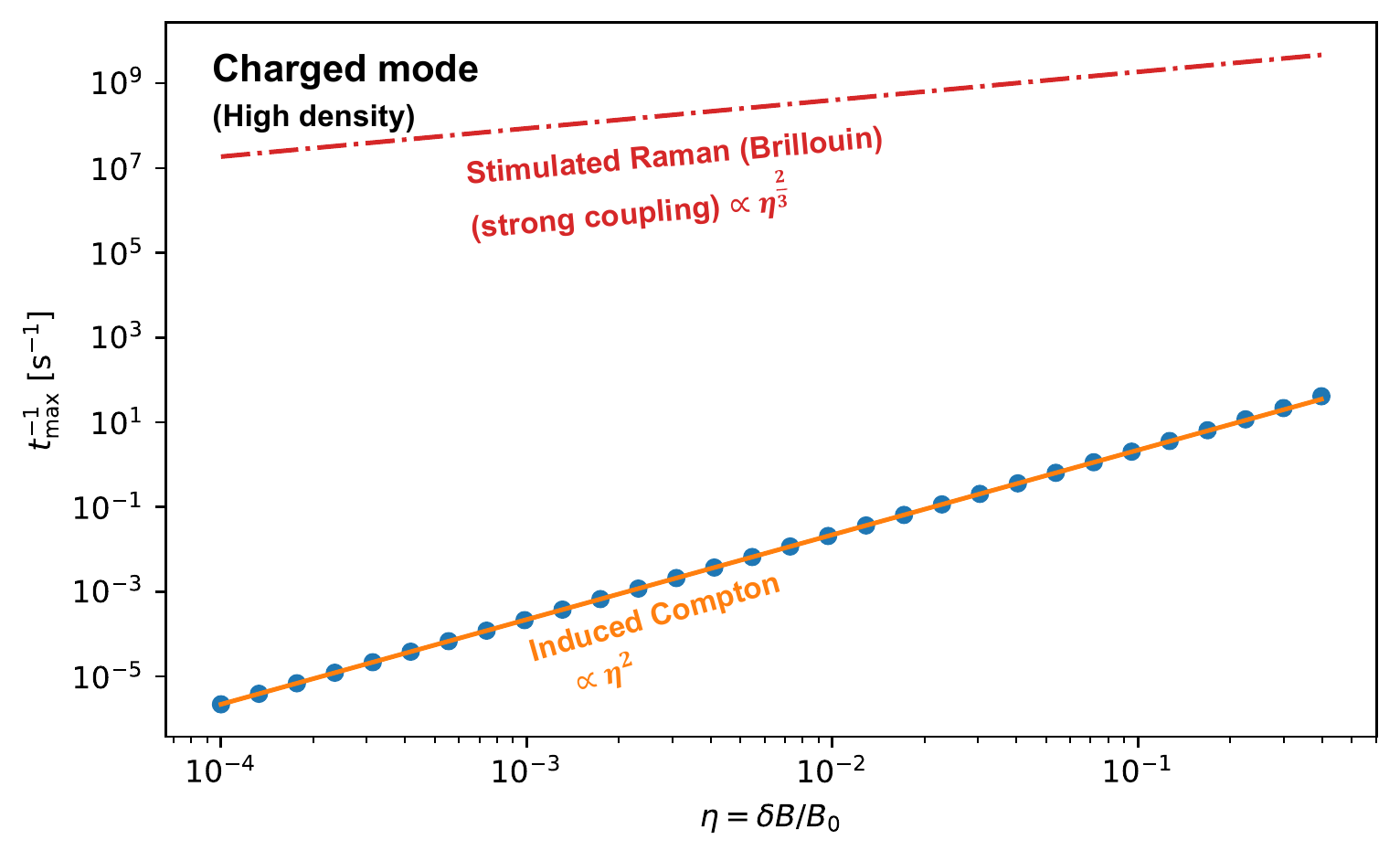}
\caption[Concept of Parametric Decay Instability]{\justifying The dependence of the maximum linear growth rate of induced scattering in the charged mode on the incident wave amplitude \eqref{eq:definition_of_eta} for the high density regime. The orange solid curve represents the growth rate of ICS, as given by Eq.~\eqref{eq:induced_Compton_Debye_screening_Brillouin}\footnote{\niShiura{As discussed in footnote~\ref{fn:growth_correction}, the plotted analytic curve includes the numerical correction to Eq.~\eqref{eq:induced_Compton_Debye_screening_Brillouin}.}}
. The red dashed-dotted line corresponds to the growth rate of SRS (SBS in the strong coupling limit), described by Eq.~\eqref{eq:induced_Brillouin_charged_maximum_growth_rate}. The blue points indicate numerical solutions of the dispersion relation in Eq.~\eqref{eq:dispersion_relation_charged_Brillouin2}.}
\label{fig:Debye_Compton_charged_growth_rate}
\end{figure*}
Fig.~\ref{fig:Debye_Compton_charged_growth_rate} shows the dependence of the maximum linear growth rate of induced scattering under the high density conditions listed in Tab.~\ref{tab:Charged_parameter}. The numerical results demonstrate that ICS with Debye screening dominates for all values of the incident wave amplitude \eqref{eq:definition_of_eta}. Analytically, 
in the high density regime, the right-hand side of the weak coupling condition in Eq.~\eqref{eq:weak_coupling_condition_for_Debye_Compton} leads to the following lower bound,
\begin{equation}
\begin{aligned}
&1<\left(\frac{\pi}{\mathrm{e}}\right)^{\frac{1}{4}}2^{\frac{1}{4}}\overset{\mathrm{Eq.~} \eqref{eq:charged_mode_instability_dense_condition_Brillouin}}{\ll}\left(\frac{\pi}{\mathrm{e}}\right)^{\frac{1}{4}}2^{\frac{3}{4}}\left(\frac{\omega_{\mathrm{p}}}{\omega_{0}}\right)^{\frac{1}{2}}\\
&\overset{\mathrm{Eq.~} \eqref{eq:charged_mode_instability_thin_condition_Brillouin_Compton}}{\ll}
\left(\frac{\pi}{\mathrm{e}}\right)^{\frac{1}{4}}
 \frac{\omega_{\mathrm{p}}}{\omega_{0}}
 \left(\frac{m_{\mathrm{e}} c^{2}}{k_{\mathrm{B}} T_{\mathrm{e}}}\right)^{\frac{1}{4}}
 \left(1+\frac{\omega_{\text{p}}^2}{\omega_{\text{c}}^2}\right)^{-\frac{1}{4}}.
\end{aligned}
\label{eq:weak_coupling_Debye_screening_henkei}
\end{equation}
Therefore, in the high density regime, ICS with Debye screening is always dominant in the linear regime of the incident wave amplitude, as described by Eq.~\eqref{eq:non_rela_neutral_and_charged_Brillouin}.

\section{Physical Interpretation}
\label{sec:physical_interpretation}
\subsection{Sidescattering in the Ordinary Mode}
\label{subsubsec:Ordinary_sidescattering}

The maximum growth of instability in the ordinary mode is realized by sidescattering, as shown in Appendix~\ref{sec:ordinary_mode_detail_derivation}. This is in clear contrast to the case without a background magnetic field, where the maximum growth occurs at $180^\circ$ backward scattering~\citep{2022ApJ...930..106G}. This difference arises because the instability is significantly suppressed when the wavevector of the density fluctuation, $\bm{k}$, is perpendicular to the background magnetic field.

Under a strong background magnetic field ($\omega_0 \ll \omega_{\mathrm{c}}$), Appendix~B of~\citet{2025PhRvD.111f3055N} demonstrates that density fluctuations propagating perpendicular to the background field do not lead to instability. There are two main reasons for this. First, in a strong background magnetic field, there are no longitudinal eigenmodes in the direction perpendicular to the field. Second, charged particles are tightly bound to the magnetic field lines, preventing sustained Landau resonance (see Sec.~55 of the book~\citep{1981phki.book.....L}).

\subsection{Transition from Weak to Strong Coupling}
\label{subsub:transition_weak_to_strong_coupling}
The behavior of induced scattering in $e^\pm$ pair plasma exhibits a clear transition from weak to strong coupling as the incident wave amplitude $a_{\mathrm{e}}$ or $\eta$ increases. This transition is observed in the ordinary, neutral, and charged modes for both the low and intermediate density regimes, as demonstrated in this work. In each case, we show that the dependence of the growth rate on the incident wave amplitude differs between the weak and strong coupling regimes.

In the weak coupling regime, ICS dominates in all three modes, except for the intermediate density regime of the charged mode. This behavior is a unique property of $e^\pm$ pair plasma and has also been noted for the case without a background magnetic field~\citep{2016PhRvL.116a5004E,2017PhRvE..96e3204S,2023MNRAS.522.2133I}. The physical reason is that, regardless of the presence of a background magnetic field, $e^\pm$ pair plasma—where the mass and temperature of the constituent particles are identical—supports only acoustic quasi-modes with strong Landau damping. As a result, SBS is suppressed, and ICS, which is characterized by Landau resonance, becomes the dominant instability. In contrast, for ion-electron plasma, ion acoustic waves can exist as linear eigenmodes with weak Landau damping, and thus SBS can dominate even in the weak coupling regime.

In the intermediate density regime of the charged mode, both ICS and SRS can be excited simultaneously in the weak coupling region. Mathematically, different solutions (branches) of the dispersion relation can be excited at the same time, and these are not mutually exclusive. Here, the ``branch" refers to the independent eigenmode solutions that emerge from the dispersion relation. Physically, ICS is excited through resonance with an acoustic quasi-mode, which is subject to significant Landau damping, while SRS is excited via resonance with a Langmuir wave, which is a linear eigenmode with much weaker Landau damping. In contrast, ICS and SBS essentially describe the same wave—an acoustic mode—treated from kinetic and fluid perspectives, respectively. Therefore, ICS and SBS always belong to the same solution branch and cannot be simultaneously excited.

As the incident wave amplitude increases and the system enters the strong coupling regime, the effects of Landau damping become relatively less significant in each mode, and the behavior approaches that expected from the fluid (strong coupling) limit. As a result, SBS or SRS becomes dominant in any mode.

\subsection{Degeneracy between SBS and SRS in the Charged Mode}
\label{subsec:degeneracy_Brillouin_Raman_charged_mode}
In the strong coupling regime, all unstable modes exhibit the same scaling for the growth rate with respect to the incident wave amplitude, specifically $\propto a_{\mathrm{e}}^{2/3}$. In particular, for the charged mode, the growth rates of SBS and SRS become identical. A similar feature is observed in ion–electron plasma~\citep{1975PhFl...18.1002F}.
\footnote{
For induced scattering in the strong coupling regime of ion–electron plasma, the growth rate given by Eq.~\eqref{eq:induced_Brillouin_charged_maximum_growth_rate} yields the SRS growth rate by substituting $\omega_{\text{p}} \to \omega_{\text{pe}}$, and the SBS growth rate by substituting $\omega_{\text{p}} \to \omega_{\text{pi}}$, where $\omega_{\text{ps}} \equiv \sqrt{4\pi e^2 n_{\text{s}} / m_{\text{s}}}$ is the plasma frequency for species $\text{s}$. For $e^\pm$ pair plasma, where $\omega_{\text{pe}} = \omega_{\text{pi}}$, it follows that the growth rates for SBS and SRS are identical.
}

Physically, in the strong coupling regime, instabilities are no longer described as resonances with linear eigenmodes but rather as nonlinear interactions with density fluctuations (quasi-modes) induced by the ponderomotive force~\citep{1975PhFl...18.1002F}. In this regime, for the ordinary, neutral, and charged modes, the respective dispersion relations—Eqs.~\eqref{eq:dispersion_relation_induced_Brillouin_nomagnetic_tenkai}, \eqref{eq:dispersion_relation_induced_Brillouin_Neutral_tenkai}, and \eqref{eq:dispersion_relation_induced_Brillouin_charged_tenkai}—are characterized by a dominant $\omega^{3}$ term and a source term proportional to $a_{\mathrm{e}}^{2}$ due to the ponderomotive force. As a result, the growth rate universally scales as $\propto(a_{\mathrm e}^{2}\omega_{0}\omega_{\mathrm p}^{2})^{1/3}$, independent of the mode.

\subsection{Excitation Conditions for SRS in the Charged Mode}
\label{subsec:physical_interpretation_raman_scattering}

SRS in the charged mode is only excited when the plasma frequency, i.e., the $e^\pm$ pair density, falls within a specific range. In this section, we first discuss the physical interpretation of the excitation condition for large-angle (backward) scattering in the intermediate density regime given by Eq.~\eqref{eq:induced_Raman_condition_Brillouin}. We then address the excitation of small-angle scattering in the low density regime, as given by Eq.~\eqref{eq:charged_mode_instability_thin_condition_Brillouin}.

\subsubsection{Large-Angle (Backward) Scattering}
The lower bound on the plasma frequency for backward scattering \eqref{eq:charged_mode_instability_thin_condition_Brillouin_Compton} originates from Eq.~\eqref{eq:plasma_density_dekai_debye_screening_Brillouin}. This criterion corresponds to the condition that the phase velocity of the Langmuir wave is much greater than the thermal velocity of electrons and positrons. If this requirement is not satisfied, the Langmuir wave undergoes strong Landau damping due to resonance with the particles, and the wave is heavily attenuated. As a result, the instability in the charged mode transitions from SRS to ICS.

On the other hand, the upper bound on the plasma frequency \eqref{eq:induced_Raman_condition_upper_limit} is set by the requirement that the growth wavenumber for SRS, given by Eq.~\eqref{eq:maximum_growth_wave_number_Raman5}, admits a real solution. This same condition can also be obtained by substituting the maximum growth wavenumber for SRS, Eqs.~\eqref{eq:maximum_growth_wave_number_Raman5} and \eqref{eq:energy_momentum_conservation_Compton_Brillouin}, into the condition $|\omega|\ll\omega_0,~\omega_1$ for the beat frequency generated by interference of the incident and scattered EM waves. Physically, this criterion indicates whether the phase velocity of the beat wave can resonate with that of the Langmuir wave. If the plasma frequency becomes too large, the phase velocity of the Langmuir wave increases, making it impossible for the beat wave to match, and the three-wave resonance condition cannot be fulfilled.

\subsubsection{Small-Angle Scattering}
In the low density regime described by Eq.~\eqref{eq:charged_mode_instability_thin_condition_Brillouin}, SRS is excited only for small-angle scattering. In this region, the phase velocity of the Langmuir wave is typically smaller than the thermal velocity of electrons and positrons, involving strong Landau damping. However, for small scattering angles, the parameter $\nu$ approaches unity, and as shown in Eq.~\eqref{eq:wave_number_approximation_x-mode}, the wavenumber $k$ of the density fluctuation decreases. Consequently, the phase velocity $\omega/k$ of the Langmuir wave increases and can exceed the thermal velocity. This leads to significant suppression of Landau damping, allowing SRS to be excited even in the low density regime, but only for small-angle scattering.

\subsection{Comparison without Magnetic Field Case}
\label{subsubsec:scattering_suppression}
The linear growth rates of each induced scattering mode in strongly magnetized $e^\pm$ pair plasma can be consistently interpreted in terms of the suppression by the gyroradius effect, characterized by the scaling with $\omega_0/\omega_{\mathrm{c}}$. This suppression arises from the differences in the dimensionless oscillation velocities of the charged particles driven by the incident EM wave. Specifically, the relevant dimensionless oscillation velocities are $a_{\mathrm{e}}$ for the ordinary mode, $a_{\mathrm{e}}(\omega_0/\omega_{\mathrm{c}})^2$ for the neutral mode, and $a_{\mathrm{e}}\omega_0/\omega_{\mathrm{c}}$ for the charged mode (see Fig.~3 and Eqs.~(38), (91), and (54) of \citep{2025PhRvD.111f3055N} for details). As shown in Eqs.~\eqref{eq:induced_scattering_growth_rate_summary_nomagnetic}, \eqref{eq:growth_rate_neutral_matome_Brillouin}, \eqref{eq:growth_rate_charged_matome_Brillouin_low}, and \eqref{eq:growth_rate_charged_matome_Brillouin_intermediate}, the difference in the order of growth rates for all modes except ICS in the intermediate and high density regimes of the charged mode \nisHiura{(as discussed below)} is determined by the scaling of this dimensionless velocity. The difference due to the subluminal effect is associated with the deviation of the EM wave phase velocity from $c$, and remains of order unity for $\omega_{\mathrm{p}}\ll\omega_{\mathrm{c}}$.

In contrast, only ICS in the intermediate and high density regimes of the charged mode exhibits a suppression stronger than that expected from the dimensionless velocity scaling. This additional suppression is caused by Debye screening and originates from the fact that the charged mode, unlike the ordinary or neutral mode, drives an instability that involves charge separation \ioka{(see Fig. \ref{fig:parametric_instability})}~\citep{2025PhRvD.111f3055N}. As a result, in the collective limit \eqref{eq:plasma_density_dekai_debye_screening_Brillouin} (i.e., in the intermediate and high density regimes), the growth of electrostatic waves is screened within the Debye length, leading to a substantial reduction in the linear growth rate.

\section{Linear Growth Rates for Broadband Incident Waves}
\label{subsec:broadband_Brillouin}
This section discusses the linear growth rates when the incident wave has a finite bandwidth $\Delta\omega$. In particular, we analytically examine how a broadband incident wave modifies the growth rates of induced scattering for the ordinary, neutral, and charged modes. In laboratory plasma, such as laser plasma, narrow-band incident waves are typically used. In contrast, astrophysical EM radiation, such as that from FRBs and pulsars, often exhibits a broadband spectrum~\citep{2021ApJ...923....1P,1994ApJ...422..304T}.

For ICS, the broadband effect has already been formulated~\citep{2025PhRvD.111f3055N}. When a monochromatic wave is assumed, as shown in Eqs.~\eqref{eq:maximum_growth_rate_no_magnetic_Compton_Brillouin}, \eqref{eq:growth_rate_induced_Compton_subdominant_Brillouin}, and \eqref{eq:maximum_growth_rate_magnetic2_Brillouin}, the linear growth rate contains a factor $(c/v_{\mathrm{th}})^2$ reflecting Doppler broadening. If the incident wave has a bandwidth $\Delta\omega$ that exceeds this Doppler width, the growth rate becomes dependent on $\Delta\omega$, with $(c/v_{\mathrm{th}})^2 \rightarrow (\omega_0/\Delta\omega)^2$. This replacement applies to ICS in the ordinary, neutral, and charged modes.

For SBS and SRS, by contrast, the resonance between the broadband incident wave (bandwidth $\Delta\omega$) and density fluctuations (with bandwidth $t_{\mathrm{coh}}^{-1}$, where $t_{\mathrm{coh}}$ is the coherence time corresponding to the growth rate) is only partial. In particular, when $\Delta\omega \gg t_{\mathrm{coh}}^{-1}$, only a fraction $t_{\mathrm{coh}}^{-1}/\Delta\omega$ of the EM wave is in resonance, reducing the efficiency of instability excitation. Thus, for a broadband incident wave, the linear growth rate is estimated as follows~\citep{1974PhFl...17..849T,1994ApJ...422..304T,kruer2019physics,2021PPCF...63i4003B,2023RvMPP...7....1Z}:
\begin{equation}
t_{\text {inc }}^{-1} \sim \frac{t_{\text {coh }}^{-2}}{\Delta \omega}
\quad \text { if } \quad \Delta \omega \gg t_{\text {coh }}^{-1}.
\label{eq:broadband_suppression_growth_rate}
\end{equation}

In this work, we also apply Eq.~\eqref{eq:broadband_suppression_growth_rate} as an approximate estimate for SBS and SRS even in the strong coupling regime. However, the precise behavior of parametric instabilities in the strong coupling regime with broadband incident waves remains a subject for future theoretical study.

\subsection{Ordinary mode}
In this section, we evaluate the linear growth rate of instabilities for the ordinary mode driven by a broadband incident wave. The growth rate of ICS for the ordinary mode has already been derived by \citet{2025PhRvD.111f3055N} and is given as follows\footnote{
\niShiura{As discussed in Appendix~\ref{sec:ordinary_mode_detail_derivation}, for the ordinary mode, only side scattering is possible for transverse waves. When this contribution is taken into account, Eq.~\eqref{eq:broad_ordinary_Compton_growth_rate} becomes smaller by a factor of 2 compared to Eq.~(113) in \citep{2025PhRvD.111f3055N}.}}:
\begin{equation}
\left(t_{\mathrm{C} \|}^{\mathrm{inc}}\right)^{-1} \sim \frac{\pi}{2} \left(a_{\mathrm{e}} \frac{\omega_{\mathrm{p}}}{\omega_0}\right)^2 
\left(\frac{\omega_0}{\Delta \omega}\right)^2 \omega_0.
\label{eq:broad_ordinary_Compton_growth_rate}
\end{equation}

For SBS in the strong coupling regime, we apply the broadband suppression factor~\eqref{eq:broadband_suppression_growth_rate} to the maximum growth rate in Eq.~\eqref{eq:induced_Brillouin_nomagnetic_maximum_growth_rate}. This gives
\begin{equation}
\left(t_{\mathrm{B} \|}^{\mathrm{inc}}\right)^{-1} \sim \frac{3}{2^{\frac{10}{3}}} 
\left(a_{\mathrm{e}} \frac{\omega_{\mathrm{p}}}{\omega_0}\right)^{\frac{4}{3}} 
\frac{\omega_0}{\Delta \omega} \omega_0.
\end{equation}

Summarizing the above, the instability growth rate for the ordinary mode in the presence of a broadband incident wave can be expressed as
\begin{equation}
\left(t_\|^{\mathrm{inc}}\right)^{-1} \sim 
\begin{cases}
\frac{\pi}{2} \left(a_{\mathrm{e}} \frac{\omega_{\mathrm{p}}}{\omega_0}\right)^2 
\left(\frac{\omega_0}{\Delta \omega}\right)^2 \omega_0, & 
a_{\mathrm{e}} \ll a_{\mathrm{e,trans}}, \\[10pt]
\frac{3}{2^{\frac{10}{3}}} \left(a_{\mathrm{e}} \frac{\omega_{\mathrm{p}}}{\omega_0}\right)^{\frac{4}{3}} 
\frac{\omega_0}{\Delta \omega} \omega_0, & 
a_{\mathrm{e,trans}} \ll a_{\mathrm{e}} \ll 1,
\end{cases}
\end{equation}
where the transition amplitude $a_{\mathrm{e,trans}}$ between ICS and SBS, at which the respective growth rates become equal, is determined as follows:
\begin{equation}
a_{\mathrm{e,trans}} \sim 8.2 \times 10^{-2} 
\frac{\omega_0}{\omega_{\mathrm{p}}} 
\left(\frac{\Delta \omega}{\omega_0}\right)^{\frac{3}{2}}.
\end{equation}

\subsection{Neutral mode}
In this section, we evaluate the linear growth rate of instabilities for the neutral mode driven by a broadband incident wave. The growth rate of ICS for the neutral mode is given by (see Eq. (119) in \citep{2025PhRvD.111f3055N})
\begin{equation}
\left(t_{\text{C}, \mathrm{neutral}}^{\mathrm{inc}}\right)^{-1} \sim \pi \frac{\omega_{\mathrm{p}}^{2} a_{\mathrm{e}}^{2}}{\omega_{0}} \left( \frac{\omega_{0}}{\omega_{\mathrm{c}}} \right)^{4}
\left( \frac{\omega_{0}}{\Delta \omega} \right)^{2}.
\end{equation}

For SBS in the strong coupling regime, the broadband suppression factor~\eqref{eq:broadband_suppression_growth_rate} is applied to the maximum growth rate in Eq.~\eqref{eq:induced_Brillouin_neutral_maximum_growth_rate}, resulting in
\begin{equation}
\left(t_{\text{B}, \mathrm{neutral}}^{\mathrm{inc}}\right)^{-1} \sim \frac{3}{2^{\frac{2}{3}}} 
\left(a_{\mathrm{e}} \frac{\omega_{\mathrm{p}}}{\omega_0}\right)^{\frac{4}{3}} 
\left(\frac{\omega_0}{\omega_{\mathrm{c}}}\right)^{\frac{8}{3}} 
\frac{\omega_0}{\Delta \omega} \omega_0.
\end{equation}

The resulting instability growth rate for the neutral mode can therefore be summarized as
\begin{equation}
\begin{aligned}
\left(t_{\mathrm{neutral}}^{\mathrm{inc}}\right)^{-1}
\sim
\begin{cases}
&\pi
\left(
a_{\mathrm{e}}
\frac{\omega_{\mathrm{p}}}{\omega_{0}}
\right)^{2}
\left(
\frac{\omega_{0}}{\omega_{\mathrm{c}}}
\right)^{4}
\left(
\frac{\omega_{0}}{\Delta\omega}
\right)^{2}\!
\omega_{0},\\
&\quad
a_{\mathrm{e}}
\frac{\omega_{0}}{\omega_{\mathrm{c}}}
\ll
\left(
a_{\mathrm{e}}
\frac{\omega_{0}}{\omega_{\mathrm{c}}}
\right)_{\mathrm{trans}},\\[2.0ex]
&\frac{3}{2^{\frac{2}{3}}}
\left(
a_{\mathrm{e}}
\frac{\omega_{\mathrm{p}}}{\omega_{0}}
\right)^{\frac{4}{3}}
\left(
\frac{\omega_{0}}{\omega_{\mathrm{c}}}
\right)^{\frac{8}{3}}
\frac{\omega_{0}}{\Delta\omega}\,
\omega_{0},\\
&\quad
\left(
a_{\mathrm{e}}
\frac{\omega_{0}}{\omega_{\mathrm{c}}}
\right)_{\mathrm{trans}}
\ll
a_{\mathrm{e}}
\frac{\omega_{0}}{\omega_{\mathrm{c}}}
\ll1.
\end{cases}
\end{aligned}
\end{equation}
The transition point between ICS and SBS in the strong coupling regime, at which the growth rates become equal, is determined as follows:
\begin{equation}
\left(a_{\mathrm{e}} \frac{\omega_0}{\omega_{\mathrm{c}}}\right)_{\mathrm{trans}} 
\sim 4.7 \times 10^{-1} 
\frac{\omega_{\mathrm{c}}}{\omega_{\mathrm{p}}} 
\left(\frac{\Delta \omega}{\omega_0}\right)^{\frac{3}{2}}.
\end{equation}

\subsection{Charged mode}
This section evaluates the growth rates for the instability of the charged mode induced by a broadband incident wave. 

\subsubsection{Induced Compton Scattering (Charged mode)}
The behavior of the ICS growth rate differs between the noncollective limit~\eqref{eq:limitation_of_plasma_frequency_Brilluin1} and the collective limit~\eqref{eq:plasma_density_dekai_debye_screening_Brillouin}. For a broadband incident wave, the growth rate for ICS has already been derived in \citet{2025PhRvD.111f3055N}.
\paragraph{Noncollective limit (low density regime)}
The ICS growth rate is expressed as (see Eq. (117) in \citep{2025PhRvD.111f3055N})
\begin{equation}
\left(t_{\text{C}, \mathrm{charged}}^{\mathrm{inc}}\right)^{-1} 
\sim 
\pi 
\frac{\omega_{\mathrm{p}}^{2} a_{\mathrm{e}}^{2}}{\omega_{0}} 
\left( \frac{\omega_{0}}{\omega_{\mathrm{c}}} \right)^{2}
\left( \frac{\omega_{0}}{\Delta \omega} \right)^{2}.
\label{eq:Compton_Broadband_charged_Brillouin}
\end{equation}
where the condition for the low density (noncollective) is given by Eq.~\eqref{eq:charged_mode_instability_thin_condition_Brillouin_Brillouin}.
\paragraph{Collective limit (intermediate and high density regimes)}
The growth rate is given by (see Eq. (116) in \citep{2025PhRvD.111f3055N})
\begin{equation}
\begin{aligned}
\left(t_{\text{C}, \mathrm{charged}}^{\mathrm{inc}}\right)^{-1}
&\sim
\frac{\pi}{4}
\left( \frac{\omega_{0}}{\omega_{\mathrm{c}}} \right)^{2}
\left( 1 + \frac{\omega_{\text{p}}^2}{\omega_{\text{c}}^2} \right)^2
\frac{\omega_{\mathrm{p}}^{2} a_{\mathrm{e}}^{2}}{\omega_{0}}
\\
&\qquad \times
\left( \frac{8 k_{\mathrm{B}} T_{\mathrm{e}}}{m_{\mathrm{e}} c^{2}} \right)^{2}
\left( \frac{\omega_{0}}{\omega_{\mathrm{p}}} \right)^{4}
\left( \frac{\omega_{0}}{\Delta \omega} \right)^{2},
\end{aligned}
\label{eq:Compton_charged_Brillouin_growth_rate_Debye}
\end{equation}
where the condition for intermediate and high density (collective) is provided by Eq.~\eqref{eq:charged_mode_instability_thin_condition_Brillouin_Compton}.

In summary, ICS for the charged mode displays qualitatively different growth behaviors in the noncollective and collective limits, as represented by Eqs.~\eqref{eq:Compton_Broadband_charged_Brillouin} and~\eqref{eq:Compton_charged_Brillouin_growth_rate_Debye}.

\subsubsection{Stimulated Brillouin Scattering (Charged mode)}
For the strong coupling regime, the growth rate of SBS (degenerate with strong-coupling SRS) can be estimated by applying the broadband suppression formula~\eqref{eq:broadband_suppression_growth_rate} to the maximum growth rate~\eqref{eq:induced_Brillouin_charged_maximum_growth_rate}. The resulting expression is given by
\begin{equation}
\left(t_{\text{B}, \mathrm{charged}}^{\mathrm{inc}}\right)^{-1} 
\sim 
\frac{3}{2^{\frac{2}{3}}} 
\left(a_{\mathrm{e}} \frac{\omega_{\mathrm{p}}}{\omega_0}\right)^{\frac{4}{3}} 
\left(\frac{\omega_0}{\omega_{\mathrm{c}}}\right)^{\frac{4}{3}} 
\frac{\omega_0}{\Delta \omega} \omega_0.
\label{eq:Brillouin_Broadband_charged_Brillouin}
\end{equation}

\subsubsection{Stimulated Raman Scattering (Charged mode)}
In the strong coupling regime, SRS is degenerate with SBS as discussed in Sec. \ref{subsec:degeneracy_Brillouin_Raman_charged_mode}. Therefore, this subsection focuses on the weak coupling regime. The effect of the incident EM wave bandwidth on the SRS growth rate depends on the plasma density regime.

\paragraph{Intermediate density regime}
For the intermediate density regime~\eqref{eq:induced_Raman_condition_Brillouin}, the maximum growth rate for the weak coupling regime~\eqref{eq:growth_rate_of_induced_Raman_weak_coupling} with the broadband correction~\eqref{eq:broadband_suppression_growth_rate} is given by
\begin{equation}
    \left(t_\text{R,weak}^{\mathrm{inc}}\right)^{-1} = a_{\mathrm{e}}^2 \omega_{\mathrm{p}} 
    \left(\frac{\omega_0}{\omega_{\mathrm{c}}}\right)^2 
    \frac{\omega_0}{\Delta \omega}.
    \label{eq:incoherent_raman_weak}
\end{equation}
This is consistent with results from the \nisHiura{magnetized }quantum approach~\citep{1998MNRAS.298.1198L}. \nIshiura{In this regime, both ICS and SRS can be driven simultaneously. However, as in the case of a monochromatic incident wave, the growth rate of SRS always exceeds that of ICS. The ratio of their growth rates can be expressed, from Eqs.~\eqref{eq:incoherent_raman_weak} and \eqref{eq:Compton_charged_Brillouin_growth_rate_Debye}, as follows:
\begin{equation}
    \frac{\left(t_{\mathrm{R,weak}}^{\mathrm{inc}}\right)^{-1}}
         {\left(t_{\text{C}, \mathrm{charged}}^{\mathrm{inc}}\right)^{-1}} 
    \sim \frac{4}{\pi}
    \left(\frac{m_{\mathrm{e}} c^{2}}{8 k_{\mathrm{B}} T_{\mathrm{e}}}\right)^{2}
    \left(\frac{\omega_{\mathrm{p}}}{\omega_{0}}\right)^{3}
    \left(1+\frac{\omega_{\mathrm{p}}^{2}}{\omega_{\mathrm{c}}^{2}}\right)^{-2}
    \frac{\Delta \omega}{\omega_{0}}.
    \label{eq:ratio_ICS_SRS_Broadband}
\end{equation}
Using the intermediate and high density regimes in Eq.~\eqref{eq:charged_mode_instability_thin_condition_Brillouin_Compton}, the right-hand side of Eq.~\eqref{eq:ratio_ICS_SRS_Broadband} satisfies the following lower bound:
\begin{equation}
\begin{aligned}
    &\frac{4}{\pi}\left(\frac{m_{\mathrm{e}} c^{2}}{8 k_{\mathrm{B}} T_{\mathrm{e}}}\right)^{2}
    \left(\frac{\omega_{\mathrm{p}}}{\omega_{0}}\right)^{3}
    \left(1+\frac{\omega_{\mathrm{p}}^{2}}{\omega_{\mathrm{c}}^{2}}\right)^{-2}
    \frac{\Delta \omega}{\omega_{0}}\\
    &\gg \frac{2}{\pi}\frac{\Delta \omega}{\omega_{0}}
    \left(\frac{2k_{\mathrm{B}} T_{\mathrm{e}}}{m_{\mathrm{e}} c^{2}}\right)^{-\frac{1}{2}}
    \left(1+\frac{\omega_{\mathrm{p}}^{2}}{\omega_{\mathrm{c}}^{2}}\right)^{-\frac{1}{2}}
    .
\end{aligned}
\end{equation}
When $v_{\mathrm{th}}/c \ll 1$, $\omega_{0} \sim \Delta \omega$, and $(1+\omega_{\mathrm{p}}^{2}/\omega_{\mathrm{c}}^{2})^{-1} \sim 1$, the right-hand side of this inequality always exceeds unity. Therefore, in the intermediate density regime, the growth rate of SRS remains larger than that of ICS even for broadband incident waves.}

\paragraph{Low density regime (small angle scattering)}
For the low density regime~\eqref{eq:charged_mode_instability_thin_condition_Brillouin_Brillouin}, where SRS is limited to small-angle scattering, the maximum growth rate~\eqref{eq:stimulated_Raman_small_angle_growth_rate} with broadband correction \eqref{eq:broadband_suppression_growth_rate} is expressed as
\begin{equation}
\begin{aligned}
\left(t_{\mathrm{R}}^{\mathrm{inc}}\right)^{-1} &\sim 9.2\times10^{-2}\, a_{\mathrm{e}}^{2} \omega_{\mathrm{p}} 
\left( \frac{\omega_{0}}{\omega_{\mathrm{c}}} \right)^{2}
\left( \frac{m_{\mathrm{e}} c^{2}}{32\, k_{\mathrm{B}} T_{\mathrm{e}}} \right) \\
& \quad \times
\left( \frac{\omega_{\mathrm{p}}}{\omega_{0}} \right)^{2}
\left( 1 + \frac{\omega_{\mathrm{p}}^{2}}{\omega_{\mathrm{c}}^{2}} \right)^{-1}
\frac{\omega_{0}}{\Delta \omega}.
\end{aligned}
\label{eq:Raman_small_angle_broadband_charged_brillouin}
\end{equation}

\subsubsection{Summary of Charged Mode}
The instability of the charged mode under broadband incident EM waves can be classified according to the coupling and resonance conditions, as in the monochromatic case, following the roadmap in Tab.~\ref{tab:roadmap_coupling}. The character of the instability further depends on the density regime—low, intermediate, or high density (see Fig. \ref{fig:charged_regime}).

\paragraph{Low density regime}
In the low density regime, as defined by Eq.~\eqref{eq:charged_mode_instability_thin_condition_Brillouin}, broadband incident waves yield the same dominant instabilities as in the monochromatic case: ICS dominates under the weak coupling condition, while SRS (degenerate with SBS) dominates in the strong coupling condition. The corresponding growth rates are given from Eqs. \eqref{eq:Compton_Broadband_charged_Brillouin} and \eqref{eq:Brillouin_Broadband_charged_Brillouin} by
\begin{equation}
\left(t_{\mathrm{charged}}^{\mathrm{inc}}\right)^{-1}\!\sim
\begin{cases}
&\pi\!\left(a_{\mathrm{e}}\frac{\omega_{\mathrm{p}}}{\omega_0}\right)^2
\!\left(\frac{\omega_0}{\omega_{\mathrm{c}}}\right)^2
\!\left(\frac{\omega_0}{\Delta\omega}\right)^2 \omega_0,\\
&\qquad a_{\mathrm{e}}\frac{\omega_0}{\omega_{\mathrm{c}}}\ll
\left(a_{\mathrm{e}}\frac{\omega_0}{\omega_{\mathrm{c}}}\right)_{\mathrm{trans}},
\\
&\frac{3}{2^{\frac{2}{3}}}\!
\left(a_{\mathrm{e}}\frac{\omega_{\mathrm{p}}}{\omega_0}\right)^{\frac{4}{3}}
\!\left(\frac{\omega_0}{\omega_{\mathrm{c}}}\right)^{\frac{4}{3}}
\frac{\omega_0}{\Delta\omega}\,\omega_0,\\
&\qquad\left(a_{\mathrm{e}}\frac{\omega_0}{\omega_{\mathrm{c}}}\right)_{\mathrm{trans}}\ll
a_{\mathrm{e}}\frac{\omega_0}{\omega_{\mathrm{c}}}\ll1.
\end{cases}
\end{equation}
Here, the transition amplitude $a_{\text{e}}\omega_0/\omega_{\text{c}}$ at which the two growth rates are equal is given by
\begin{equation}
\left(a_{\mathrm{e}} \frac{\omega_0}{\omega_{\mathrm{c}}}\right)_{\mathrm{trans}} 
\sim 4.7 \times 10^{-1} 
\frac{\omega_0}{\omega_{\mathrm{p}}} 
\left(\frac{\Delta \omega}{\omega_0}\right)^{\frac{3}{2}}.
\end{equation}

Additionally, SRS for small-angle scattering can be excited, with a growth rate given by Eq.~\eqref{eq:Raman_small_angle_broadband_charged_brillouin}. The ratio of the growth rates for SRS and ICS, as expressed in Eqs.~\eqref{eq:Raman_small_angle_broadband_charged_brillouin} and \eqref{eq:Compton_Broadband_charged_Brillouin}, is estimated as follows:
\begin{equation}
\begin{aligned}
\frac{\left(t_{\mathrm{R}}^{\mathrm{inc}}\right)^{-1}}
     {\left(t_{\text{C},\,\mathrm{charged}}^{\mathrm{inc}}\right)^{-1}}
&\sim 2.9 \times 10^{-2}\,
   \left(\frac{m_{\mathrm{e}} c^{2}}{32\,k_{\mathrm{B}} T_{\mathrm{e}}}\right)
   \frac{\omega_{\mathrm{p}}}{\omega_{0}} \\[0.5ex]
&\quad\times
   \left(1+\frac{\omega_{\mathrm{p}}^{2}}{\omega_{\mathrm{c}}^{2}}\right)^{-1}
   \frac{\Delta\omega}{\omega_{0}} .
\end{aligned}
\end{equation}
This result matches the ratio of SRS to ICS in ion–electron plasma without magnetic fields, except for two differences \citep{2021Univ....7...56L}. First, the geometry imposed by the magnetic field significantly restricts the allowed scattering angles, suppressing the SRS growth rate by a factor $2.9 \times 10^{-2}$. Second, the subluminal effect introduces a correction factor $\left(1 + \omega_{\mathrm{p}}^2 / \omega_{\mathrm{c}}^2\right)$ of order unity when $\omega_{\mathrm{p}} \ll \omega_{\mathrm{c}}$.

\paragraph{Intermediate density regime}
\nIshiura{In the intermediate density regime in Eq.~\eqref{eq:induced_Raman_condition_Brillouin}, broadband incident waves yield the same dominant instabilities as in the monochromatic case: SRS dominates under the weak coupling condition, while SRS (degenerate with SBS) dominates in the strong coupling condition. The linear growth rates are given from Eqs. \eqref{eq:incoherent_raman_weak} and \eqref{eq:Brillouin_Broadband_charged_Brillouin} by
\begin{equation}
\left(t_{\mathrm{charged}}^{\mathrm{inc}}\right)^{-1} \sim 
\begin{aligned}
\begin{cases}
a_{\mathrm{e}}^2 \omega_{\mathrm{p}} 
    \left(\frac{\omega_0}{\omega_{\mathrm{c}}}\right)^2 
    \frac{\omega_0}{\Delta \omega},  \\
\quad a_{\mathrm{e}} \frac{\omega_0}{\omega_{\mathrm{c}}} \ll 
\left(a_{\mathrm{e}} \frac{\omega_0}{\omega_{\mathrm{c}}}\right)_{\mathrm{trans}}, \\[1.0ex]
\frac{3}{2^{\frac{2}{3}}} 
\left(a_{\mathrm{e}} \frac{\omega_{\mathrm{p}}}{\omega_0}\right)^{\frac{4}{3}} 
\left(\frac{\omega_0}{\omega_{\mathrm{c}}}\right)^{\frac{4}{3}} 
\frac{\omega_0}{\Delta \omega} \, \omega_0,  \\
\quad \left(a_{\mathrm{e}} \frac{\omega_0}{\omega_{\mathrm{c}}}\right)_{\mathrm{trans}} \ll 
a_{\mathrm{e}} \frac{\omega_0}{\omega_{\mathrm{c}}} \ll 1.
\end{cases}
\end{aligned}
\end{equation}
Here, the transition amplitude $a_{\text{e}}\omega_0/\omega_{\text{c}}$ is determined from the condition that the two growth rates are equal, and is expressed as  
\begin{equation}
\left(a_{\mathrm{e}} \frac{\omega_0}{\omega_{\mathrm{c}}}\right)_{\mathrm{trans}}
\simeq 2.6~
\left(\frac{\omega_{\mathrm{p}}}{\omega_0}\right)^{\frac{1}{2}}.
\end{equation}
}

\paragraph{High density regime}
Finally, in the high density regime in Eq.~\eqref{eq:charged_mode_instability_dense_condition_Brillouin}, the dominant instability is ICS with Debye screening, with the growth rate given by Eq.~\eqref{eq:Compton_charged_Brillouin_growth_rate_Debye}.

\section{Conclusion}
In this study, we have systematically derived the linear growth rates and dominant regimes of induced scattering instabilities for the ordinary, neutral, and charged modes in strongly magnetized $e^\pm$ pair plasma. When the incident EM wave is polarized parallel to the background magnetic field, the ordinary mode is excited. When the polarization is perpendicular, both the neutral and charged modes are excited \ioka{(see Fig. \ref{fig:parametric_instability})}. In the ordinary and neutral modes, ICS and SBS occur, while in the charged mode, SRS can also be excited. Notably, in strongly magnetized $e^\pm$ pair plasma, SRS arises in the charged mode, whereas it is absent in the non-magnetized case~\citep{2016PhRvL.116a5004E,2017PhRvE..96e3204S,2023MNRAS.522.2133I}.

As summarized in Tab.~\ref{tab:roadmap_coupling}, the dominant instability regime for each mode can be classified by the coupling and resonance conditions. The coupling condition is categorized into weak and strong coupling, as defined in Eqs.~\eqref{eq:weak_coupling_condition} and~\eqref{eq:strong_coupling_condition}. In the weak coupling regime, instabilities are governed by Landau resonance or linear eigenmode resonance, while in the strong coupling regime, instabilities originate from nonlinear interactions with density fluctuations induced by the ponderomotive force of the incident and scattered waves. For the ordinary and neutral modes, ICS dominates in the weak coupling regime, and SBS dominates in the strong coupling regime. For the charged mode, SRS dominates only if the resonance with a Langmuir wave without significant Landau damping is established under weak coupling. Moreover, the charged mode exhibits different dominant instabilities in the low, intermediate, and high density regimes, as illustrated in Fig.~\ref{fig:charged_regime}.

The linear growth rates of induced scattering in each instability mode are determined by the dimensionless oscillatory velocity driven in electrons and positrons by the incident wave. For the ordinary mode, the control parameter is $a_{\mathrm{e}}$; for the neutral mode, it is $a_{\mathrm{e}}(\omega_0/\omega_{\mathrm{c}})^2$; and for the charged mode, it is $a_{\mathrm{e}}\omega_0/\omega_{\mathrm{c}}$. In the ordinary mode, the growth rates of ICS and SBS are of the same order as in the non-magnetized case, as electrons and positrons can move freely along the field (see Eq. \eqref{eq:induced_scattering_growth_rate_summary_nomagnetic}). For the neutral mode, one must replace $a_{\mathrm{e}}$ with $a_{\mathrm{e}}(\omega_0/\omega_{\mathrm{c}})^2$, yielding a suppression factor of $(\omega_0/\omega_{\mathrm{c}})^4$ for ICS and $(\omega_0/\omega_{\mathrm{c}})^{4/3}$ for SBS in the growth rate (see Eq. \eqref{eq:growth_rate_neutral_matome_Brillouin}.). Similarly, for the charged mode, the relevant parameter is $a_{\mathrm{e}}\omega_0/\omega_{\mathrm{c}}$. The suppression factors are $(\omega_0/\omega_{\mathrm{c}})^2$ for ICS, $(\omega_0/\omega_{\mathrm{c}})^{2/3}$ for SBS, and $(\omega_0/\omega_{\mathrm{c}})$ for SRS (See Eqs. \eqref{eq:growth_rate_charged_matome_Brillouin_low}, \eqref{eq:growth_rate_charged_matome_Brillouin_intermediate}, and \eqref{eq:growth_rate_charged_matome_Brillouin_high}.) \footnote{For SRS, we compare with results for ion–electron plasma, since SRS does not arise in non-magnetized $e^\pm$ pair plasma.}. However, for ICS in the charged mode in the intermediate and high density regimes, additional suppression due to Debye screening \eqref{eq:suppression_effect_Debye_screening_Brillouin} appears~\citep{2025PhRvD.111f3055N} (see Eq. \eqref{eq:growth_rate_charged_matome_Brillouin_high}.). Furthermore, both the neutral and charged modes exhibit an order-unity correction $\left(1 + \omega_{\mathrm{p}}^2 / \omega_{\mathrm{c}}^2\right)$ when $\omega_{\mathrm{p}} \ll \omega_{\mathrm{c}}$ from the subluminal effect, which reflects the deviation of the EM wave phase velocity from the speed of light.

\rei{In companion papers (\citet{2026arXiv260101169K,2026arXiv260118865N}), we demonstrate that the analytical formulae derived in this work are consistent with results obtained from PIC simulations, thereby validating our theoretical framework.}
\ioka{In companion PIC papers \citet{2026arXiv260101169K,2026arXiv260118865N}, we tested representative parameter regimes of the neutral and charged modes within the present framework and found agreement with the analytical formulae. The setup consists of one-dimensional PIC simulations. A circularly polarized Alfv\'en wave is imposed initially, and the subsequent time evolution of the scattered wave is followed. \citet{2026arXiv260101169K} showed that, in parameter regimes where ICS is dominant in the neutral and charged modes, the analytical linear growth rates agree with the simulation results. \citet{2026arXiv260118865N} showed that, in the neutral mode, varying $a_\mathrm{e}$ reproduces the continuous transition from weak coupling ICS to strong coupling SBS. The linear growth rate was also confirmed to be consistent with the present theory. Both papers further examined the nonlinear evolution of the instability.}

The unified framework for induced scattering developed in this paper is applicable to both the emission and propagation of FRBs. In particular, it can be used to evaluate whether low-frequency EM pulses in magnetar magnetospheres can convert into FRBs via free electron laser or Compton scattering mechanism~\citep{2021ApJ...922..166L,2022ApJ...925...53Z,2024ApJ...972..124Q}, or whether FRBs can escape the magnetospheric plasma without significant induced scattering. By comparing the timescales of induced scattering with FRB durations in various regions of the magnetosphere, one can quantitatively assess the potential locations of FRB production and attenuation.

Future work should consider the competition with other parametric instabilities not addressed here. In particular, it will be important to examine nonlinear interactions of Alfvén and fast-mode (X-mode) waves~\citep{1998PhRvD..57.3219T,2019ApJ...881...13L,2019MNRAS.483.1731L,2023ApJ...957..102G}, as well as modulation instabilities. A unified description of these processes will require relaxing the assumption of two transverse waves and one longitudinal wave in Eqs. \eqref{eq:transverse_condition_Brillouin} and \eqref{eq:longitudinal_condition_Brillouin}, and extending the electromagnetic potential coupling terms for higher-order (four-wave or more) interactions.

Finally, comprehensive understanding of EM wave–plasma interactions throughout all regions traversed by FRBs is also crucial. The present theory is primarily valid in the inner magnetosphere, where the wave amplitude is linear ($a_{\mathrm{e}}\omega_0/\omega_{\mathrm{c}}<1$). In contrast, in the outer magnetosphere or beyond it, nonlinear effects may dominate, making numerical approaches such as PIC simulations useful for future studies.

\begin{acknowledgments}
We gratefully acknowledge insightful discussions with Masanori Iwamoto and Wataru Ishizaki on the derivational steps and the physical interpretation. We are especially grateful to Wataru Ishizaki for carefully reading the manuscript and for constructive advice. RN is supported by JST SPRING, Grant No. JPMJSP2110, and JSPS KAKENHI, Grant No. 25KJ1562. KI is supported by MEXT/JSPS KAKENHI Grant No.23H01172, 23H05430, 23H04900, 22H00130. SK is supported by MEXT/JSPS KAKENHI Grant No.22H00130 and 23K20038. \niShiura{The authors thank the Yukawa Institute for Theoretical Physics at Kyoto University, where this work was further developed during the YITP-W-25-08 on "Exploring Extreme Transients: Frontiers in the Early Universe and Time-Domain Astronomy".}
\end{acknowledgments}
\appendix
\section{Treatment of Circularly Polarized Incident Waves}
\label{ap:linear_to_circular}
Throughout this study, we have derived the linear growth rate of induced scattering under the assumption that the incident wave is linearly polarized. For circular polarization, however, the \nisHiura{peak amplitude} differs, and a correction must be applied to the linear growth rate. The incident wave is expressed \nisHiura{from Eq. \eqref{eq:A_wave_general}} as
\begin{equation}
    \bm{A}_{\text{w}0}(\bm{r}, t) = A_{0}\bm{\epsilon}_0 \mathrm{e}^{\mathrm{i}\left(\bm{k}_{0} \cdot \bm{r} - \omega_{0} t+\psi_0\right)} + \text{c.c.}
\end{equation}
For linear polarization, the polarization vector is, e.g., $\bm{\epsilon}_0 = (0,1,0)$, and the peak amplitude of the incident wave is given by $\left|\bm{A}_{\text{w}0}\right|_{\mathrm{max}} = 2A_0$. In contrast, for circular polarization, the polarization vector is written as $\bm{\epsilon}_0 = (1/\sqrt{2})\,(0,1,\pm\mathrm{i})$, where the $+$ sign corresponds to left-handed and the $-$ sign to right-handed circular polarization. In this case, the amplitude becomes $\left|\bm{A}_{\text{w}0}\right| = \sqrt{2}A_0$. Thus, the strength parameter for circular polarization is given by \nisHiura{(c.f. Eq. \eqref{eq:strength_parameter_no_magnetic_Brillouin} for linear polarization)}
\begin{equation}
    \nisHiura{a_{\mathrm{e}} \equiv 
\frac{e\left|\bm{A}_{\mathrm{w}0}\right|_{\mathrm{max}}}{m_{\mathrm{e}} c^2}
\xrightarrow[\;]{\text{circ.\ pol.}}
\frac{\sqrt{2} e A_0}{m_{\mathrm{e}} c^2}.}
\end{equation}
Accordingly, the incident wave amplitude $a_{\mathrm{e}}$ and $\eta$, defined in Eqs.~\eqref{eq:strength_parameter_no_magnetic_Brillouin} and \eqref{eq:definition_of_eta}, appearing in the linear growth rate should be replaced as
\begin{equation}
    a_{\mathrm{e}}\rightarrow \sqrt{2}a_{\mathrm{e}}^{\text{circ}},\qquad \eta\rightarrow \sqrt{2}\eta^{\text{circ}}.
\end{equation}

\section{Detailed Analytical Derivation for the Ordinary Mode}
\label{sec:ordinary_mode_detail_derivation}
This section provides a detailed analytical derivation of the growth rates of SBS and ICS for the ordinary mode in strongly magnetized $e^\pm$ pair plasma. Sec.~\ref{subsec:strong_coupling_Brillouin_ordinary_derivation} presents the derivation of the approximate dispersion relation \eqref{eq:dispersion_relation_induced_Brillouin_nomagnetic_ap} and Sec. \ref{subsec:Derivation_of_SBS_ordinary_growth_rate} calculates the maximum linear growth rate for SBS under the strong coupling condition \eqref{eq:strong_coupling_condition}. Subsequently, Sec.~\ref{subsec:angle_dependence_ordinary_Brillouin} analytically evaluates the angular dependence of the SBS growth rate and derives the angular configuration that maximizes the instability.

\subsection{Derivation of dispersion relation for Strong Coupling SBS (Ordinary mode)}
\label{subsec:strong_coupling_Brillouin_ordinary_derivation}
This section presents the detailed derivation of the approximate dispersion relation~\eqref{eq:dispersion_relation_induced_Brillouin_nomagnetic_ap} and the maximum linear growth rate~\eqref{eq:induced_Brillouin_nomagnetic_maximum_growth_rate} for SBS in the strong coupling regime \eqref{eq:strong_coupling_condition}. 
The dispersion relation~\eqref{eq:dispersion_relation_no_magnetic_Brillouin} can be approximated by expanding the plasma dispersion function~\eqref{eq:plasma_dispersion_function_Brillouin} for large arguments as follows,
\begin{equation}
    1 + \zeta Z(\zeta) \sim -\frac{1}{2\zeta^2}.
    \label{eq:plasma_dispersion_function_yuurikannsuu_Brillouin}
\end{equation}
Under the assumption that the beat frequency is much smaller than the frequencies of the incident and scattered waves ($|\omega| \ll \omega_0 \sim \omega_1$), the left-hand side of Eq.~\eqref{eq:dispersion_relation_no_magnetic_Brillouin} can be rewritten as~\footnote{The detailed procedure is given by
\[
\begin{aligned}
&-2 \omega_{0}\Biggl\{\omega - \frac{c^{2}\left(\bm{k}^{2} + 2\bm{k}_{0}\cdot\bm{k}\right)}{2\omega_{0}}\Biggr\}\\
&= -2\omega_{0}\omega + c^{2} k_{1}^{2} - 2c^{2}\bm{k}_{1}\cdot\bm{k}_{0} + c^{2}k_{0}^{2} + 2c^{2}\bm{k}_{0}\cdot\left(\bm{k}_{1} - \bm{k}_{0}\right)\\[1mm]
&= -2\omega_{0}\left(\omega_{1}-\omega_{0}\right) + c^{2} k_{1}^{2} - c^{2}k_{0}^{2}\\[1mm]
&= -2\omega_{0}\omega_{1} + c^{2} k_{1}^{2} + \omega_{0}^{2} + \omega_{\mathrm{p}}^{2}\\[1mm]
&= c^{2} k_{1}^{2} - \omega_{1}^{2} + \omega_{\mathrm{p}}^{2} + \omega^{2}\\[1mm]
&= c^{2} k_{1}^{2} - \omega_{1}^{2} + \omega_{\mathrm{p}}^{2} + \mathcal{O}\left(\left(\frac{\omega}{\omega_{1}}\right)^2\right),
\end{aligned}
\]
where, from the third to the fourth line, $c^{2} k_{0}^{2} - \omega_{0}^{2} + \omega_{\mathrm{p}}^{2}=0$ is used.},
\begin{equation}
\begin{aligned}
c^{2}k_{1}^{2} - \omega_{1}^{2} + \omega_{\mathrm{p}}^{2}
&= -2\omega_{0}
\Biggl\{
\omega 
- \frac{c^{2}\!\left(k^{2} + 2\,\bm{k}_{0}\!\cdot\!\bm{k}\right)}{2\,\omega_{0}}
\Biggr\} \\
&\quad + 
\mathcal{O}\!\left[\left(\frac{\omega}{\omega_{1}}\right)^{2}\right].
\end{aligned}
\label{eq:right_hand_side_of_dispersion_relation_Brillouin}
\end{equation}
By substituting this result and Eq.~\eqref{eq:definition_of_zeta_Brillouin} into the original dispersion relation~\eqref{eq:dispersion_relation_no_magnetic_Brillouin}, we obtain
\begin{equation}
    \omega^{2}\left\{\omega - \frac{c^{2}\left(k^{2} + 2 \bm{k}_{0}\cdot\bm{k}\right)}{2 \omega_{0}}\right\} 
    = \frac{a_{\text{e}}^{2} \omega_{\mathrm{p}}^{2} \mu^{2}c^2 k_{\|}^{2}}{8 \omega_{0}}.
    \label{eq:dispersion_relation_induced_Brillouin_nomagnetic_ap}
\end{equation}

\subsection{Derivation of the linear growth rate for the Strong Coupling SBS (Ordinary mode)}
\label{subsec:Derivation_of_SBS_ordinary_growth_rate}
Starting from the approximate dispersion relation for strong-coupling SBS,
Eq.~\eqref{eq:dispersion_relation_induced_Brillouin_nomagnetic_ap},
we expand it to obtain
\begin{equation}
\omega^{3}
- \frac{c^{2}\!\left(k^{2}+2\,\bm{k}_{0}\!\cdot\!\bm{k}\right)}{2\omega_{0}}\,
\omega^{2}
- \frac{a_{\mathrm{e}}^{2}\,\omega_{\mathrm{p}}^{2}\,c^{2}\,k_{\|}^{2}\,\mu^{2}}
       {8\,\omega_{0}}
= 0.
\label{eq:dispersion_relation_induced_Brillouin_nomagnetic_tenkai}
\end{equation}
Under the strong coupling condition,
Eq.~\eqref{eq:strong_coupling_condition},
the $\omega^{3}$ term dominates the $\omega^{2}$ term\footnote{\Nishiura{Under the strong coupling condition, $\gamma \gg k_{\|} v_{\text{th}}$, and the assumption that the thermal frequency is comparable to the density fluctuation frequency, $k_{\|} v_{\text{th}} \sim |\text{Re}~\omega|$, the $\omega^3$ term dominates over the $\omega^2$ term. Setting $\omega = \text{Re}~\omega + \text{i}\gamma$, the cubic term can be estimated as follows,
\begin{equation}
    \left|\omega^3\right| = \left|(\text{Re}~\omega + \text{i}\gamma)^3\right| \sim \gamma^3.
    \nonumber
\end{equation}
The quadratic term can be evaluated as shown in
\begin{equation}
\begin{aligned}
    \left|\frac{c^{2}\left(k^{2} - 2 \bm{k}_{0}\cdot\bm{k}\right)}{2 \omega_{0}}\omega^{2}\right|
    &= \left|\frac{c^2(k_1^2 - k_0^2)}{2\omega_0}\right| \left|\omega^2\right| \\
    &\sim \left|\text{Re}~\omega\right| \gamma^2.
\end{aligned}
\nonumber
\end{equation}
Therefore, when $\gamma \gg |\text{Re}~\omega|$, the cubic term is much larger than the quadratic term.}
}.
Because
$k^{2}+2\,\bm{k}_{0}\!\cdot\!\bm{k}=k_{1}^{2}-k_{0}^{2}\le 0$ for a Stokes wave,
the source term attains its maximum when
\begin{equation}
    k_{1}=k_{0}\quad\Longrightarrow\quad
    k=\sqrt{2\!\left(1-\nu\right)}\,k_{0}.
\label{eq:induced_Brillouin_nomagnetic_wave_vector_ap}
\end{equation}
Substituting Eq.~\eqref{eq:induced_Brillouin_nomagnetic_wave_vector_ap}
into Eq.~\eqref{eq:dispersion_relation_induced_Brillouin_nomagnetic_tenkai}
gives
\begin{equation}
-\omega^{3}\simeq
-\frac{a_{\mathrm{e}}^{2}\,\omega_{\mathrm{p}}^{2}\,c^{2}\,k_{0}^{2}\,\mu^{2}
      (1-\nu)\cos^{2}\theta_{kB}}
     {4\,\omega_{0}}.
\end{equation}
Using the vacuum dispersion relation $\omega_{0}\simeq c k_{0}$ yields
\begin{equation}
-\omega \simeq
\mathrm{e}^{\pm\frac{\pi}{3}\mathrm{i}}
\left[
\frac{a_{\mathrm{e}}^{2}\,\omega_{\mathrm{p}}^{2}\,\omega_{0}}{2}\,
\frac{\mu^{2}(1-\nu)\cos^{2}\theta_{kB}}{2}
\right]^{\frac{1}{3}},
\end{equation}
so that
\begin{equation}
\mathrm{Im}\,\omega(\mu,\nu,\cos\theta_{kB}) =
\frac{\sqrt{3}}{2}
\left[
\frac{a_{\mathrm{e}}^{2}\,\omega_{\mathrm{p}}^{2}\,\omega_{0}}{2}\,
\frac{\mu^{2}(1-\nu)\cos^{2}\theta_{kB}}{2}
\right]^{\frac{1}{3}}.
\label{eq:induced_Brillouin_nomagnetic_imaginary_part}
\end{equation}
The maximum growth occurs, for instance, at
\begin{equation}
\begin{aligned}
\left(\mu, \nu, \cos \theta_{k B}\right)= & \left(\frac{1}{\sqrt{2}}, \pm \frac{1}{\sqrt{2}}, \pm\frac{\sqrt{2 \pm \sqrt{2}}}{2}\right), \\
& \left(\frac{1}{\sqrt{2}}, 0, \pm\frac{1}{2} \right),
\end{aligned}
\label{eq:maximum_growth_angle_condition}
\end{equation}
as demonstrated in Appendix~\ref{subsec:angle_dependence_ordinary_Brillouin}.
\nisHiura{All} sets correspond to an EM wave linearly polarised parallel
to $\bm{B}_{0}$ and incident perpendicular to the field:
the first and second give $135^{\circ}$ backward \Nishiura{or $45^\circ$ forward~}scattering within the
$\bm{k}_{0}$–$\bm{B}_{0}$ plane, whereas the third and forth give $90^{\circ}$
scattering perpendicular to that plane.

The maximum growth rate of the scattered-wave energy is therefore
\begin{equation}
\left(t_{\mathrm{B}\parallel}^{\max}\right)^{-1}
= \frac{\sqrt{3}}{2^{\frac{4}{3}}}
\left(
\frac{a_{\mathrm{e}}^{2}\,\omega_{\mathrm{p}}^{2}\,\omega_{0}}{2}
\right)^{\frac{1}{3}}.
\label{eq:induced_Brillouin_nomagnetic_maximum_growth_rate_ap}
\end{equation}
The corresponding wavenumber follows from
Eqs.~\eqref{eq:induced_Brillouin_nomagnetic_wave_vector_ap}
and \eqref{eq:maximum_growth_angle_condition}\Nishiura{, 
where the angular parameter is set for the representative case of $90^\circ$ sidescattering, as}
\begin{equation}
    k_{\max}\simeq\sqrt{2}\,k_{0}.
\end{equation}

\subsection{Maximum Growth Angle Parameters for SBS (Ordinary mode)}
\label{subsec:angle_dependence_ordinary_Brillouin}
The maximum value of the angular dependence $\mu^{2}(1-\nu)\cos^{2}\theta_{kB}$, which appears in the linear growth rate of SBS \eqref{eq:induced_Brillouin_nomagnetic_imaginary_part}, can be derived analytically.
\Nishiura{In this study, the following simplified set of assumptions is adopted for the polarization and propagation direction of the incident and scattered waves when evaluating the maximum growth angle.
\begin{enumerate}[label=(\roman*)]
    \item The incident wave is assumed to be an O-mode wave that propagates perpendicular to the background magnetic field, with its electric field component along the magnetic field direction.
    \item The scattered wave is assumed to be a transverse wave propagating obliquely with respect to the background magnetic field. However, in magnetized $e^\pm$ pair plasma, a transverse wave propagating obliquely to the magnetic field is not an exact linear eigenmode. Therefore, the angle parameter for maximum growth obtained under this assumption may differ by a factor from the true value.
\end{enumerate}
To obtain the exact angle parameter for maximum growth, one must relax the condition that both the incident and scattered waves are transverse, as expressed in Eq.~\eqref{eq:transverse_condition_Brillouin}. This would require a more elaborate formulation.
}

First, set the coordinate system so that the incident wave vector, the electric field component, and the background magnetic field are given by
\begin{equation}
\bm{k}_0 \equiv k_0 \bm{e}_z, \qquad
\bm{E}_0 \equiv E_0\bm{e}_x, \qquad
\bm{B}_0 \equiv B_0 \bm{e}_x.
\label{eq:ordinary_zahyousextutei}
\end{equation}
The direction of the scattered wave vector is expressed in spherical coordinates as
\begin{equation}
\hat{\bm{k}}_1 \equiv \bm{e}_r = \sin\theta \cos\phi\, \bm{e}_x + \sin\theta \sin\phi\, \bm{e}_y + \cos\theta\, \bm{e}_z.
\label{eq:ordinary_zahyousextutei2}
\end{equation}
The wave vector for the density fluctuation is given by
\begin{equation}
\begin{aligned}
\bm{k} &= \bm{k}_1 - \bm{k}_0 \\
&\simeq k_0 \left\{ \sin\theta \cos\phi\, \bm{e}_x + \sin\theta \sin\phi\, \bm{e}_y - (1-\cos\theta)\, \bm{e}_z \right\},
\end{aligned}
\label{eq:ordinary_zahyousextutei3}
\end{equation}
where we have used the approximation $k_0 \simeq k_1$ since $\omega_0\sim\omega_1\gg\omega_{\text{p}}$.
From this, the angle parameters $\cos\theta_{kB}$ and $\nu$ are given by
\begin{equation}
\cos\theta_{kB} = \frac{\sin\theta \cos\phi}{\sqrt{2(1-\cos\theta)}}, \qquad
\nu \simeq \cos\theta.
\end{equation}

For the polarization of the scattered wave, we consider both $\bm{e}_\theta$ and $\bm{e}_\phi$ components due to the transverse condition. For the $\bm{e}_\theta$ polarization,
\begin{equation}
\hat{\bm{E}}_1 \equiv \bm{e}_\theta = \cos\theta \cos\phi\, \bm{e}_x + \cos\theta \sin\phi\, \bm{e}_y - \sin\theta\, \bm{e}_z,
\end{equation}
the angular parameter becomes
\begin{equation}
\nisHiura{\mu = \left|\hat{\bm{E}}_0 \cdot \hat{\bm{E}}_1\right| = \left|\cos\theta \cos\phi\right|.}
\end{equation}
Thus, the angular dependence of the growth rate is given by
\begin{equation}
\mu^{2}(1-\nu)\cos^{2}\theta_{kB} = \frac{1}{2} \cos^{2}\theta \sin^{2}\theta \cos^{4}\phi \leq \frac{1}{8}.
\end{equation}
Equality is achieved for
\begin{equation}
(\theta, \phi) = \left(\frac{\pi}{4}, 0 \right), \left(\frac{3\pi}{4}, 0 \right), \left(\frac{\pi}{4}, \pi \right), \left(\frac{3\pi}{4}, \pi \right),
\end{equation}
with the corresponding angle parameters
\begin{equation}
\nisHiura{\begin{aligned}
(\mu, \nu, \cos\theta_{kB}) = & \left(\frac{1}{\sqrt{2}}, \pm \frac{1}{\sqrt{2}}, \pm\frac{\sqrt{2 \pm \sqrt{2}}}{2}\right), \\
& \left(  \frac{1}{\sqrt{2}}, \mp \frac{1}{\sqrt{2}}, \pm\frac{\sqrt{2 \mp \sqrt{2}}}{2}\right),
\end{aligned}}
\label{eq:theta_polarization_ordinary_angular_parameter}
\end{equation}
which represent $135^\circ$ backscattering\Nishiura{~or $45^\circ$ forwardscattering}.

For the $\bm{e}_\phi$ polarization,
\begin{equation}
\hat{\bm{E}}_1 \equiv \bm{e}_\phi = -\sin\phi\, \bm{e}_x + \cos\phi\, \bm{e}_y,
\end{equation}
the angular parameter is
\begin{equation}
\nisHiura{\mu = \left|-\sin\phi\right|}.
\end{equation}
In this case, the angular dependence is
\begin{equation}
\mu^{2}(1-\nu)\cos^{2}\theta_{kB} = \frac{1}{2} \sin^{2}\phi \cos^{2}\phi \sin^{2}\theta \leq \frac{1}{8},
\end{equation}
and equality holds for
\begin{equation}
\begin{aligned}
   (\theta, \phi) = &\left(\frac{\pi}{2}, \frac{\pi}{4} \right),~ \left(\frac{\pi}{2}, \frac{3\pi}{4} \right),\\
   &\left(\frac{\pi}{2}, \frac{5\pi}{4} \right),~\left(\frac{\pi}{2}, \frac{7\pi}{4} \right),
\end{aligned}
\end{equation}
with angle parameters
\begin{equation}
\nisHiura{(\mu, \nu, \cos\theta_{kB}) = \left(\frac{1}{\sqrt{2}}, 0, \pm\frac{1}{2} \right),}
\label{eq:phi_polarization_ordinary_angular_parameter}
\end{equation}
which correspond to $90^\circ$ sidescattering.

In summary, for both $\bm{e}_\theta$ and $\bm{e}_\phi$ polarizations, the angular dependence in the SBS growth rate can reach its maximum. Therefore, any transverse-polarized scattered wave, represented by a linear combination of these, achieves the maximal growth rate.

\section{Detailed Analytic Derivation for the Neutral Mode}
\label{sec:neutral_mode_detail_derivation}
In this section, we present a detailed analytic derivation of the approximate dispersion relation \eqref{eq:dispersion_relation_induced_Brillouin_Neutral_ap} and the maximum linear growth rate \eqref{eq:induced_Brillouin_neutral_maximum_growth_rate} for SBS in the strong coupling regime for the neutral mode. In the dispersion relation \eqref{eq:dispersion_relation_neutral_Brillouin2}, we use the large-argument asymptotic form of the plasma dispersion function, as given in Eq.~\eqref{eq:plasma_dispersion_function_yuurikannsuu_Brillouin}. Assuming the frequency of the beat wave is much smaller than the incident and scattered wave frequencies ($|\omega| \ll \omega_0 \sim \omega_1$), the left-hand side of Eq.~\eqref{eq:dispersion_relation_neutral_Brillouin2} can be transformed, following the same procedure as for the ordinary mode, as follows\Nishiura{~(see Eq. \eqref{eq:right_hand_side_of_dispersion_relation_Brillouin} for details)\footnote{
\nIshiura{In the algebraic manipulation, we use the dispersion relation for the incident wave, $c^{2} k_{0}^{2} - \omega_{0}^{2} - \omega_{\mathrm{p}}^{2} (\omega_0/\omega_{\mathrm{c}})^{2} = 0$, or equivalently $\omega_{0} = k_{0} v_{\mathrm{A}}$.}
}}:
\begin{equation}
\begin{aligned}
&c^{2} k_{1}^{2} - \omega_{1}^{2} - \omega_{\mathrm{p}}^{2}\left(\frac{\omega_{1}}{\omega_{\mathrm{c}}}\right)^{2} \\
&= -2 \omega_{0}\left\{\omega\frac{c^2}{v_{\text{A}}^2} - \frac{c^{2}\left(k^{2} + 2 \bm{k}_{0}\cdot\bm{k}\right)}{2 \omega_{0}}\right\} + \mathcal{O}\left(\left(\frac{\omega}{\omega_{1}}\right)^2\right).
\end{aligned}
\label{eq:right_hand_side_of_dispersion_relation_Brillouin_Neutral}
\end{equation}
Using this result, the characteristic SBS dispersion relation for the neutral mode is expressed as
\begin{equation}
 \omega^{2}\left\{\omega\frac{c^2}{v_{\text{A}}^2} - \frac{c^{2}\left(k^{2} + 2 \bm{k}_{0} \cdot \bm{k}\right)}{2 \omega_{0}}\right\} = \frac{\omega_{\mathrm{p}}^{2} a_{\mathrm{e}}^{2} \mu^{2} c^{2} k_{\|}^{2}}{8 \omega_{0}}\left(\frac{\omega_{0}}{\omega_{\mathrm{c}}}\right)^{4}.
 \label{eq:dispersion_relation_induced_Brillouin_Neutral_ap}
\end{equation}
\rei{which is equivalent to Eq.~\eqref{eq:dispersion_relation_induced_Brillouin_Neutral}.}

To derive the linear growth rate of the scattered wave from the dispersion relation \eqref{eq:dispersion_relation_induced_Brillouin_Neutral_ap}, expand the equation as follows:
\begin{equation}
\begin{aligned}
\omega^{3}\frac{c^2}{v_{\text{A}}^2} 
- \frac{c^{2}\left(k^{2} + 2 \bm{k}_{0} \cdot \bm{k}\right)}{2 \omega_{0}} \omega^{2}
- \frac{a_{\text{e}}^{2} \omega_{\mathrm{p}}^{2} c^{2} k_{\|}^{2} \mu^{2}}{8 \omega_{0}}\left(\frac{\omega_{0}}{\omega_{\mathrm{c}}}\right)^{4} = 0.
\end{aligned}
\label{eq:dispersion_relation_induced_Brillouin_Neutral_tenkai}
\end{equation}
Under the strong coupling condition \eqref{eq:strong_coupling_condition}, as in the ordinary mode\Nishiura{~(see Eq. \eqref{eq:dispersion_relation_induced_Brillouin_nomagnetic_tenkai})}, the $\omega^3$ term dominates over the $\omega^2$ term. The maximum growth occurs when the second term in Eq. \eqref{eq:dispersion_relation_induced_Brillouin_Neutral_tenkai} is zero. As in the ordinary mode, this is satisfied for
\begin{equation}
k_1 = k_0 \quad\Longrightarrow\quad k = \sqrt{2(1-\nu)} k_0.
\label{eq:induced_Brillouin_neutral_wave_vector}
\end{equation}
Substituting Eq.~\eqref{eq:induced_Brillouin_neutral_wave_vector} into Eq.~\eqref{eq:dispersion_relation_induced_Brillouin_Neutral_tenkai} gives
\begin{equation}
-\omega^{3}\frac{c^2}{v_{\text{A}}^2} 
\sim -\frac{a_{\text{e}}^{2}\omega_{\mathrm{p}}^{2}c^{2}(1-\nu)k_{0}^{2}\mu^{2}\cos^{2}\theta_{kB}}{4\omega_{0}}\left(\frac{\omega_{0}}{\omega_{\mathrm{c}}}\right)^{4},
\label{eq:neutral_strong_dispersion_relation_Brillouin}
\end{equation}
where $k_\| = k |\cos\theta_{kB}|$ and $v_{\text{A}}$ is the Alfvén speed \eqref{eq:Alfven_velocity_Brillouin}. Using the incident wave dispersion relation \(\omega_{0}\simeq k_{0}v_{\text{A}}\), we find
\begin{equation}
\begin{aligned}
-\omega \sim e^{\pm \frac{\pi}{3}\text{i}}
\left(\frac{a_{\mathrm{e}}^{2}\omega_{\mathrm{p}}^{2}\omega_{0}}{2}\frac{\mu^{2}(1-\nu)\cos^{2}\theta_{kB}}{2}\right)^{\frac{1}{3}}\left(\frac{\omega_{0}}{\omega_{\mathrm{c}}}\right)^{\frac{4}{3}}.\\
\Rightarrow \text{Im}~\omega 
= \frac{\sqrt{3}}{2} 
\left(\frac{a_{\mathrm{e}}^{2}\omega_{\mathrm{p}}^{2}\omega_{0}}{2}
\frac{\mu^{2}(1-\nu)\cos^{2}\theta_{kB}}{2}\right)^{\frac{1}{3}}
\left(\frac{\omega_{0}}{\omega_{\mathrm{c}}}\right)^{\frac{4}{3}}.
\end{aligned}
\label{eq:induced_Brillouin_neutral_imaginary_part}
\end{equation}
The maximum growth is achieved for
\begin{equation}
\nisHiura{\left(\mu,~\nu,~ \cos \theta_{kB}\right) = (1,~-1,~\pm1),}
\label{eq:maximum_growth_angle_condition_neutral_Brillouin_appendix}
\end{equation}
as in Eq.~\eqref{eq:maximum_growth_angle_condition_neutral_Brillouin}. The linear growth rate of the scattered wave energy at maximum is therefore
\begin{equation}
\left(t_{\mathrm{B},\text{neutral}}^{\max}\right)^{-1}  = \sqrt{3}\left(\frac{a_{\text{e}}^{2}\omega_{\mathrm{p}}^{2}\omega_{0}}{2}\right)^{\frac{1}{3}}\left(\frac{\omega_{0}}{\omega_{\mathrm{c}}}\right)^{\frac{4}{3}}.
\end{equation}
The corresponding maximum growth wave number is, from Eqs.~\eqref{eq:induced_Brillouin_neutral_wave_vector} and \eqref{eq:maximum_growth_angle_condition_neutral_Brillouin_appendix},
\begin{equation}
    k_{\text{max}}\simeq2k_0.
\end{equation}

\section{Detailed Analytical Derivation for the Charged Mode}
\label{sec:charged_mode_detail_derivation}
This section presents the analytical derivations for the growth rates and dispersion relations of various induced scattering processes in the charged mode. Sec.~\ref{subsec:strong_coupling_Brillouin_charged} describes the derivation of the dispersion relation and the maximum growth rate for SBS in the strong coupling regime. Sec.~\ref{subsec:weak_coupling_Raman_charged} discusses the growth rate of SRS under weak coupling conditions in the intermediate and high density regimes \eqref{eq:induced_Raman_condition_Brillouin}. Sec.~\ref{subsec:raman_compton_competition_charged} addresses the competition between SRS and ICS in the intermediate density regime, and derives the dominant mode. Sec.~\ref{subsec:small_angle_Raman_charged} details the derivation of the growth rate for small-angle SRS in the low density regime \eqref{eq:charged_mode_instability_thin_condition_Brillouin}. Finally, Sec.~\ref{subsec:small_angle_Raman_angle_max_charged} provides an analytical discussion of the conditions that maximize the angular dependence for small-angle SRS.

\subsection{Derivation of Strong-Coupling SBS (Charged mode)}
\label{subsec:strong_coupling_Brillouin_charged}
The dispersion relation and the linear growth rate for strong-coupling SBS (which is degenerate with SRS in this regime) in the charged mode can be derived analytically as follows. First, in the dispersion relation~\eqref{eq:dispersion_relation_charged_Brillouin2}, the plasma dispersion function \eqref{eq:definition_of_plasma_dispersion_function} in the large argument limit is approximated by Eq.~\eqref{eq:plasma_dispersion_function_yuurikannsuu_Brillouin}. Additionally, when the beat frequency is much smaller than the frequencies of the incident and scattered waves ($|\omega| \ll \omega_0 \sim \omega_1$), the left-hand side of Eqs.~\eqref{eq:dispersion_relation_neutral_Brillouin2} and \eqref{eq:dispersion_relation_charged_Brillouin2} can be rewritten in the form of Eq.~\eqref{eq:right_hand_side_of_dispersion_relation_Brillouin_Neutral}. Using these approximations and following the same procedure as for the neutral mode, the dispersion relation for SBS in the charged mode is given by
\begin{equation}
\begin{aligned}
&\left\{
\omega \frac{c^2}{v_{\mathrm{A}}^2}
- \frac{c^2\left(k^2 + 2 \bm{k} \cdot \bm{k}_0\right)}{2 \omega_0}
\right\} \\
&\quad = 
\frac{1}{4} \frac{a_{\mathrm{e}}^2 \omega_{\mathrm{p}}^2}{\omega_0}
\left(\frac{c}{v_{\mathrm{th}}}\right)^2
\left(\frac{\omega_0}{\omega_{\mathrm{c}}}\right)^2
\left(1 - \mu^2\right)
\left| \bm{n} \cdot \hat{\bm{B}_0} \right|^2 \\
&\hspace{2.5em} \times 
\frac{1}{2 \zeta^2}
\frac{1}{1 - \dfrac{\omega_{\mathrm{p}}^2}{k^2 v_{\mathrm{th}}^2} \dfrac{1}{\zeta^2}}.
\end{aligned}
\label{eq:charged_Brillouin_tochyuu}
\end{equation}
Substituting the definition of $\zeta$ from Eq.~\eqref{eq:definition_of_zeta_Brillouin} and further simplification yields the following dispersion relation for SBS in the charged mode:
\begin{equation}
\begin{aligned}
&\left(\omega_{\mathrm{p}}^{2}\cos^{2} \theta_{kB}-\omega^{2} \right)
\left\{\omega\frac{c^2}{v_{\text{A}}^2} - \frac{c^{2}\left(k^{2} + 2 \bm{k}_{0} \cdot \bm{k}\right)}{2 \omega_{0}}\right\} \\
&= -\frac{1}{8} \frac{a_{\mathrm{e}}^{2} \omega_{\mathrm{p}}^{2} c^{2} k^{2}}{\omega_{0}}
\left(\frac{\omega_{0}}{\omega_{\mathrm{c}}}\right)^{2}\cos^{2} \theta_{kB}\left(1 - \mu^{2}\right)\left|\bm{n} \cdot \hat{\bm{B}}_{0}\right|^{2}.
\end{aligned}
\label{eq:dispersion_relation_neutral_tochuu_Raman}
\end{equation}

The linear growth rate can be derived by expanding Eq.~\eqref{eq:dispersion_relation_neutral_tochuu_Raman} as follows:
\begin{equation}
\begin{aligned}
&\omega^3 \frac{c^2}{v_{\text{A}}^2}
   - \frac{c^2\left(k^2+2 \boldsymbol{k}_0 \cdot \boldsymbol{k}\right)}{2 \omega_0} \omega^2 \\
&\quad
   - \omega \frac{c^2}{v_{\text{A}}^2} \omega_{\mathrm{p}}^2 \cos^{2} \theta_{kB}
   + \omega_{\mathrm{p}}^2 \cos^{2} \theta_{kB} \frac{c^2\left(k^2+2 \boldsymbol{k}_0 \cdot \boldsymbol{k}\right)}{2 \omega_0} \\
&\quad
   - \frac{1}{8} \frac{a_{\mathrm{e}}^2 \omega_{\mathrm{p}}^2 c^2 k^2}{\omega_0}
     \left(\frac{\omega_0}{\omega_{\mathrm{c}}}\right)^2
     \cos^{2} \theta_{kB} (1-\mu^2)
     \left|\boldsymbol{n} \cdot \hat{\boldsymbol{B}}_0\right|^2 = 0.
\end{aligned}
\label{eq:dispersion_relation_induced_Brillouin_charged_tenkai}
\end{equation}
In the strong coupling regime, the $\omega^3$ term dominates over the other terms. In particular, the assumptions $\omega^2 \gg \omega_{\mathrm{p}}^2 \cos^2\theta_{kB}$ and the relative magnitudes of the remaining terms allow us to neglect the subdominant terms. Consequently, Eq.~\eqref{eq:dispersion_relation_induced_Brillouin_charged_tenkai} takes the similar form as the strong coupling dispersion relations for the ordinary and neutral modes (see Eqs.~\eqref{eq:dispersion_relation_induced_Brillouin_nomagnetic_tenkai} and \eqref{eq:dispersion_relation_induced_Brillouin_Neutral_tenkai}). 

Following the same procedure as in the ordinary and neutral modes, the imaginary part of the density fluctuation frequency is given by
\begin{equation}
\begin{aligned}
     \text{Im}~&\omega = \frac{\sqrt{3}}{2} \left(\frac{\omega_{0}}{\omega_{\mathrm{c}}}\right)^{\frac{2}{3}}\\
     &\times\left(\frac{a_{\mathrm{e}}^{2}\omega_{\mathrm{p}}^{2}\omega_{0}}{2}\frac{(1-\mu^{2})(1-\nu)\left|\bm{n} \cdot \hat{\bm{B}}_{0}\right|^{2}\cos^{2}\theta_{kB}}{2}\right)^{\frac{1}{3}}.   
\end{aligned}
\end{equation}
This maximum growth rate is achieved under the condition $(\mu,\,\nu,\,\cos\theta_{kB},\,\left|\bm{n} \cdot \hat{\bm{B}}_{0}\right|)=(0,-1,\pm1,1)$, as described in Eq.~\eqref{eq:maximum_growth_angle_condition_charged_Brillouin}. The corresponding linear growth rate for the scattered wave energy is expressed as
\begin{equation}
\left(t_{\mathrm{B},\text{charged}}^{\max}\right)^{-1}  = \sqrt{3}\left(\frac{a_{\text{e}}^{2}\omega_{\mathrm{p}}^{2}\omega_{0}}{2}\right)^{\frac{1}{3}}\left(\frac{\omega_{0}}{\omega_{\mathrm{c}}}\right)^{\frac{2}{3}}.
\end{equation}
The maximum growth wavenumber is \nIshiura{(c.f. Eq. \eqref{eq:induced_Brillouin_neutral_wave_vector})}
\begin{equation}
    k_{\text{max}}\simeq2k_0.
\end{equation}

\subsection{Derivation of SRS in the Intermediate and High Density Regimes (Charged mode)}
\label{subsec:weak_coupling_Raman_charged}
In the intermediate and high density regimes, SRS becomes the dominant instability when the resonance condition and the condition for negligible Landau damping of the Langmuir wave~\eqref{eq:resonance_condition_raman} are satisfied in the dispersion relation for SBS~ \eqref{eq:dispersion_relation_neutral_tochuu_Raman}. \nIshiura{The applicability of the same dispersion relation as SBS is justified by the resonance condition, which implies $|\text{Re}~\omega| \gg k_{\|} v_{\text{th}}$, so that the $|\zeta|\gg 1$ expansion to the dispersion relation of the charged mode \eqref{eq:dispersion_relation_charged_Brillouin2} is valid.}

To derive the SRS growth rate under the weak coupling condition \eqref{eq:weak_coupling_condition}, we express the longitudinal plasma wave frequency as
\begin{equation}
\omega = -\omega_{\mathrm{p}}|\cos \theta_{kB}| + \text{i} \gamma.
\label{eq:omega_density_fluctuation_decomposition_Brillouin}
\end{equation}
The weak coupling condition~\eqref{eq:weak_coupling_condition} combined with the resonance condition~\eqref{eq:resonance_condition_raman} is given by
\begin{equation}
\gamma \ll \omega_{\mathrm{p}}|\cos \theta_{kB}|.
\label{eq:weak_coupling_condition_induced_Raman_scattering1}
\end{equation}
Substituting Eq.~\eqref{eq:omega_density_fluctuation_decomposition_Brillouin} into the SBS dispersion relation~\eqref{eq:dispersion_relation_neutral_tochuu_Raman}, we obtain
\begin{equation}
\begin{aligned}
&\left(2 \text{i} \gamma \omega_{\mathrm{p}}|\cos \theta_{kB}| + \gamma^{2}\right)\\
&\times\left\{-\omega_{\mathrm{p}}|\cos \theta_{kB}|\frac{c^2}{v_{\text{A}}^2} + \text{i} \gamma\frac{c^2}{v_{\text{A}}^2} - \frac{c^{2}\left(k^{2} + 2 \bm{k}_{0} \cdot \bm{k}\right)}{2 \omega_{0}}\right\} \\
&+ \frac{1}{8} \frac{a_{\mathrm{e}}^{2} \omega_{\mathrm{p}}^{2} c^{2} k^{2}}{\omega_{0}}\left(\frac{\omega_{0}}{\omega_{\mathrm{c}}}\right)^{2}\cos^{2} \theta_{kB}\left(1 - \mu^{2}\right)\left|\bm{n} \cdot \hat{\bm{B}}_{0}\right|^{2} = 0.
\end{aligned}
\end{equation}
Under the weak coupling condition~\eqref{eq:weak_coupling_condition_induced_Raman_scattering1}, the $\gamma^3$ term can be neglected, so the equation simplifies to
\begin{equation}
\begin{aligned}
&-\left\{3 \omega_{\mathrm{p}}|\cos \theta_{kB}|\frac{c^2}{v_{\text{A}}^2} 
+ \frac{c^{2}\left(k^{2} + 2 \bm{k}_{0} \cdot \bm{k}\right)}{2 \omega_{0}}\right\} 
\gamma^{2} \\
&+ \frac{1}{8} \frac{a_{\mathrm{e}}^{2} \omega_{\mathrm{p}}^{2} c^{2} k^{2}}{\omega_{0}}
\left(\frac{\omega_{0}}{\omega_{\mathrm{c}}}\right)^{2}\cos^2 \theta_{kB}
\left(1 - \mu^{2}\right)\left|\bm{n} \cdot \hat{\bm{B}}_{0}\right|^{2} \\
-& \text{i}\left\{2 \gamma \omega_{\mathrm{p}}^2\cos^2 \theta_{kB}\frac{c^2}{v_{\text{A}}^2} 
+ 2 \gamma \omega_{\mathrm{p}}|\cos \theta_{kB}| 
\frac{c^{2}\left(k^{2} + 2 \bm{k}_{0} \cdot \bm{k}\right)}{2 \omega_{0}}\right\} = 0.
\end{aligned}
\label{eq:weak_Raman_tenkai}
\end{equation}
The growth wavenumber is obtained by setting the imaginary part to zero, while the real part determines the linear growth rate. From the imaginary part, we find
\begin{equation}
\frac{c^{2}\left(k^{2} + 2 \bm{k}_{0} \cdot \bm{k}\right)}{2 \omega_{0}} = -\omega_{\mathrm{p}}|\cos \theta_{kB}|\frac{c^2}{v_{\text{A}}^2}.
\label{eq:Raman_imaginary_zero}
\end{equation}
Using Eq.~\eqref{eq:Raman_imaginary_zero} and setting the real part of Eq.~\eqref{eq:weak_Raman_tenkai} to zero, the growth rate is given by
\begin{equation}
\gamma^{2} = \frac{1}{16} \frac{a_{\mathrm{e}}^{2} \omega_{\mathrm{p}} v_{\text{A}}^2 k^{2}}{\omega_{0} } |\cos \theta_{kB}|\left(\frac{\omega_{0}}{\omega_{\mathrm{c}}}\right)^{2}\left(1 - \mu^{2}\right)\left|\bm{n} \cdot \hat{\bm{B}}_{0}\right|^{2}.
\end{equation}
Applying the wavenumber approximation from Eq.~\eqref{eq:wave_number_approximation_x-mode}, the linear growth rate becomes
\begin{equation}
\gamma^{2} \sim \frac{1}{8} \omega_{0} a_{\mathrm{e}}^{2}\left(\frac{\omega_{0}}{\omega_{\mathrm{c}}}\right)^{2} \omega_{\mathrm{p}} |\cos \theta_{kB}|(1 - \nu)\left(1 - \mu^{2}\right)\left|\bm{n} \cdot \hat{\bm{B}}_{0}\right|^{2}.
\label{eq:growth_rate_Raman_totyuu_kakudo}
\end{equation}
This growth rate is maximized under the condition $(\mu,\,\nu,\,\cos\theta_{kB},\,|\bm{n} \cdot \hat{\bm{B}}_{0}|)=(0,-1,\pm1,1)$, as given by Eq.~\eqref{eq:maximum_growth_angle_condition_charged_Raman}. Physically, this corresponds to a situation where the incident EM wave is scattered by $180^\circ$ (backward scattering) with a $90^\circ$ rotation of the scattered polarization with respect to the incident wave. The maximum linear growth rate is then
\begin{equation}
\left(t_{\text{R}}^{\text{max}}\right)^{-1} \equiv 2 \gamma = a_{\mathrm{e}} \frac{\omega_{0}}{\omega_{\mathrm{c}}}\left(\omega_{0} \omega_{\mathrm{p}}\right)^{\frac{1}{2}}.
\end{equation}
The wavenumber for maximum growth is found from Eq.~\eqref{eq:Raman_imaginary_zero} by solving the quadratic equation
\begin{equation}
k^{2} - 2 k_{0} k + \frac{2 \omega_{0} \omega_{\mathrm{p}}}{v_{\text{A}}^2} \simeq 0.
\end{equation}
This yields
\begin{equation}
k_{\pm} \simeq k_{0}\left(1 \pm \sqrt{1 - 2 \frac{\omega_{\mathrm{p}}}{\omega_{0}}}\right).
\end{equation}
For backward scattering, the wavenumber of the scattered wave is given by $k_1 = k - k_0$. When $k = k_{-}$, the resulting $k_1$ becomes negative, which is unphysical. Therefore, we adopt $k_{\text{max}} = k_{+}$ to ensure that $k_1 > 0$ is satisfied. 

\subsection{Competition between SRS and ICS in the Intermediate Density Regime (Charged mode)}
\label{subsec:raman_compton_competition_charged}
\niShiura{In the intermediate density regime, as defined by Eq.~\eqref{eq:induced_Raman_condition_Brillouin}, both ICS and SRS can be excited. However, SRS consistently exhibits a higher growth rate than ICS. The ratio of the maximum linear growth rates for Debye-screened ICS~\eqref{eq:induced_Compton_Debye_screening_Brillouin} and SRS~\eqref{eq:growth_rate_of_induced_Raman_weak_coupling} is
\begin{equation}
\begin{aligned}
&\frac{\left(t_{\mathrm{C}, \text{charged}}^{\text{max}}\right)^{-1}}{\left(t_{\text{R}}^{\text{max}}\right)^{-1}}
= \sqrt{\frac{32\, \text{e}}{\pi}}\,
\frac{k_{\mathrm{B}} T_{\mathrm{e}}}{m_{\mathrm{e}} c^{2}} 
a_{\mathrm{e}}\frac{\omega_{0}}{\omega_{\mathrm{c}}}
\left(\frac{\omega_{\mathrm{p}}}{\omega_{0}}\right)^{-\frac{5}{2}}
\left(1+\frac{\omega_{\text{p}}^2}{\omega_{\text{c}}^2}\right)   \\
&\overset{\eqref{eq:weak_coupling_condition_for_Debye_Compton}}{\ll}4\sqrt{2}\left(\frac{\mathrm{e}}{\pi}\right)^{\frac{1}{4}}\left(\frac{k_{\mathrm{B}} T_{\mathrm{e}}}{m_{\mathrm{e}} c^{2}}\right)^{\tfrac{3}{4}}\left(\frac{\omega_{\mathrm{p}}}{\omega_{0}}\right)^{-\tfrac{3}{2}}\left(1+\frac{\omega_{\text{p}}^2}{\omega_{\text{c}}^2}\right)^{\tfrac{3}{4}} \\
&\overset{\eqref{eq:charged_mode_instability_thin_condition_Brillouin_Compton}}{\ll}\left(\frac{2\mathrm{e}}{\pi}\right)^{\tfrac{1}{4}}\sim1.1.
\end{aligned}
\end{equation}
Thus, provided that the weak coupling condition~\eqref{eq:weak_coupling_condition_for_Debye_Compton} and the intermediate- and high-density condition~\eqref{eq:charged_mode_instability_thin_condition_Brillouin_Compton} are well satisfied, the ratio remains well below unity, 
\[
\left(t_{\mathrm{C}, \text{charged}}^{\text{max}}\right)^{-1} \ll \left(t_{\text{R}}^{\text{max}}\right)^{-1},
\] 
showing that the growth rate of ICS is always smaller than that of SRS.}

\subsection{Derivation of Small-Angle SRS in the Low Density Regime (Charged mode)}
\label{subsec:small_angle_Raman_charged}
This section derives the linear growth rate for small-angle SRS of the charged mode in the low density regime, as defined by Eq.~\eqref{eq:charged_mode_instability_thin_condition_Brillouin}. We assume a Langmuir wave propagating at an angle $\theta_{kB}$ with respect to the background magnetic field,
\begin{equation}
\omega = -\omega_{\mathrm{p}} |\cos \theta_{kB}|.
\end{equation}
To neglect Landau damping for the Langmuir wave, its phase velocity must be much larger than the thermal velocity. This requirement is expressed as
\begin{equation}
k \ll \frac{1}{4}\lambda_{\mathrm{De}}^{-1}\left|\cos \theta_{kB}\right|=\frac{\sqrt{2}\omega_{\mathrm{p}}}{4 v_{\mathrm{th}}} |\cos \theta_{kB}|.
\label{eq:upper_limit_wave_number_small_angle}
\end{equation}
The factor of $4$ represents a conventional factor \citep{1994ApJ...422..304T,2021Univ....7...56L}. This inequality defines an upper limit for the wave number of the density fluctuation:
\begin{equation}
k_{\text{L}} \equiv \frac{\sqrt{2}\omega_{\mathrm{p}}}{4 v_{\mathrm{th}}} |\cos \theta_{kB}|.
\end{equation}
\ioka{Using the energy-momentum conservation condition for density fluctuations in
Eq.~\eqref{eq:energy_momentum_conservation_Compton_Brillouin} and the dispersion
relations for the incident and scattered waves, $\omega_0 \simeq k_0 v_{\mathrm A}$
and $\omega_1 \simeq k_1 v_{\mathrm A}$, the wavenumber of the density fluctuation is
given by
\begin{equation}
\begin{aligned}
k^{2} &= \frac{1}{v_{\mathrm A}^2}\left\{\left(\omega_{1} - \omega_{0}\right)^{2} 
+ 2(1 - \nu) \omega_{0} \omega_{1}\right\} \\
&= \frac{\omega_{0}^{2}}{v_{\mathrm A}^2}
\left[
2(1-\nu)
+ \mathcal{O}\left(\frac{\omega^2}{\omega_{0}\omega_{1}}\right)
\right].
\end{aligned}
\label{eq:wave_number_approximation_x-mode}    
\end{equation}}For $180^{\circ}$ backward scattering, Eq.~\eqref{eq:wave_number_approximation_x-mode} gives $k \sim 2\omega_{0}/v_{\mathrm{A}}$. In this case, the upper limit on the allowed incident wave frequency, $\omega_{0\text{L}}$, is thus
\begin{equation}
\omega_{0\text{L}} \equiv \frac{1}{2} k_{\text{L}} v_{\mathrm{A}} 
= \frac{\sqrt{2}v_{\mathrm{A}}}{8 v_{\mathrm{th}}} \omega_{\mathrm{p}} |\cos \theta_{kB}|.
\end{equation}
SRS is limited to sidescattering when $\omega_0 > \omega_{0\text{L}}$.

When only small-angle SRS is allowed, the angular parameter $\nu$ is subject to the following constraint (see Eq. \eqref{eq:wave_number_approximation_x-mode}):
\begin{equation}
k_{\text{L}}^{2} = 4 \frac{\omega_{0\text{L}}^{2}}{v_{\mathrm{A}}^{2}} 
> 2(1-\nu) \frac{\omega_{0}^{2}}{v_{\mathrm{A}}^{2}}.
\end{equation}
This provides an upper bound on $1-\nu$, which determines the maximum scattering angle. The physical interpretation is discussed in Sec.~\ref{subsec:physical_interpretation_raman_scattering}. The maximal value is given by
\begin{equation}
\begin{aligned}
(1-\nu)_{\mathrm{max}}
&= 2\left(\frac{\omega_{0\text{L}}}{\omega_{0}}\right)^{2} \\
&= \frac{m_{\mathrm{e}} c^{2}}{32\,k_{\mathrm{B}} T_{\mathrm{e}}}
   \left(\frac{\omega_{\mathrm{p}}}{\omega_{0}}\right)^{2}
   \left(1 + \frac{\omega_{\mathrm{p}}^{2}}{\omega_{\mathrm{c}}^{2}}\right)^{-1}
   \cos^{2} \theta_{kB}.
\end{aligned}
\label{eq:max_1-nu_Raman}
\end{equation}

By substituting $(1-\nu)_{\mathrm{max}}$ into Eq.~\eqref{eq:growth_rate_Raman_totyuu_kakudo}, the growth rate can be written as
\begin{equation}
\gamma^{2} =
\frac{1}{8} \omega_{0} a_{\mathrm{e}}^{2}
\left(\frac{\omega_{0}}{\omega_{\mathrm{c}}}\right)^{2}
\omega_{\mathrm{p}}
\frac{m_{\mathrm{e}} c^{2}}{32 k_{\mathrm{B}} T_{\mathrm{e}}}
\left(\frac{\omega_{\mathrm{p}}}{\omega_{0}}\right)^{2}
\left(1 + \frac{\omega_{\mathrm{p}}^{2}}{\omega_{\mathrm{c}}^{2}}\right)^{-1}
f,
\label{eq:growth_rate_small_raman_f}
\end{equation}
where $f$ is a function containing only the angular dependence of the growth rate, defined as
\begin{equation}
f \equiv |\cos \theta_{kB}|^3
\left(1 - \mu^{2}\right)
\left| \bm{n} \cdot \hat{\bm{B}}_{0} \right|^{2}.
\label{eq:angular_dependence_small_angle_Raman}
\end{equation}
The maximization condition for $f$ is analyzed in the following section.

\subsection{Maximum Growth Angular Parameter for Small-Angle SRS (Charged mode)}
\label{subsec:small_angle_Raman_angle_max_charged}
The maximum value of the angular dependence $f$ in Eq.~\eqref{eq:angular_dependence_small_angle_Raman} appearing in the linear growth rate for small-angle SRS can be derived analytically.
\nIshiura{In this study, the following simplified set of assumptions is adopted for the polarization and propagation direction of the incident and scattered waves when evaluating the maximum growth angle.  
\begin{enumerate}[label=(\roman*)]
    \item The incident wave is assumed to be an X-mode EM wave, with its electric field component perpendicular to the background magnetic field, and propagating at an arbitrary angle, $\theta_0$, to the background  magnetic field.  
    \item The scattered wave is assumed to be a transverse wave whose propagation direction is nearly identical to that of the incident wave (later defined as $\theta \ll 1$). In magnetized $e^\pm$ pair plasma, such a transverse wave is not a strict linear eigenmode. Therefore, the angle parameter for maximum growth obtained under this assumption may differ by a factor from the true value.
\end{enumerate}
}

The coordinate system is defined as follows. The wave vector and electric field of the incident EM wave are given by
\begin{equation}
\bm{k}_{0} = k_{0} \bm{e}_{z},\qquad \bm{E}_{0} = E_{0} \bm{e}_{y},
\end{equation}
and the background magnetic field is expressed as
\begin{equation}
\bm{B}_{0} = B_{0} \sin\theta_{0} \bm{e}_{x} + B_{0} \cos\theta_{0} \bm{e}_{z}.
\end{equation}
The scattered wave vector is parameterized as
\begin{equation}
\bm{k}_{1} = k_{1} \bm{e}_{r} = k_{1} ( \sin\theta \cos\phi\, \bm{e}_{x} + \sin\theta \sin\phi\, \bm{e}_{y} + \cos\theta\, \bm{e}_{z} ),
\end{equation}
and the wave vector of the density fluctuation is
\begin{equation}
\bm{k} = k ( \cos\phi\, \bm{e}_{x} + \sin\phi\, \bm{e}_{y} )+\mathcal{O}\left(\frac{\omega}{\omega_1},~\theta^2\right).
\end{equation}
Thus, the angular parameter $\cos\theta_{kB}$ can be written as
\begin{equation}
\cos\theta_{kB} = \cos\phi\, \sin\theta_{0}.
\label{eq:costheta_kB_angleparameter}
\end{equation}

We next consider the two cases for the polarization of the scattered wave, corresponding to $\bm{e}_\phi$ and $\bm{e}_\theta$, for the small angle scattering $\theta\ll1$. For the $\bm{e}_\phi$-polarized scattered wave, the electric field and angular parameters are given by
\begin{equation}
\nisHiura{\begin{aligned}
\bm{E}_{1} &= E_{1} \bm{e}_{\phi} = E_{1} ( -\sin\phi\, \bm{e}_{x} + \cos\phi\, \bm{e}_{y} ),\\
\mu &= \left|\cos\phi\right|,    
\end{aligned}}
\end{equation}
\begin{equation}
\bm{n}=\frac{\bm{E}_{1} \times \bm{E}_{0}}{E_{1} E_{0}} = -\sin\phi\, \bm{e}_{z} \quad \Rightarrow \quad \left| \bm{n} \cdot \hat{\bm{B}_{0}} \right| = \left|\cos\theta_{0}\sin \phi\right|.
\end{equation}
The corresponding angular dependence of the growth rate in Eq. \eqref{eq:angular_dependence_small_angle_Raman} is then expressed as
\begin{equation}
f_{\phi} = |\cos^{3}\phi\, \sin^{3}\theta_{0}\,| \sin^{4}\phi\, \cos^{2}\theta_{0} \leq \frac{864}{8575\sqrt{35}}.
\label{eq:fphi_raman}
\end{equation}
The maximum is achieved for
\begin{equation}
\theta_{0} = \arcsin\sqrt{\frac{3}{5}},\quad \phi = \arccos\sqrt{\frac{3}{7}}.
\end{equation}
Thus, the growth rate is maximized when the incident wave propagates at $\sim 51^\circ$ relative to the background magnetic field and the scattered wave is $\bm{e}_\phi$-polarized.

For the $\bm{e}_\theta$-polarized case, the electric field and angular parameters are given by
\begin{equation}
\nisHiura{\begin{aligned}
 \bm{E}_{1}& = E_{1} \bm{e}_{\theta} = E_{1} ( \cos\theta \cos\phi\, \bm{e}_{x} + \cos\theta \sin\phi\, \bm{e}_{y} - \sin\theta\, \bm{e}_{z} ),\\
 \mu &= \left|\cos\theta\sin\phi\right|=\left|\sin\phi\right|+\mathcal{O}(\theta^2),   
\end{aligned}}
\end{equation}
\begin{equation}
\begin{aligned}
\bm{n}&=\frac{\bm{E}_{1} \times \bm{E}_{0}}{E_{1} E_{0}} = \sin\theta\, \bm{e}_{x} + \cos\theta\cos\phi\, \bm{e}_{z}\\ &\Rightarrow \quad \left| \bm{n} \cdot \hat{\bm{B}}_{0} \right| =\cos\theta_0\cos\phi+\mathcal{O}(\theta^2).    
\end{aligned}
\end{equation}
The corresponding angular dependence in Eq. \eqref{eq:angular_dependence_small_angle_Raman} is then
\begin{equation}
f_{\theta} = |\cos^{7}\phi\, \sin^{3}\theta_{0}|\, \cos^{2}\theta_0 \leq \frac{6}{25} \sqrt{\frac{3}{5}},
\label{eq:f_theta_small_raman}
\end{equation}
where the maximum is realized when
\begin{equation}
\phi =0,\quad \theta_{0} = \arcsin\sqrt{\frac{3}{5}}.
\label{eq:phi_polarization_small_angle}
\end{equation}
That is, for $\bm{e}_\theta$ polarization, the growth rate is maximized for an incident angle of $51^\circ$ relative to the magnetic field.

Comparing Eqs. \eqref{eq:fphi_raman} and \eqref{eq:f_theta_small_raman}, $\bm{e}_\theta$ polarization yields a larger growth rate. Therefore, substituting $f_{\theta}$ in Eq.~\eqref{eq:f_theta_small_raman} into the linear growth rate~\eqref{eq:growth_rate_small_raman_f}, the maximum linear growth rate of the scattered wave energy is given by
\begin{equation}
    \left(t_{\text{R}}^{\text{max}}\right)^{-1}\sim0.30\, a_{\mathrm{e}}\, \frac{\omega_0}{\omega_{\mathrm{c}}}
\left( \omega_0 \omega_{\mathrm{p}} \right)^{\frac{1}{2}}
\left( \frac{m_{\mathrm{e}} c^2}{32\, k_{\mathrm{B}} T_{\mathrm{e}}} \right)^{\frac{1}{2}}
\frac{\omega_{\mathrm{p}}}{\omega_0}
\left( 1 + \frac{\omega_{\mathrm{p}}^2}{\omega_{\mathrm{c}}^2} \right)^{-\frac{1}{2}}.
\label{eq:SRS_small_angele_Unified}
\end{equation}
as illustrated in Eq.~\eqref{eq:stimulated_Raman_growth_rate_small_angle}.

\nocite{*}
\newpage
\bibliographystyle{apsrev4-2}
\bibliography{apssamp}

\end{document}